\documentclass[a4paper,runningheads]{llncs}

\ifdefined\ifdraft%
\let\oldifdraft\ifdraft
\let\ifdraft\relax
\usepackage{templatetools}
\let\ifdraft\oldifdraft
\else%
    \usepackage{templatetools}
\fi%

\IfPackageNotLoaded{inputenc}{\usepackage[utf8]{inputenc}}

\IfPackageNotLoaded{fontenc}{\usepackage[T1]{fontenc}}

\IfPackageNotLoaded{babel}{\usepackage[english]{babel}}

\IfPackageNotLoaded{color}{\usepackage{color}}

\IfPackageNotLoaded{xcolor}{\usepackage[dvipsnames,hyperref]{xcolor}}

\IfPackageNotLoaded{graphicx}{\usepackage{graphicx}}

\IfPackageNotLoaded{geometry}{\usepackage[pass]{geometry}}

\IfPackageNotLoaded{rotating}{\usepackage{rotating}}

\IfPackageNotLoaded{lmodern}{\usepackage{lmodern}}

\IfPackageNotLoaded{float}{\usepackage{float}}

\IfPackageNotLoaded{bm}{\usepackage{bm}}

\IfPackageNotLoaded{xparse}{\usepackage{xparse}}

\IfPackageNotLoaded{ifthen}{\usepackage{ifthen}}
\IfPackageNotLoaded{ifdraft}{\usepackage{ifdraft}}

\IfPackageNotLoaded{calc}{\usepackage{calc}}

\IfPackageNotLoaded{silence}{\usepackage{silence}}

\IfPackageNotLoaded{etoolbox}{\usepackage{etoolbox}}

\IfPackageNotLoaded{proof}{\usepackage{proof}}

\IfPackageNotLoaded{mathpartir}{\usepackage{mathpartir}}
\IfPackageNotLoaded{mathtools}{\usepackage{mathtools}}
\IfPackageNotLoaded{amsmath}{%
    \let\subseteq\relax
    \usepackage{amsmath}
}

\IfPackageNotLoaded{amsthm}{%

    \usepackage{amsthm}
}

\IfPackageNotLoaded{thmtools}{\usepackage{thmtools}}
\IfPackageNotLoaded{thm-restate}{\usepackage{thm-restate}}

\IfPackageNotLoaded{longtable}{\usepackage{longtable}}

\IfPackageNotLoaded{enumitem}{\usepackage[inline]{enumitem}}
\SetLabelAlign{parright}{\parbox[t]{\labelwidth}{\raggedleft#1}}
\setlist[description]{itemsep=0.5ex,labelsep=2ex,parsep=0ex,listparindent=0ex}%
\setlist[itemize]{topsep=0.5pt,itemsep=0.5pt}%
\setlist[enumerate]{topsep=0.5pt,itemsep=0.5pt}%

\IfPackageNotLoaded{tasks}{\usepackage{tasks}}
\settasks{style=enumerate,ref=\arabic*,after-item-skip=0.5ex}

\IfPackageNotLoaded{algorithm2e}{\usepackage[ruled,vlined,linesnumbered,commentsnumbered,resetcount]{algorithm2e}}

\IfPackageNotLoaded{listings}{\usepackage{listings}}

\IfPackageNotLoaded{wrapfig}{\usepackage{wrapfig}}

\IfPackageNotLoaded{xfrac}{\usepackage{xfrac}}

\IfPackageNotLoaded{delarray}{\usepackage{delarray}}

\IfPackageNotLoaded{tikz}{\usepackage{tikz}}
\newcounter{tmkcount}
\usetikzlibrary{shapes,trees,calc,tikzmark,positioning}
\tikzset{
    use tikzmark/.style={
            remember picture,
            overlay,
            execute at end picture={
                    \stepcounter{tmkcount}
                },
        },
    tikzmark suffix={-\thetmkcount}
}

\IfPackageNotLoaded{titlesec}{%
    \let\llncssubparagraph\subparagraph
    \let\subparagraph\paragraph
    \usepackage{titlesec}
    \let\subparagraph\llncssubparagraph
}

\IfPackageNotLoaded{multicol}{\usepackage{multicol}}
\IfPackageNotLoaded{subfig}{\usepackage[caption=false]{subfig}}
\IfPackageNotLoaded{amsfonts}{\usepackage{amsfonts}}

\IfPackageNotLoaded{amstext}{\usepackage{amstext}}

\IfPackageNotLoaded{mathabx}{%
    \let\olddegree\degree
    \let\degree\relax
    \usepackage{mathabx}
    \let\degree\olddegree
}

\IfPackageNotLoaded{amssymb}{\usepackage{amssymb}}
\let\emptyset\varnothing

\IfPackageNotLoaded{amsbsy}{\usepackage{amsbsy}}

\IfPackageNotLoaded{stmaryrd}{\usepackage{stmaryrd}}

\IfPackageNotLoaded{AnonymousPro}{\usepackage[ttdefault=true]{AnonymousPro}}

\IfPackageNotLoaded{orcidlink}{\usepackage{orcidlink}}

\IfPackageNotLoaded{hyperref}{\usepackage{hyperref}}
\hypersetup{
    linktocpage=false,
    colorlinks=true,
    bookmarksopen=false,
    linkcolor=blue,
    anchorcolor=red,
    citecolor=blue,
    filecolor=magenta,
    urlcolor=blue,
    citebordercolor={0 0 0},
    filebordercolor={1 0 1},
    linkbordercolor={0 0 0},
    menubordercolor={0 0 0},
    urlbordercolor={0 0 0},
    pdfstartpage=1
}

\IfPackageNotLoaded{bookmark}{\usepackage[depth=5,atend]{bookmark}}

\IfPackageNotLoaded{cleveref}{\usepackage{cleveref}}

\IfPackageNotLoaded{tcolorbox}{\usepackage[theorems,xparse]{tcolorbox}}

\IfPackageNotLoaded{soul}{\usepackage{soul}}
\let\oldst\st
\let\st\relax
\IfPackageNotLoaded{ulem}{\usepackage{ulem}}
\IfPackageNotLoaded{doi}{\usepackage{doi}}

\usetikzlibrary{arrows,shapes,automata,backgrounds,petri,positioning,fit,calc,
  decorations.markings,shadows,fadings,patterns,decorations.pathreplacing}

\newcommand{\tcolorNew}[0]{green!10}
\newcommand{\tcolorMod}[0]{orange!10}
\newcommand{\tcolorOld}[0]{gray!0}

\SetSymbolFont{stmry}{bold}{U}{stmry}{b}{n}
\SetSymbolFont{stmry}{bold}{U}{stmry}{m}{n}

\titlespacing{\subsection}{0pt}{2ex plus1ex minus1ex}{1ex plus0.5ex minus0.5ex}
\titlespacing{\subsubsection}{0pt}{1.5ex plus0.5ex minus0.5ex}{1ex plus1ex minus1pt}
\titlespacing{\paragraph}{0pt}{0.5ex}{2ex plus1ex minus1pt}

\makeatletter
\let\oldendrestatable\endthmt@restatable
\renewcommand\endthmt@restatable{\oldendrestatable\!}
\makeatother

\makeatletter
\def\restatable@space@setup{%
  \restatable@preskip=1ex plus0.5ex minus0.5ex%
  \restatable@postskip=2ex plus1ex minus1ex%
}
\makeatother

\makeatletter
\def\definition@space@setup{%
  \definition@preskip=2ex plus1ex minus1ex%
  \definition@postskip=2ex plus1ex minus1ex%
}
\makeatother

\makeatletter
\def\theorem@space@setup{%
  \theorem@preskip=2ex plus1ex minus1ex%
  \theorem@postskip=1ex plus0.5ex minus0.5ex%
}
\makeatother

\makeatletter
\def\lemma@space@setup{%
  \lemma@preskip=2ex plus1ex minus1ex%
  \lemma@postskip=1ex plus0.5ex minus0.5ex%
}
\makeatother

\makeatletter
\def\proposition@space@setup{%
  \proposition@preskip=2ex plus1ex minus1ex%
  \proposition@postskip=1ex plus0.5ex minus0.5ex%
}
\makeatother

\makeatletter
\def\note@space@setup{%
  \note@preskip=2ex plus1ex minus1ex%
  \note@postskip=2ex plus1ex minus1ex%
}
\makeatother

\let\oldcite=\cite
\renewcommand\cite[2][]{\ifthenelse{\equal{#2}{NEEDED}}{\!{\highlight{\textbf{[?]}}}}{\ifthenelse{\equal{#1}{}}{\oldcite{#2}}{\oldcite[#1]{#2}}}}

\let\oldref=\ref
\renewcommand\ref[1]{\ifthenelse{\equal{#1}{NEEDED}}{\!{\highlight[green]{\textbf{[?]}}}}{\oldref{#1}}}

\let\oldcref=\cref
\renewcommand\cref[1]{\ifthenelse{\equal{#1}{NEEDED}}{\!{\highlight[green]{\textbf{[?]}}}}{\oldcref{#1}}}

\NewDocumentCommand{\Strike}{s m}{%
	\IfBooleanTF{#1}{\sout{#2}}{\oldst{#2}}
}

\NewDocumentCommand{\RFC}{s}{\texttt{RFC}\IfBooleanT{#1}{\ {5321}}}
\newcommand{\picalc}[0]{$\pi$-calculus}

\NewDocumentCommand{\isWF}{s}{%
    \IfBooleanTF{#1}{$}{~\text{is\ }}
    \mathtt{WF}%
    \IfBooleanTF{#1}{$}{}
}

\NewDocumentCommand{\isFE}{s t! t?}{%
    \IfBooleanTF{#1}{$}{~\text{is\ }}
    \mathtt{FE}\IfBooleanTF{#2}{\TypSend}{\IfBooleanT{#3}{\TypRecv}}%
    \IfBooleanTF{#1}{$}{}
}
\NewDocumentCommand{\isLE}{s t! t?}{%
    \IfBooleanTF{#1}{$}{~\text{is\ }}
    \mathtt{LE}\IfBooleanTF{#2}{\TypSend}{\IfBooleanT{#3}{\TypRecv}}%
    \IfBooleanTF{#1}{$}{}
}

\crefname{task}{condition}{conditions}
\crefformat{task}{condition~#2#1#3}
\crefmultiformat{task}{conditions~#2#1#3}{ and~#2#1#3}{, #2#1#3}{ and~#2#1#3}
\crefrangeformat{task}{conditions~#3#1#4 to~#5#2#6}

\Crefname{task}{Condition}{Conditions}
\Crefformat{task}{Condition~#2#1#3}
\Crefmultiformat{task}{Conditions~#2#1#3}{ and~#2#1#3}{, #2#1#3}{ and~#2#1#3}
\Crefrangeformat{task}{Conditions~#3#1#4 to~#5#2#6}

\newlist{inductivecase}{enumerate}{4}

\setlist[inductivecase,1]{label={\itshape{Case~\arabic*.}},ref={\arabic*},topsep=1ex,itemsep=1ex,parsep=0.5ex,leftmargin=5ex,itemindent=4ex,left=0ex}

\setlist[inductivecase,2]{label={\itshape{\roman*}.},ref={\arabic{inductivecasei}.\roman*},topsep=0.5ex,itemsep=0.75ex,parsep=0.5ex,leftmargin=4ex,itemindent=0ex,left=-2ex}

\setlist[inductivecase,3]{label={\itshape{\alph*}.},ref={\arabic{inductivecasei}.\roman{inductivecaseii}.\alph*},topsep=0.5ex,itemsep=0.75ex,parsep=0.5ex,leftmargin=4ex,itemindent=0ex,left=-2ex}

\crefname{inductivecasei}{case}{cases}
\crefformat{inductivecasei}{case~#2#1#3}
\crefmultiformat{inductivecasei}{cases~#2#1#3}{ and~#2#1#3}{, #2#1#3}{ and~#2#1#3}
\crefrangeformat{inductivecasei}{cases~#3#1#4 to~#5#2#6}

\crefname{inductivecaseii}{subcase}{subcases}
\crefformat{inductivecaseii}{subcase~#2#1#3}
\crefmultiformat{inductivecaseii}{subcases~#2#1#3}{ and~#2#1#3}{, #2#1#3}{ and~#2#1#3}
\crefrangeformat{inductivecaseii}{subcases~#3#1#4 to~#5#2#6}

\crefname{inductivecaseiii}{subcase}{subcases}
\crefformat{inductivecaseiii}{subcase~#2#1#3}
\crefmultiformat{inductivecaseiii}{subcases~#2#1#3}{ and~#2#1#3}{, #2#1#3}{ and~#2#1#3}
\crefrangeformat{inductivecaseiii}{subcases~#3#1#4 to~#5#2#6}

\Crefname{inductivecasei}{Case}{Cases}
\Crefformat{inductivecasei}{Case~#2#1#3}
\Crefmultiformat{inductivecasei}{Cases~#2#1#3}{ and~#2#1#3}{, #2#1#3}{ and~#2#1#3}
\Crefrangeformat{inductivecasei}{Cases~#3#1#4 to~#5#2#6}

\Crefname{inductivecaseii}{Subcase}{Subcases}
\Crefformat{inductivecaseii}{Subcase~#2#1#3}
\Crefmultiformat{inductivecaseii}{Subcases~#2#1#3}{ and~#2#1#3}{, #2#1#3}{ and~#2#1#3}
\Crefrangeformat{inductivecaseii}{Subcases~#3#1#4 to~#5#2#6}

\Crefname{inductivecaseiii}{Subcase}{Subcases}
\Crefformat{inductivecaseiii}{Subcase~#2#1#3}
\Crefmultiformat{inductivecaseiii}{Subcases~#2#1#3}{ and~#2#1#3}{, #2#1#3}{ and~#2#1#3}
\Crefrangeformat{inductivecaseiii}{Subcases~#3#1#4 to~#5#2#6}

\NewDocumentCommand{\NewCase}{e| o}{\IfValueT{#2}{\IfValueT{#1}{#1~}{#2}:}\hspace{1.5ex}}

\newcounter{proof}
\AtBeginEnvironment{lemma}{\setcounter{proof}{\thelemma}}

\crefalias{proof}{lemma}

\makeatletter
\def\colorstart{\edef\colorend{\pdfliteral{\current@color}}}
\makeatother
\newcommand{\highlight}[2][magenta]{\color{#1}\colorstart {#2} \colorend}

\NewDocumentCommand{\TODO}{s o}{%
   \IfBooleanT{#1}{$} 
   {\highlight[red]{ToDo\IfNoValueF{#2}{\footnote{\highlight[red]{ToDo: #2}}}}}\!
   \IfBooleanT{#1}{$} 
}
\NewDocumentCommand{\REDO}{s o}{%
   \IfBooleanT{#1}{$} 
   {\highlight[orange]{Redo\IfNoValueF{#2}{\footnote{\highlight[orange]{Redo: #2}}}}}\!
   \IfBooleanT{#1}{$} 
}
\NewDocumentCommand{\NOTE}{s o}{%
   \IfBooleanT{#1}{$} 
   {\highlight[teal]{Note\IfNoValueF{#2}{\footnote{\highlight[teal]{Note: #2}}}}}\!
   \IfBooleanT{#1}{$} 
}
\NewDocumentCommand{\DONE}{s o}{%
   \IfBooleanT{#1}{$} 
   {\highlight[green]{Done\IfNoValueF{#2}{\footnote{\highlight[green]{Done: #2}}}}}\!
   \IfBooleanT{#1}{$} 
}

\setlist[itemize]{topsep=0.5pt,itemsep=0.5pt}
\setlist[enumerate]{topsep=0.5pt,itemsep=0.5pt}

\NewDocumentEnvironment{inline}{s O{(\arabic{*})} t/ t+ t| t; t. t> O{\;$\rightarrow$\,}}{%
   \begin{enumerate*}[%
         label={\IfBooleanT{#1}{#2}},%
         before=\unskip{\IfBooleanF{#3}{:~}},%
         itemjoin=\discretionary{\IfBooleanTF{#8}{#9}{;}}{}{\hbox{\IfBooleanTF{#8}{#9}{;}}},%
         itemjoin*=\discretionary{
            \hbox{\IfBooleanTF{#8}{#9}{\IfBooleanTF{#6}{;}{,}}}%
         }{
            \hbox{\IfBooleanTF{#4}{~and}{\IfBooleanT{#5}{~or}}\IfBooleanTF{#6}{,}{}}%
         }{
            \hbox{\IfBooleanTF{#8}{#9}{\IfBooleanTF{#6}{;}{,}\IfBooleanTF{#4}{~and}{\IfBooleanT{#5}{~or}}\IfBooleanTF{#6}{,}{}}}%
         },%
         after=\unskip{\IfBooleanTF{#7}{\!}{.}}%
      ]%
      }{
   \end{enumerate*}\!
}

\newcounter{claim}
\let\oldclaim\claim
\let\oldendclaim\endclaim
\renewenvironment{claim}{%
   \setlength{\topsep}{0pt}%
   \setlength{\leftmargin}{1cm}%
   \setlength{\rightmargin}{1cm}%
   \setlength{\listparindent}{\parindent}%
   \setlength{\itemindent}{\parindent}%
   \setlength{\parsep}{\parskip}%
   \stepcounter{claim}%
   \oldclaim[\theclaim]%
}{%
   \oldendclaim%
}%
\AtBeginEnvironment{lemma}{\setcounter{claim}{0}}

\NewDocumentEnvironment{minieq}{s o}{%
\let\minieqoldabovedisplayskip\abovedisplayskip%
\let\minieqoldpartopsep\partopsep%
\setlength{\abovedisplayskip}{7pt plus2pt minus4pt}%
\setlength{\belowdisplayskip}{7pt plus2pt minus4pt}%
\setlength{\partopsep}{0pt}
\IfBooleanT{#1}{\begin{equation*}}\IfBooleanF{#1}{\begin{equation}}%
            }{%
            \IfBooleanF{#1}{\end{equation}}\IfBooleanT{#1}{\end{equation*}}%
\IfNoValueF{#2}{\hfill\ \emph{Note:}\;{\footnotesize #2}}%
\setlength{\abovedisplayskip}{\minieqoldabovedisplayskip}%
\setlength{\partopsep}{\minieqoldpartopsep}%
\!%
}

\NewDocumentEnvironment{desc}{s e1 e2 e3 e4}{%
\IfBooleanTF{#1}
{ 
   \begin{description}[itemsep=1.25ex,labelsep=2ex,parsep=0.5ex,listparindent=4ex]
}{ 
   \begin{description}[itemsep=0.5ex,labelsep=2ex,parsep=0ex,listparindent=0ex]
      }
      }{%
   \end{description}
}

\NewTotalTCBox{\ColBoxText}{O{\tcolorOld} m}{%
   size=title,tcbox raise base,nobeforeafter,top=2pt,bottom=0pt,left=2pt,right=2pt,boxsep=0mm,colback=#1,colframe=black!0}{{\rule{0pt}{1ex}#2}}

\NewDocumentCommand{\ColBoxNew}{s m}{%
   \IfBooleanTF{#1}{
      \ColBoxText[\tcolorNew]{#2}%
   }{
      \tcboxmath[size=title,boxsep=0mm,colback=\tcolorNew,colframe=black!0]{\begin{array}[c]{c}#2\end{array}}
   }%
}

\NewDocumentCommand{\ColBoxMod}{s m}{%
   \IfBooleanTF{#1}{
      \ColBoxText[\tcolorMod]{#2}%
   }{
      \tcboxmath[size=title,boxsep=0mm,colback=\tcolorMod,colframe=black!0]{\begin{array}[c]{c}#2\end{array}}
   }%
}

\NewDocumentCommand{\ColBoxOld}{s m}{%
   \IfBooleanTF{#1}{
      \ColBoxText[\tcolorOld]{#2}%
   }{
      \tcboxmath[size=title,boxsep=0mm,colback=\tcolorOld,colframe=black!0]{\begin{array}[c]{c}#2\end{array}}
   }%
}

\DeclareMathOperator{\mathOpEntails}{\vdash}
\DeclareMathOperator{\mathOpImplies}{\implies}

\DeclareMathOperator{\mathOpSetOr}{\bigvee}
\DeclareMathOperator{\mathOpSetUnion}{\bigcup}
\DeclareMathOperator{\mathOpImpEach}{\Longleftrightarrow}

\DeclareMathOperator{\mathOpManyAct}{\;{\longrightarrow}^{\ast}\;}

\DeclareMathOperator{\mathOpTypeChoice}{\mathtt{c}}
\DeclareMathOperator{\mathOpConst}{\delta}
\DeclareMathOperator{\mathOpReSet}{\lambda}

\DeclareMathOperator{\mathOpRecDef}{\mu}
\DeclareMathOperator{\mathOpTypRecLabel}{\alpha}
\DeclareMathOperator{\mathOpCfgRecLabel}{\mathtt{\lowercase{t}}}

\DeclareMathOperator{\mathOpProgAction}{\ell}
\DeclareMathOperator{\mathOpSiltAction}{\tau}

\DeclareMathOperator{\mathOpPrcScope}{\nu}
\DeclareMathOperator{\mathOpPrcSorts}{\boldsymbol{\DataType}}
\DeclareMathOperator{\mathOpRedTimers}{\boldsymbol{\theta}}
\DeclareMathOperator{\mathOpPrcVals}{\boldsymbol{\ValClocks}}

\DeclareMathOperator{\mathOpRecPrcMsg}{\mathbf{\uppercase{t}}}
\DeclareMathOperator{\mathOpRecPrcTimers}{\boldsymbol{\Const}}

\DeclareMathOperator{\mathOpTypComm}{\mbox{\tiny{$\square$}}}
\DeclareMathOperator{\mathOpTypSend}{!}
\DeclareMathOperator{\mathOpTypRecv}{?}

\DeclareMathOperator{\mathOpPrcRecv}{\vartriangleright}
\DeclareMathOperator{\mathOpPrcSend}{\vartriangleleft}
\DeclareMathOperator{\mathOpPrcTyped}{\triangleright}

\DeclareMathOperator{\mathOpTRed}{\rightsquigarrow}
\DeclareMathOperator{\mathOpARed}{\rightharpoonup}
\DeclareMathOperator{\mathOpPrcRed}{\longrightarrow}

\DeclareMathOperator{\mathOpTypEnv}{\gamma}

\DeclareMathOperator{\mathOpValClocks}{\nu}

\DeclareMathOperator{\mathOpQueue}{\mathbf{\mathtt{\uppercase{m}}}}

\DeclareMathOperator{\mathOpRatSet}{{\mathbb{\uppercase{r}}}_{\geq0}}

\DeclareMathOperator{\mathOpClockSet}{\mathbb{\uppercase{x}}}
\DeclareMathOperator{\mathOpGuardSet}{\mathbb{\uppercase{g}}\mkTup[\ClockSet]}

\DeclareMathOperator{\mathOpTimerSet}{\mathbb{\uppercase{t}}}

\DeclareMathOperator{\mathOpFuncDelay}{\mathtt{delay}}
\DeclareMathOperator{\mathOpFuncWait}{\mathtt{Wait}}
\DeclareMathOperator{\mathOpFuncNEQ}{\mathtt{NEQueue}}

\DeclareMathOperator{\mathOpPrcFail}{\mathtt{failed}}
\DeclareMathOperator{\mathOpPrcErr}{\mathtt{error}}
\DeclareMathOperator{\mathOpTTPrcDef}{\mathtt{def}}
\DeclareMathOperator{\mathOpTTPrcDefIn}{\mathtt{in}}
\DeclareMathOperator{\mathOpTTPrcAft}{\mathtt{after}}
\DeclareMathOperator{\mathOpTTPrcSet}{\mathtt{set}}

\DeclareMathOperator{\mathOpTTPrcIf}{\mathtt{if}}
\DeclareMathOperator{\mathOpTTPrcThen}{\mathtt{then}}
\DeclareMathOperator{\mathOpTTPrcElse}{\mathtt{else}}

\DeclareMathOperator{\mathOpTTIf}{{\mathOpTTPrcIf}-{\mathOpTTPrcThen}-{\mathOpTTPrcElse}}
\DeclareMathOperator{\mathOpTTIff}{{\mathOpTTPrcIf}}

\DeclareMathOperator{\mathOpTypeTrue}{\mathtt{\lowercase{true}}}
\DeclareMathOperator{\mathOpTypeFalse}{\mathtt{\lowercase{false}}}

\DeclareMathOperator{\mathOpTypeEnd}{\mathbf{\mathtt{\lowercase{end}}}}

\DeclareMathOperator{\mathOpBTypeNat}{\mathtt{Nat}}
\DeclareMathOperator{\mathOpBTypeBool}{\mathtt{Bool}}
\DeclareMathOperator{\mathOpBTypeString}{\mathtt{String}}
\DeclareMathOperator{\mathOpBTypeNone}{\mathtt{None}}

\newcommand\texOpTypeS{\uppercase{s}}
\newcommand\texOpDataType{\uppercase{t}}

\newcommand\texOpPrcLabel{\lowercase{l}}
\newcommand\texOpPrcVal{\lowercase{v}}
\newcommand\texOpMsgLabel{\lowercase{l}}
\newcommand\texOpMsgData{\uppercase{t}}

\newcommand\textOpPrcEnd{\texttt{0}}
\newcommand\texOpPrc{\uppercase{p}}
\newcommand\texOpQrc{\uppercase{q}}

\newcommand\texOpPrcP{\lowercase{p}}
\newcommand\texOpPrcQ{\lowercase{q}}

\newcommand\texOpSesP{\texOpPrcP}
\newcommand\texOpSesQ{\texOpPrcQ}

\newcommand\texOpPrcAfter{\lowercase{n}}

\newcommand\texOpRecVar{\uppercase{x}}
\newcommand\texOpRecSetMsg{\mathbf{\uppercase{\scriptstyle v}}}
\newcommand\texOpRecSetTimers{\mathbf{\uppercase{\scriptstyle t}}}
\newcommand\texOpRecSetRoles{\mathbf{\uppercase{\scriptstyle r}}}

\newcommand\texOpRecEnv{\uppercase{a}}

\newcommand\texOpVarEnv{\Gamma}
\newcommand\texOpSesEnv{\Delta}
\newcommand\texOpSesSet{\Theta}
\newcommand\texOpSesDom{domain}
\newcommand\texOpBalSes{Bal}

\newcommand\texOpPrcTime{\Phi}

\newcommand\texOpValTime{\lowercase{t}}

\newcommand\texOpCfgIso{\mathbf{\lowercase{s}}}
\newcommand\texOpCfgSoc{\mathbf{\uppercase{s}}}

\newcommand\texOpQHead{\lowercase{h}}
\newcommand\texOpMsg{\lowercase{m}}

\newcommand\texOpQVal{\lowercase{v}}
\newcommand\texOpQLabel{\lowercase{a}}

\newcommand\texOpSetI{\uppercase{i}}
\newcommand\texOpSetJ{\uppercase{j}}
\newcommand\texOpSetK{\uppercase{k}}
\newcommand\texOpJudgeTyp{\texOpRecEnv;\mathOpConst\mathOpEntails\texOpTypeS}

\newcommand\texOpJudgePrc{\texOpVarEnv\mathOpEntails\texOpPrc\mathOpPrcTyped\texOpTypeS}

\newcommand\texOpClx{\lowercase{x}}
\newcommand\texOpCly{\lowercase{y}}
\newcommand\texOpClz{\lowercase{z}}
\newcommand\texOpTix{\lowercase{x}}
\newcommand\texOpTiy{\lowercase{y}}
\newcommand\texOpTiz{\lowercase{z}}

\newcommand\texOpMsgType{\lowercase{a}}

\newcommand\texOpPrcMsg{\lowercase{b}}

\NewDocumentCommand{\FmtMathOp}{m o m o o e_ e^}{%
   \IfValueT{#2}{\IfBooleanT{#2}{$}}
   {\IfValueTF{#6}{{#1}_{#6}}{\IfValueTF{#7}{{#1}^{#7}}{{#1}}}%
      \IfBooleanT{#3}{'\IfValueT{#4}{\IfBooleanT{#4}{'\IfValueT{#5}{\IfBooleanT{#5}{'}}}}}}%
   \IfNoValueF{#2}{\IfBooleanT{#2}{$}}
}

\NewDocumentCommand{\FmtMacro}{m m o o e_ e^}{%
   {#1\IfValueTF{#5}{_{#5}}{\IfValueT{#6}{^{#6}}}\IfBooleanT{#2}{'\IfValueT{#3}{\IfBooleanT{#3}{'\IfValueT{#4}{\IfBooleanT{#4}{'}}}}}}%
}

\NewDocumentCommand{\Entails}{s t'}{\FmtMathOp{\mathOpEntails}[#1]{#2}}
\NewDocumentCommand{\Implies}{s t'}{\FmtMathOp{\mathOpImplies}[#1]{#2}}

\NewDocumentCommand{\PrcEnd}{s t'}{\FmtMathOp{\textOpPrcEnd}[#1]{#2}}

\NewDocumentCommand{\SetOr}{s t'}{\FmtMathOp{\mathOpSetOr}[#1]{#2}}
\NewDocumentCommand{\SetUnion}{s t'}{\FmtMathOp{\mathOpSetUnion}[#1]{#2}}
\NewDocumentCommand{\EntailsEach}{s t'}{\FmtMathOp{\mathOpImpEach}[#1]{#2}}
\NewDocumentCommand{\ManyAct}{s t'}{\FmtMathOp{\mathOpManyAct}[#1]{#2}}

\NewDocumentCommand{\TypeS}{s t'}{\FmtMathOp{\texOpTypeS}[#1]{#2}}
\NewDocumentCommand{\DataType}{s t'}{\FmtMathOp{\texOpDataType}[#1]{#2}}

\NewDocumentCommand{\PrcLabel}{s t'}{\FmtMathOp{\texOpPrcLabel}[#1]{#2}}
\NewDocumentCommand{\PrcVal}{s t'}{\FmtMathOp{\texOpPrcVal}[#1]{#2}}
\NewDocumentCommand{\MsgLabel}{s t'}{\FmtMathOp{\texOpMsgLabel}[#1]{#2}}
\NewDocumentCommand{\MsgData}{s t'}{\FmtMathOp{\texOpMsgData}[#1]{#2}}

\NewDocumentCommand{\TypeChoice}{s t'}{\FmtMathOp{\mathOpTypeChoice}[#1]{#2}}
\NewDocumentCommand{\Const}{s t'}{\FmtMathOp{\mathOpConst}[#1]{#2}}
\NewDocumentCommand{\RSet}{s t'}{\FmtMathOp{\mathOpReSet}[#1]{#2}}

\NewDocumentCommand{\RecDef}{s t'}{\FmtMathOp{\mathOpRecDef}[#1]{#2}}
\NewDocumentCommand{\TypRecLabel}{s t'}{\FmtMathOp{\mathOpTypRecLabel}[#1]{#2}}
\NewDocumentCommand{\CfgRecLabel}{s t'}{\FmtMathOp{\mathOpCfgRecLabel}[#1]{#2}}

\NewDocumentCommand{\ProgAction}{s t'}{\FmtMathOp{\mathOpProgAction}[#1]{#2}}
\NewDocumentCommand{\SiltAction}{s t'}{\FmtMathOp{\mathOpSiltAction}[#1]{#2}}

\NewDocumentCommand{\Prc}{s t'}{\FmtMathOp{\texOpPrc}[#1]{#2}}
\NewDocumentCommand{\Qrc}{s t'}{\FmtMathOp{\texOpQrc}[#1]{#2}}

\NewDocumentCommand{\PrcP}{s t'}{\FmtMathOp{\texOpPrcP}[#1]{#2}}
\NewDocumentCommand{\PrcQ}{s t'}{\FmtMathOp{\texOpPrcQ}[#1]{#2}}

\NewDocumentCommand{\SesP}{s t'}{\FmtMathOp{\texOpSesP}[#1]{#2}}
\NewDocumentCommand{\SesQ}{s t'}{\FmtMathOp{\texOpSesQ}[#1]{#2}}

\NewDocumentCommand{\PrcScope}{s t'}{\FmtMathOp{\mathOpPrcScope}[#1]{#2}}
\NewDocumentCommand{\PrcSorts}{s t'}{\FmtMathOp{\mathOpPrcSorts}[#1]{#2}}
\NewDocumentCommand{\RedTimers}{s t'}{\FmtMathOp{\mathOpRedTimers}[#1]{#2}}
\NewDocumentCommand{\PrcVals}{s t'}{\FmtMathOp{\mathOpPrcVals}[#1]{#2}}
\NewDocumentCommand{\PrcAfter}{s t'}{\FmtMathOp{\texOpPrcAfter}[#1]{#2}}

\NewDocumentCommand{\RecVar}{s t'}{\FmtMathOp{\texOpRecVar}[#1]{#2}}
\NewDocumentCommand{\RecSetChan}{s t'}{\FmtMathOp{\texOpRecSetChan}[#1]{#2}}
\NewDocumentCommand{\RecSetMsg}{s t'}{\FmtMathOp{\texOpRecSetMsg}[#1]{#2}}
\NewDocumentCommand{\RecSetRoles}{s t'}{\FmtMathOp{\texOpRecSetRoles}[#1]{#2}}
\NewDocumentCommand{\RecSetTimers}{s t'}{\FmtMathOp{\texOpRecSetTimers}[#1]{#2}}

\NewDocumentCommand{\RecPrcMsg}{s t'}{\FmtMathOp{\mathOpRecPrcMsg}[#1]{#2}}
\NewDocumentCommand{\RecPrcTimers}{s t'}{\FmtMathOp{\mathOpRecPrcTimers}[#1]{#2}}

\NewDocumentCommand{\TypComm}{s t'}{\FmtMathOp{\mathOpTypComm}[#1]{#2}}
\NewDocumentCommand{\TypSend}{s t'}{\FmtMathOp{\mathOpTypSend}[#1]{#2}}
\NewDocumentCommand{\TypRecv}{s t'}{\FmtMathOp{\mathOpTypRecv}[#1]{#2}}

\NewDocumentCommand{\PrcRecv}{s t'}{\FmtMathOp{\mathOpPrcRecv}[#1]{#2}}
\NewDocumentCommand{\PrcSend}{s t'}{\FmtMathOp{\mathOpPrcSend}[#1]{#2}}
\NewDocumentCommand{\PrcTyped}{s t'}{\FmtMathOp{\mathOpPrcTyped}[#1]{#2}}

\NewDocumentCommand{\TRed}{s t'}{\FmtMathOp{\mathOpTRed}[#1]{#2}}
\NewDocumentCommand{\ARed}{s t'}{\FmtMathOp{\mathOpARed}[#1]{#2}}
\NewDocumentCommand{\PrcRed}{s t'}{\FmtMathOp{\mathOpPrcRed}[#1]{#2}}

\NewDocumentCommand{\TypEnv}{s t'}{\FmtMathOp{\mathOpTypEnv}[#1]{#2}}
\NewDocumentCommand{\RecEnv}{s t'}{\FmtMathOp{\texOpRecEnv}[#1]{#2}}

\NewDocumentCommand{\VarEnv}{s t'}{\FmtMathOp{\texOpVarEnv}[#1]{#2}}
\NewDocumentCommand{\SesEnv}{s t'}{\FmtMathOp{\texOpSesEnv}[#1]{#2}}
\NewDocumentCommand{\SesSet}{s t'}{\FmtMathOp{\texOpSesSet}[#1]{#2}}
\NewDocumentCommand{\SesDom}{s t'}{\FmtMathOp{\texOpSesDom}[#1]{#2}}
\NewDocumentCommand{\BalSes}{s t'}{\FmtMathOp{\texOpBalSes}[#1]{#2}}

\NewDocumentCommand{\PrcTime}{s t'}{\FmtMathOp{\texOpPrcTime}[#1]{#2}}

\NewDocumentCommand{\ValClocks}{s t'}{\FmtMathOp{\mathOpValClocks}[#1]{#2}}
\NewDocumentCommand{\ValTime}{s t'}{\FmtMathOp{\texOpValTime}[#1]{#2}}

\NewDocumentCommand{\IsoCfg}{s t'}{\FmtMathOp{\texOpCfgIso}[#1]{#2}}
\NewDocumentCommand{\SocCfg}{s t'}{\FmtMathOp{\texOpCfgSoc}[#1]{#2}}

\NewDocumentCommand{\Queue}{s t'}{\FmtMathOp{\mathOpQueue}[#1]{#2}}
\NewDocumentCommand{\QHead}{s t'}{\FmtMathOp{\texOpQHead}[#1]{#2}}
\NewDocumentCommand{\Msg}{s t'}{\FmtMathOp{\texOpMsg}[#1]{#2}}

\NewDocumentCommand{\QVal}{s t'}{\FmtMathOp{\texOpQVal}[#1]{#2}}
\NewDocumentCommand{\QLabel}{s t'}{\FmtMathOp{\texOpQLabel}[#1]{#2}}

\NewDocumentCommand{\SetI}{s t'}{\FmtMathOp{\texOpSetI}[#1]{#2}}
\NewDocumentCommand{\SetJ}{s t'}{\FmtMathOp{\texOpSetJ}[#1]{#2}}
\NewDocumentCommand{\SetK}{s t'}{\FmtMathOp{\texOpSetK}[#1]{#2}}

\NewDocumentCommand{\RatSet}{s t'}{\FmtMathOp{\mathOpRatSet}[#1]{#2}}

\NewDocumentCommand{\ClockSet}{s t'}{\FmtMathOp{\mathOpClockSet}[#1]{#2}}
\NewDocumentCommand{\GuardSet}{s t'}{\FmtMathOp{\mathOpGuardSet}[#1]{#2}}

\NewDocumentCommand{\TimerSet}{s t'}{\FmtMathOp{\mathOpTimerSet}[#1]{#2}}

\NewDocumentCommand{\JudgeTyp}{s t'}{\FmtMathOp{\texOpJudgeTyp}[#1]{#2}}

\NewDocumentCommand{\JudgePrc}{s t'}{\FmtMathOp{\texOpJudgePrc}[#1]{#2}}

\NewDocumentCommand{\FuncDelay}{s t'}{\FmtMathOp{\mathOpFuncDelay}[#1]{#2}}
\NewDocumentCommand{\FuncWait}{s t'}{\FmtMathOp{\mathOpFuncWait}[#1]{#2}}
\NewDocumentCommand{\FuncNEQ}{s t'}{\FmtMathOp{\mathOpFuncNEQ}[#1]{#2}}

\NewDocumentCommand{\PrcFail}{s t'}{\FmtMathOp{\mathOpPrcFail}[#1]{#2}}
\NewDocumentCommand{\PrcErr}{s t'}{\FmtMathOp{\mathOpPrcErr}[#1]{#2}}
\NewDocumentCommand{\TTPrcDef}{s t'}{\FmtMathOp{\mathOpTTPrcDef}[#1]{#2}}
\NewDocumentCommand{\TTPrcDefIn}{s t'}{\FmtMathOp{\mathOpTTPrcDefIn}[#1]{#2}}
\NewDocumentCommand{\TTPrcAft}{s t'}{\FmtMathOp{\mathOpTTPrcAft}[#1]{#2}}
\NewDocumentCommand{\TTPrcSet}{s t'}{\FmtMathOp{\mathOpTTPrcSet}[#1]{#2}}
\NewDocumentCommand{\TTPrcIf}{s t'}{\FmtMathOp{\mathOpTTPrcIf}[#1]{#2}}
\NewDocumentCommand{\TTPrcThen}{s t'}{\FmtMathOp{\mathOpTTPrcThen}[#1]{#2}}
\NewDocumentCommand{\TTPrcElse}{s t'}{\FmtMathOp{\mathOpTTPrcElse}[#1]{#2}}

\NewDocumentCommand{\TTIf}{s t'}{\FmtMathOp{\mathOpTTIf}[#1]{#2}}
\NewDocumentCommand{\TTIff}{s t'}{\FmtMathOp{\mathOpTTIff}[#1]{#2}}

\NewDocumentCommand{\TypeTrue}{s t'}{\FmtMathOp{\mathOpTypeTrue}[#1]{#2}}
\NewDocumentCommand{\TypeFalse}{s t'}{\FmtMathOp{\mathOpTypeFalse}[#1]{#2}}
\NewDocumentCommand{\TypeEnd}{s t'}{\FmtMathOp{\mathOpTypeEnd}[#1]{#2}}

\NewDocumentCommand{\BTypeNat}{s t'}{\FmtMathOp{\mathOpBTypeNat}[#1]{#2}}
\NewDocumentCommand{\BTypeBool}{s t'}{\FmtMathOp{\mathOpBTypeBool}[#1]{#2}}
\NewDocumentCommand{\BTypeString}{s t'}{\FmtMathOp{\mathOpBTypeString}[#1]{#2}}
\NewDocumentCommand{\BTypeNone}{s t'}{\FmtMathOp{\mathOpBTypeNone}[#1]{#2}}

\NewDocumentCommand{\Clx}{s t'}{\FmtMathOp{\texOpClx}[#1]{#2}}
\NewDocumentCommand{\Cly}{s t'}{\FmtMathOp{\texOpCly}[#1]{#2}}
\NewDocumentCommand{\Clz}{s t'}{\FmtMathOp{\texOpClz}[#1]{#2}}

\NewDocumentCommand{\Tix}{s t'}{\FmtMathOp{\texOpTix}[#1]{#2}}
\NewDocumentCommand{\Tiy}{s t'}{\FmtMathOp{\texOpTiy}[#1]{#2}}
\NewDocumentCommand{\Tiz}{s t'}{\FmtMathOp{\texOpTiz}[#1]{#2}}

\NewDocumentCommand{\MsgType}{s t'}{\FmtMathOp{\texOpMsgType}[#1]{#2}}

\NewDocumentCommand{\PrcMsg}{s t'}{\FmtMathOp{\texOpPrcMsg}[#1]{#2}}

\NewDocumentCommand{\RuleLbl}{m m o}{%
   \IfNoValueTF{#3}{%
      \IfBooleanTF{#1}{%
         \RName*[#2]%
      }{%
         \RName[#2]%
      }%
   }{%
      \IfBooleanTF{#1}{%
         \RName*[#2][#3]%
      }{%
         \RName[#2][#3]%
      }%
   }%
}

\NewDocumentCommand{\LblTypEnd}{s}{%
   \RuleLbl{#1}{end}%
}
\NewDocumentCommand{\LblTypVar}{s}{%
   \RuleLbl{#1}{var}%
}
\NewDocumentCommand{\LblTypRec}{s}{%
   \RuleLbl{#1}{rec}%
}
\NewDocumentCommand{\LblTypChoice}{s}{%
   \RuleLbl{#1}{choice}%
}

\NewDocumentCommand{\LblCfgIsoUnfold}{s}{%
   \RuleLbl{#1}{unfold}%
}
\NewDocumentCommand{\LblCfgIsoTick}{s}{%
   \RuleLbl{#1}{tick}%
}
\NewDocumentCommand{\LblCfgIsoInteract}{s}{%
   \RuleLbl{#1}{act}%
}

\NewDocumentCommand{\LblCfgSocSend}{s}{%
   \RuleLbl{#1}{snd}%
}
\NewDocumentCommand{\LblCfgSocRecv}{s}{%
   \RuleLbl{#1}{rcv}%
}
\NewDocumentCommand{\LblCfgSocEnqu}{s}{%
   \RuleLbl{#1}{que}%
}
\NewDocumentCommand{\LblCfgSocTime}{s}{%
   \RuleLbl{#1}{time}%
}

\NewDocumentCommand{\LblCfgSysLComm}{s}{%
   \RuleLbl{#1}{com\text{-}l}%
}
\NewDocumentCommand{\LblCfgSysLPar}{s}{%
   \RuleLbl{#1}{par\text{-}l}%
}
\NewDocumentCommand{\LblCfgSysWait}{s}{%
   \RuleLbl{#1}{wait}%
}

\NewDocumentCommand{\LblCfgSysRComm}{s}{%
   \RuleLbl{#1}{com\text{-}r}%
}
\NewDocumentCommand{\LblCfgSysRPar}{s}{%
   \RuleLbl{#1}{par\text{-}r}%
}

\NewDocumentCommand{\LblPrcRedARed}{s}{%
   \RuleLbl{#1}{ARed}%
}
\NewDocumentCommand{\LblPrcRedAStr}{s}{%
   \RuleLbl{#1}{Str}%
}
\NewDocumentCommand{\LblPrcRedScope}{s}{%
   \RuleLbl{#1}{Scope}%
}
\NewDocumentCommand{\LblPrcRedTRed}{s}{%
   \RuleLbl{#1}{TRed}%
}
\NewDocumentCommand{\LblPrcRedTStr}{s}{%
   \RuleLbl{#1}{TStr}%
}
\NewDocumentCommand{\LblPrcRedDelay}{s}{%
   \RuleLbl{#1}{Delay}%
}
\NewDocumentCommand{\LblPrcRedPar}{s}{%
   \RuleLbl{#1}{Par}%
}

\NewDocumentCommand{\LblPrcRedParR}{s}{%
   \RuleLbl{#1}{ParR}%
}

\NewDocumentCommand{\LblPrcRedParL}{s}{%
   \RuleLbl{#1}{ParL}%
}

\NewDocumentCommand{\LblPrcRedDef}{s}{%
   \RuleLbl{#1}{Def}%
}
\NewDocumentCommand{\LblPrcRedDet}{s}{%
   \RuleLbl{#1}{Det}%
}
\NewDocumentCommand{\LblPrcRedSend}{s}{%
   \RuleLbl{#1}{Send}%
}
\NewDocumentCommand{\LblPrcRedRecv}{s}{%
   \RuleLbl{#1}{Recv}%
}
\NewDocumentCommand{\LblPrcRedRec}{s}{%
   \RuleLbl{#1}{Rec}%
}

\NewDocumentCommand{\LblPrcRedIfThenElse}{s}{%
   \RuleLbl{#1}{If\text{-}Then\text{-}Else}%
}

\NewDocumentCommand{\LblPrcRedIf}{s}{%
   \RuleLbl{#1}{IfT}%
}
\NewDocumentCommand{\LblPrcRedElse}{s}{%
   \RuleLbl{#1}{IfF}%
}

\NewDocumentCommand{\LblPrcRedSetTimer}{s}{%
   \RuleLbl{#1}{Set}%
}

\NewDocumentCommand{\LblPrcTypEnd}{s}{%
   \RuleLbl{#1}{End}%
}
\NewDocumentCommand{\LblPrcTypEmptQ}{s}{%
   \RuleLbl{#1}{EmptQ}%
}
\NewDocumentCommand{\LblPrcTypTimer}{s}{%
   \RuleLbl{#1}{Timer}%
}
\NewDocumentCommand{\LblPrcTypDQue}{s}{%
   \RuleLbl{#1}{DQue}%
}
\NewDocumentCommand{\LblPrcTypWeak}{s}{%
   \RuleLbl{#1}{Weak}%
}
\NewDocumentCommand{\LblPrcTypVQue}{s}{%
   \RuleLbl{#1}{VQue}%
}
\NewDocumentCommand{\LblPrcTypPar}{s}{%
   \RuleLbl{#1}{Par}%
}
\NewDocumentCommand{\LblPrcTypDelTime}{s}{%
   \RuleLbl{#1}{Del}[{\text{-}\ValTime}]%
}
\NewDocumentCommand{\LblPrcTypDelDelta}{s}{%
   \RuleLbl{#1}{Del}[{\text{-}\Const}]%
}
\NewDocumentCommand{\LblPrcTypVar}{s}{%
   \RuleLbl{#1}{Var}%
}
\NewDocumentCommand{\LblPrcTypVal}{s}{%
   \RuleLbl{#1}{Val}%
}
\NewDocumentCommand{\LblPrcTypBranch}{s}{%
   \RuleLbl{#1}{Branch}%
}
\NewDocumentCommand{\LblPrcTypVRecv}{s}{%
   \RuleLbl{#1}{VRecv}%
}
\NewDocumentCommand{\LblPrcTypDRecv}{s}{%
   \RuleLbl{#1}{DRecv}%
}
\NewDocumentCommand{\LblPrcTypIfTrue}{s}{%
   \RuleLbl{#1}{IfTrue}%
}
\NewDocumentCommand{\LblPrcTypIfFalse}{s}{%
   \RuleLbl{#1}{IfFalse}%
}
\NewDocumentCommand{\LblPrcTypVSend}{s}{%
   \RuleLbl{#1}{VSend}%
}
\NewDocumentCommand{\LblPrcTypDSend}{s}{%
   \RuleLbl{#1}{DSend}%
}
\NewDocumentCommand{\LblPrcTypRes}{s}{%
   \RuleLbl{#1}{Res}%
}
\NewDocumentCommand{\LblPrcTypRec}{s}{%
   \RuleLbl{#1}{Rec}%
}

\NewDocumentCommand{\LblSesRedSend}{s}{%
   {\mkReg[{\SesEnv\mathtt{-Send}}]}%
}
\NewDocumentCommand{\LblSesRedRed}{s}{%
   {\mkReg[{\SesEnv\mathtt{-Red}}]}%
}
\NewDocumentCommand{\LblSesRedSilent}{s}{%
   {\mkReg[{\SesEnv\mathtt{-Silent}}]}%
}

\NewDocumentCommand{\DefineMap}{s}{%
    \IfBooleanT{#1}{$}
    \IfBooleanT{#1}{$}
}

\NewDocumentCommand{\mkFunc}{s t* O{???} O{}}{%
    \IfBooleanT{#1}{$}
    \IfBooleanTF{#2}{\mathtt{#3}}{#3}
    \left(#4\right)%
    \IfBooleanT{#1}{$}
}

\NewDocumentCommand{\Par}{s O{0} O{0}}{%
    \IfBooleanT{#1}{$}
    {#2}\,\mid{#3}%
    \IfBooleanT{#1}{$}
}

\NewDocumentCommand{\Compat}{s o o t'}{%
    \IfBooleanT{#1}{$}
    \IfValueTF{#2}{#2}{%
        \IfBooleanTF{#4}{\VSoc'_1}{\VSoc_1}%
    }\!~\!\bot\!~\!\IfValueTF{#3}{#3}{%
        \IfBooleanTF{#4}{\VSoc'_2}{\VSoc_2}%
    }%
    \IfBooleanT{#1}{$}
}

\NewDocumentCommand{\Dual}{s O{\TypeS} t' e_}{%
    \IfBooleanT{#1}{$}
    \overline{{#2\IfBooleanT{#3}{'}\IfValueT{#4}{_{#4}}}\mathstrut}
    \IfBooleanT{#1}{$}
}

\NewDocumentCommand{\mkReg}{s O{\emptyset}}{%
    \IfBooleanT{#1}{$}
    \left[\,#2\,\right]%
    \IfBooleanT{#1}{$}
}

\NewDocumentCommand{\mkSet}{s O{\emptyset}}{%
    \IfBooleanT{#1}{$}
    \left\{\,#2\,\right\}%
    \IfBooleanT{#1}{$}
}

\NewDocumentCommand{\mkTup}{s t; O{\emptyset} o o o o o}{%
    \IfBooleanT{#1}{$}
    ({#3}%
    \IfNoValueF{#4}{\IfBooleanTF{#2}{;}{,}{#4}}%
    \IfNoValueF{#5}{\IfBooleanTF{#2}{;}{,}{#5}}%
    \IfNoValueF{#6}{\IfBooleanTF{#2}{;}{,}{#6}}%
    \IfNoValueF{#7}{\IfBooleanTF{#2}{;}{,}{#7}}%
    \IfNoValueF{#8}{\IfBooleanTF{#2}{;}{,}{#8}}%
    )\!%
    \IfBooleanT{#1}{$}
}

\NewDocumentCommand{\Subst}{s O{?} O{?}}{%
    \IfBooleanT{#1}{$}
    \mkReg[\sfrac{#2}{#3}]%
    \IfBooleanT{#1}{$}
}

\NewDocumentCommand{\Buff}{O{1.25}}{%
    \rule{0pt}{#1ex}%
}

\NewDocumentCommand{\RName}{s O{???} O{}}{%
    \IfBooleanT{#1}{$}
    {\mkReg[{\mathtt{#2}}{#3}]}%
    \IfBooleanT{#1}{$}
}

\NewDocumentCommand{\InSet}{s t^ O{\emptyset} t' O{0} t' o t'}{%
    \IfBooleanT{#1}{$}
    {\mkSet[#3]\IfBooleanT{#4}{'}}%
    \IfBooleanTF{#2}{^}{_}%
    {%
    \AsSet[\FmtMathOp{#5}{#6}][\FmtMathOp{#7}{#8}]%
    }%
    \IfBooleanT{#1}{$}
}

\NewDocumentCommand{\AsSet}{s O{0} o}{%
    \IfBooleanT{#1}{$}
    \IfNoValueTF{#3}{\uppercase{#2}}{\lowercase{#2}\in\uppercase{#3}}%
    \IfBooleanT{#1}{$}
}

\NewDocumentCommand{\Sat}{s o t' t: o t'}{%
    \IfBooleanT{#1}{$}
    {\IfValueTF{#2}{#2\IfBooleanT{#3}{'}\,\models\,\IfValueTF{#5}{#5\IfBooleanT{#6}{'}}{\Const\IfBooleanT{#6}{'}}}{\ValClocks\IfBooleanT{#3}{'}\,\models\,\IfValueTF{#5}{#5\IfBooleanT{#6}{'}}{\Const\IfBooleanTF{#6}{'}{\IfBooleanT{#3}{'}}}}}%
    \IfBooleanT{#1}{$}
}

\NewDocumentCommand{\CommAction}{s O{\MsgType} t' e_}{{\TypComm\IfValueT{#4}{_{#4}}}{{#2}\IfBooleanT{#3}{'}\IfValueT{#4}{_{#4}}}}

\NewDocumentCommand{\RecvAction}{s O{\MsgType} t' e_}{{\TypRecv\IfValueT{#4}{_{#4}}}{{#2}\IfBooleanT{#3}{'}\IfValueT{#4}{_{#4}}}}

\NewDocumentCommand{\SendAction}{s O{\MsgType} t' e_}{{\TypSend\IfValueT{#4}{_{#4}}}{{#2}\IfBooleanT{#3}{'}\IfValueT{#4}{_{#4}}}}

\NewDocumentCommand{\CommMsg}{s O{\Msg} t' e_}{\CommAction[#2]\IfBooleanT{#3}{'}\IfValueT{#4}{_{#4}}}

\NewDocumentCommand{\RecvMsg}{s O{\Msg} t' e_}{\RecvAction[#2]\IfBooleanT{#3}{'}\IfValueT{#4}{_{#4}}}

\NewDocumentCommand{\SendMsg}{s O{\Msg} t' e_}{\SendAction[#2]\IfBooleanT{#3}{'}\IfValueT{#4}{_{#4}}}

\NewDocumentCommand{\Act}{s t| O{\ProgAction}}{%
   \IfBooleanT{#1}{$}
   \IfBooleanTF{#2}
   {
      \stackrel{#3}{\not\rightarrow}%
   }{
      \xrightarrow{#3}%
   }%
   \IfBooleanT{#1}{$}
}

\NewDocumentCommand{\Trans}{s m t| e: t* o}{%
   \IfBooleanT{#1}{$}
   {#2}\,%
   \IfValueTF{#4}{
      \IfBooleanTF{#3}
      {
         {\stackrel{\ListToSpacedTrans{#4}}{\not\rightarrow}}%
         \IfBooleanT{#5}{^{\ast}}%
      }{
         {\xrightarrow{\ListToSpacedTrans{#4}}}%
         \IfBooleanT{#5}{^{\ast}}%
      }%
   }{
      {\IfBooleanTF{#3}{\not\rightarrow}{\rightarrow}}%
      \IfBooleanT{#5}{^{\ast}}%
   }%
   \IfValueT{#6}{\,{#6}}%
   \IfBooleanT{#1}{$}
}

\NewDocumentCommand\ListToSpacedTrans{>{\SplitList{,}}m}
{%
   {\ProcessList{#1}{\HeadSpacedTrans}}%
}

\newcommand\HeadSpacedTrans[1]{%
   {#1}%
   \let\HeadSpacedTrans\TailSpacedTrans%
}

\newcommand\TailSpacedTrans[1]{%
   \!\;{#1}%
}

\NewDocumentCommand{\ApplyReset}{s O{\RSet}}{%
   \IfBooleanT{#1}{$}
   #2\mapsto0%
   \IfBooleanT{#1}{$}
}

\NewDocumentCommand{\Resets}{s E1{{\Const}} E2{{\RSet}}}{%
   \IfBooleanT{#1}{$}
   #2\mkReg[\ApplyReset[#3]]%
   \IfBooleanT{#1}{$}
}

\NewDocumentCommand{\ReSet}{s O{\Const} E:{\RSet} e_}{%
   \IfBooleanT{#1}{$}
   {#2\mkReg[{#3\IfValueT{#4}{_{#4}}}\mapsto0]}%
   \IfBooleanT{#1}{$}
}

\NewDocumentCommand{\FutureEn}{s E1{\IsoCfg} t/ O{\TypComm}}{%
   \IfBooleanT{#1}{$}
   {#2}\stackrel{#4}{\IfBooleanT{#3}{\not}\Rightarrow}
   \IfBooleanT{#1}{$}
}

\NewDocumentCommand{\simplechoice}{E_{i} o}{ \{ \TypeChoice_{#1}.S_{#1} \}_{\lowercase{#1}\in \IfNoValueTF{#2}{\uppercase{#1}}{\uppercase{#2}}} }

\NewDocumentCommand{\Past}{s t' O{\Const}}{%
   \IfBooleanT{#1}{$}
   \downarrow\!\FmtMathOp{#3}{#2}%
   \IfBooleanT{#1}{$}
}

\NewDocumentCommand{\DefMsgType}{s O{\MsgLabel} t' e_ O{\DataType} t' e_}{%
   \IfBooleanT{#1}{$}
   \IfNoValueTF{#4}{\FmtMathOp{#2}{#3}}{\FmtMathOp{#2}{#3}_{#4}}
   \left\langle
   \IfNoValueTF{#7}{\FmtMathOp{#5}{#6}}{\FmtMathOp{#5}{#6}_{#7}}
   \right\rangle
   \IfBooleanT{#1}{$}
}

\NewDocumentCommand{\TypRecDef}{s O{\RecDef} O{\TypRecLabel} e: t. O{\TypeS}}{%
   \IfBooleanT{#1}{$}
   {#2}{{#3}\IfValueT{#4}{^{#4}}}%
   \IfBooleanTF{#5}{.}{.#6}
   \IfBooleanT{#1}{$}
}

\NewDocumentCommand{\TypRecCall}{s O{\TypRecLabel} e:}{%
   \IfBooleanT{#1}{$}
   {{#2}\IfValueT{#3}{^{#3}}}%
   \IfBooleanT{#1}{$}
}

\NewDocumentCommand{\TypJudge}{s E1{{\RecEnv}} t| E2{{\Const}} o O{i} E3{{\TypInteract}}}{%
   \IfBooleanT{#1}{$}
   {#2};\,
   \IfNoValueTF{#5}{
      \IfBooleanTF{#3}{\Past[#4]}{#4}
      \;\Entails\,
      {#7}
   }{
      \IfBooleanTF{#3}{
         \Past[{\SetOr_{\hspace{-0.5ex}\vspace{0.5ex}\scriptstyle{\AsSet[#5][#6]}}}{#4_{#5}}]
      }{
         {\SetOr_{\hspace{-0.5ex}\vspace{0.5ex}\scriptstyle{\AsSet[#5][#6]}}}{#4_{#5}}
      }
      \;\Entails\,
      {#7[#5][#6]}
   }
   \IfBooleanT{#1}{$}
}

\NewDocumentCommand{\TypCond}{s t' e_ E1{\Const} E2{\RSet}}{%
   \IfBooleanT{#1}{$}
   \mkTup[\FmtMathOp{#4}{#2}_{#3}][\FmtMathOp{#5}{#2}_{#3}]
   \IfBooleanT{#1}{$}
}

\NewDocumentCommand{\TypAct}{s t' e_ E1{\TypComm} E2{\MsgType}}{%
   \IfBooleanT{#1}{$}
   \FmtMathOp{#4}{#2}_{#3}\FmtMathOp{\MsgLabel}{#2}_{#3}\left\langle\FmtMathOp{\DataType}{#2}_{#3}\right\rangle
   \IfBooleanT{#1}{$}
}

\NewDocumentCommand{\TypInteract}{s t| E1{{}} E2{{}} E3{{\TypeS}} t_ O{i} O{i} t~}{%
\IfBooleanT{#1}{$}
\IfBooleanTF{#2}{
\IfBooleanTF{#6}{
\IfBooleanTF{#9}{
\Dual[{\TypAct_{#7}#3}{\TypCond_{#7}#4}\,.{#5_{#7}}]%
}{
{\TypAct_{#7}#3}{\TypCond_{#7}#4}\,.{#5_{#7}}%
}%
}{
\IfBooleanTF{#9}{
   \Dual[{\TypAct#3}{\TypCond#4}\,.#5]%
}{
   {\TypAct#3}{\TypCond#4}\,.#5%
}%
}%
}{
\IfBooleanTF{#9}{
\InSet[\Dual[{\TypAct_{#7}#3}{\TypCond_{#7}#4}\,.{#5_{#7}}]][#7][#8]%
}{
\InSet[{\TypAct_{#7}#3}{\TypCond_{#7}#4}\,.{#5_{#7}}][#7][#8]%
}%
}%
\IfBooleanT{#1}{$}
}

\NewDocumentCommand{\TmpTypBTDelegate}{s O{\Const} t' e_ O{\TypeS} t' e_}{%
   \IfBooleanT{#1}{$}
   \mkTup[%
      \IfNoValueTF{#4}{\FmtMathOp{#2}{#3}}{\FmtMathOp{#2}{#3}_{#4}}
   ][%
      \IfNoValueTF{#7}{\FmtMathOp{#5}{#6}}{\FmtMathOp{#5}{#6}_{#7}}
   ]\,
   \IfBooleanT{#1}{$}
}

\NewDocumentCommand{\TmpTypFutureSat}{s E1{\TypEnv} t' e_ E2{{1{\Const}2{\RSet}}}}{%
\IfBooleanT{#1}{$}
{\Resets#5}%
\subseteq%
\IfNoValueTF{#4}{\FmtMathOp{#2}{#3}}{\FmtMathOp{#2}{#3}_{#4}}%
\IfBooleanT{#1}{$}
}

\NewDocumentCommand{\CfgRecDef}{s O{\RecDef} O{\CfgRecLabel}}{%
    \IfBooleanT{#1}{$}
    {#2}{#3}
    \IfBooleanT{#1}{$}
}

\NewDocumentCommand{\CfgIso}{s e_ E1{{\ValClocks}} t' E2{{\TypeS}} t'}{%
    \IfBooleanT{#1}{$}
    \mkTup[{\FmtMathOp{#3}{#4}_{#2}}][{\FmtMathOp{#5}{#6}_{#2}}]
    \IfBooleanT{#1}{$}
}

\NewDocumentCommand{\CfgSoc}{s e_ E1{{\ValClocks}} t' E2{{\TypeS}} t' E3{{\Queue}} t'}{%
    \IfBooleanT{#1}{$}
    \mkTup[{\FmtMathOp{#3}{#4}_{#2}}][{\FmtMathOp{#5}{#6}_{#2}}][{\FmtMathOp{#7}{#8}_{#2}}]
    \IfBooleanT{#1}{$}
}

\NewDocumentCommand{\CfgSys}{s E1{{\SocCfg_1}} E2{{\SocCfg_2}}}{%
    \IfBooleanT{#1}{$}
    \Par[#2][#3]
    \IfBooleanT{#1}{$}
}

\NewDocumentCommand{\PrcCalcErase}{O{\Prc}}{\mathtt{erase}(#1)}
\NewDocumentCommand{\PrcCalcOn}{m}{\lowercase{#1}}
\NewDocumentCommand{\PrcCalcSend}{m o E.{\Prc}}{%
~\PrcSend~%
\begin{array}[t]{l}
{#1}{\IfValueT{#2}{%
   {#2}
}}.{#3}%
\end{array}
}
\NewDocumentCommand{\PrcCalcRecv}{s e< t| m o E:{\Prc} e_ o}{%
\IfValueT{#2}{
   ^{#2}
}~\PrcRecv~
{\IfBooleanTF{#1}{
   \begin{array}[t]{ll}%
      \IfBooleanTF{#3}{
      \ListToArray{#4}%
      }{
      \IfValueTF{#7}{
      {{#4}_{\lowercase{#7}}}%
      \IfValueT{#5}{({#5}_{\lowercase{#7}})}:%
      {{#6}_{\lowercase{#7}}}%
      }{
      {#4}\IfValueT{#5}{({#5})}:{#6}%
      }%
      }%
   \end{array}
}{
   \begin{array}[t]\{{ll}\}%
      \IfBooleanTF{#3}{
      \ListToArray{#4}%
      }{
      \IfValueTF{#7}{
      {{#4}_{\lowercase{#7}}}%
      \IfValueT{#5}{({#5}_{\lowercase{#7}})}:%
      {{#6}_{\lowercase{#7}}}%
      }{
      {#4}\IfValueT{#5}{({#5})}:{#6}%
      }%
      }%
   \end{array}
}}%
\IfValueT{#8}{
   _{\lowercase{#7}\in\uppercase{#8}}%
}\!
}
\NewDocumentCommand{\PrcCalcOption}{m o E:{\Prc} e_}{%
\IfValueTF{#4}{%
{{{#1}_{#4}}~{\IfValueT{#2}{({{#2}}_{#4})}}:{{#3}_{#4}}}%
}{%
{{{#1}}~{\IfValueT{#2}{({#2})}}:{{#3}}}%
}%
}
\NewDocumentCommand{\PrcCalcAfter}{E<{n} E:{\Qrc}}{%
\mathtt{after\ }{#1}:{#2}
}

\NewDocumentCommand{\PrcCalcIf}{s m}{{\mathtt{if\ }\IfBooleanTF{#1}{#2}{(#2)}}}
\NewDocumentCommand{\PrcCalcThen}{m}{{\mathtt{then}:{#1}}}
\NewDocumentCommand{\PrcCalcElse}{m}{{\mathtt{else}:{#1}}}

\NewDocumentCommand{\PrcCalcTimePass}{E_{\ValTime} m}{{{\Phi_{#1}}\begin{array}[t]({l}){#2}\end{array}}}

\ProvideDocumentCommand{\Erase}{}{}
\ProvideDocumentCommand{\On}{}{}
\ProvideDocumentCommand{\Send}{}{}
\ProvideDocumentCommand{\Recv}{}{}
\ProvideDocumentCommand{\Option}{}{}
\ProvideDocumentCommand{\After}{}{}
\ProvideDocumentCommand{\TimePass}{}{}
\ProvideDocumentCommand{\If}{}{}
\ProvideDocumentCommand{\Then}{}{}
\ProvideDocumentCommand{\Else}{}{}

\NewDocumentCommand{\PCalc}{s m}{%
   \let\TempErase\Erase
   \let\TempOn\On
   \let\TempSend\Send
   \let\TempRecv\Recv
   \let\TempOption\Option
   \let\TempAfter\After
   \let\TempTimePass\TimePass
   \let\TempIf\If
   \let\TempThen\Then
   \let\TempElse\Else
   \let\Erase\PrcCalcErase
   \let\On\PrcCalcOn
   \let\Send\PrcCalcSend
   \let\Recv\PrcCalcRecv
   \let\Option\PrcCalcOption
   \let\After\PrcCalcAfter
   \let\TimePass\PrcCalcTimePass
   \let\If\PrcCalcIf
   \let\Then\PrcCalcThen
   \let\Else\PrcCalcElse
   \IfBooleanTF{#1}{${#2}$}{{#2}}
   \let\Erase\TempErase
   \let\On\TempOn
   \let\Send\TempSend
   \let\Recv\TempRecv
   \let\Option\TempOption
   \let\After\TempAfter
   \let\TimePass\TempTimePass
   \let\If\TempIF
   \let\Then\TempThen
   \let\Else\TempElse
}

\NewDocumentEnvironment{processcalculus}{s}{
\let\TempErase\Erase
\let\TempOn\On
\let\TempSend\Send
\let\TempRecv\Recv
\let\TempOption\Option
\let\TempAfter\After
\let\TempTimePass\TimePass
\let\TempIf\If
\let\TempThen\Then
\let\TempElse\Else
\let\Erase\PrcCalcErase
\let\On\PrcCalcOn
\let\Send\PrcCalcSend
\let\Recv\PrcCalcRecv
\let\Option\PrcCalcOption
\let\After\PrcCalcAfter
\let\TimePass\PrcCalcTimePass
\let\If\PrcCalcIf
\let\Then\PrcCalcThen
\let\Else\PrcCalcElse
\IfBooleanTF{#1}{\begin{minieq}*}{\begin{minieq}}%
      }{
   \end{minieq}
   \let\Erase\TempErase
   \let\On\TempOn
   \let\Send\TempSend
   \let\Recv\TempRecv
   \let\Option\TempOption
   \let\After\TempAfter
   \let\TimePass\TempTimePass
   \let\If\TempIF
   \let\Then\TempThen
   \let\Else\TempElse
}

\NewDocumentCommand{\PFuncTime}{s E_{\ValTime} O{\Prc}}{%
   \IfBooleanT{#1}{$}
   \PrcTime_{#2}\left({#3}\right)\,%
   \IfBooleanT{#1}{$\!}
}

\NewDocumentCommand{\PFuncWait}{s O{\Prc}}{%
   \IfBooleanT{#1}{$}
   \FuncWait({#2})%
   \IfBooleanT{#1}{$\!}
}

\NewDocumentCommand{\PFuncFQ}{s O{\SesEnv}}{%
   \IfBooleanT{#1}{$}
   \mathtt{fq}({#2})%
   \IfBooleanT{#1}{$\!}
}

\NewDocumentCommand{\PFuncNEQ}{s O{\Prc}}{%
   \IfBooleanT{#1}{$}
   \FuncNEQ({#2})%
   \IfBooleanT{#1}{$\!}
}

\NewDocumentCommand{\PFuncDelay}{s O{\Const} t. O{\Prc}}{%
   \IfBooleanT{#1}{$}
   \FuncDelay({#2})%
   \IfBooleanTF{#3}{.}{.#4}
   \IfBooleanT{#1}{$\!}
}

\NewDocumentCommand{\PFuncWF}{s O{\Prc}}{%
   \IfBooleanT{#1}{$}
   \mathtt{wf}\left({#2}\right)\,%
   \IfBooleanT{#1}{$\!}
}

\NewDocumentCommand{\TmpPrcJudge}{s E1{{\VarEnv}} E2{{\Prc}} t| E3{{1{\SesEnv}2{1{\SesP}}}}}{%
\IfBooleanT{#1}{$}
{#2}\;\Entails\,%
{#3}\;\PrcTyped\;%
\IfBooleanTF{#4}{#5}{\PSesEnvTypeSession#5}%
\IfBooleanT{#1}{$\!}
}

\NewDocumentCommand{\PSesEnvTypeSession}{s E1{{\SesEnv}} E2{{1{\SesP}}}}{%
   \IfBooleanT{#1}{$}
   {#2},\,{\PSesRole#3}%
   \IfBooleanT{#1}{$\!}
}

\NewDocumentCommand{\PSesRole}{s E1{{\SesP}} E2{{1{\ValClocks}2{\TypeS}}}}{%
\IfBooleanT{#1}{$}
{#2}:{\CfgIso#3}\,%
\IfBooleanT{#1}{$\!}
}

\NewDocumentCommand{\PSesEnvTRead}{s t| O{\SesEnv}}{%
\IfBooleanT{#1}{$}
{#3}+{\ValTime'}
\IfBooleanTF{#2}{\Act|[\TypRecv\Msg]}{\Act[\TypRecv\Msg]}%
\IfBooleanT{#1}{$\!}
}

\NewDocumentCommand{\PCalcSend}{s E1{{\SesP}} E2{{\PrcVal}} t. O{\Prc}}{%
   \IfBooleanT{#1}{$}
   {#2}\,\PrcSend\,{\PrcLabel}{#3}%
   \IfBooleanTF{#4}{.}{.#5}
   \IfBooleanT{#1}{$\!}
}

\NewDocumentCommand{\PCalcRecv}{s E1{{\SesP}} E2{{\PrcAfter}} E3{{\PrcMsg}} E4{{\Prc}} O{i} O{I} t| O{\Qrc}}{%
\IfBooleanT{#1}{$}
{#2}^{#3}\PrcRecv\,\InSet[{\PrcLabel}_{#6}{#4}_{#6}:{#5}_{#6}][#6][#7]%
\IfBooleanF{#8}{\;\TTPrcAft\,{#9}}
\IfBooleanT{#1}{$\!}
}

\NewDocumentCommand{\PCalcRecvSingle}{s E1{{\SesP}} E2{{\PrcAfter}} E3{{\PrcMsg}} E4{{\Prc}} e_}{%
\IfBooleanT{#1}{$}
{#2}^{#3}\PrcRecv\,{\PrcLabel}\IfValueT{#6}{_{#6}}{#4}\IfValueT{#6}{_{#6}}:{#5}\IfValueT{#6}{_{#6}}%
\IfBooleanT{#1}{$\!}
}

\NewDocumentCommand{\PCalcEndPoints}{s E1{{\PrcP}} E2{{\PrcQ}}}{%
   \IfBooleanT{#1}{$}
   {#2}{#3}%
   \IfBooleanT{#1}{$\!}
}

\NewDocumentCommand{\PCalcScope}{s E1{{1{\PrcP}2{\PrcQ}}} t| O{\Prc}}{%
\IfBooleanT{#1}{$}
\left({\PrcScope}\PCalcEndPoints#2\right)%
\IfBooleanF{#3}{#4}
\IfBooleanT{#1}{$\!}
}

\NewDocumentCommand{\PCfgQueue}{s E1{{1{\PrcP}2{\PrcQ}}} E2{{\QHead}} o}{%
\IfBooleanT{#1}{$}
{\PCalcEndPoints#2}:{#3}%
\IfNoValueF{#4}{; #4}\,
\IfBooleanT{#1}{$\!}
}

\NewDocumentCommand{\PCalcBuffer}{s E1{{1{\PrcP}2{\PrcQ}}} E2{{\QHead}} o}{%
\IfBooleanT{#1}{$}
{\PCalcEndPoints#2}:{#3}%
\IfNoValueF{#4}{\cdot #4}\,
\IfBooleanT{#1}{$\!}
}

\NewDocumentCommand{\PCalcSetTimer}{s E1{\Tix} t. O{\Prc}}{%
   \IfBooleanT{#1}{$}
   \TTPrcSet\left({#2}\right)%
   \IfBooleanTF{#3}{.}{.#4}
   \IfBooleanT{#1}{$\!}
}

\NewDocumentCommand{\PCalcIf}{s O{\Const} E2{{\Prc}} E3{{\Qrc}}}{%
\IfBooleanT{#1}{$}
\TTPrcIf\;{#2}\;\TTPrcThen\;{#3}\;\TTPrcElse\;{#4}\;%
\IfBooleanT{#1}{$\!}
}

\NewDocumentCommand{\PSesEnvTypeBuffer}{s E1{{\SesEnv}} E2{{2{\Queue}}}}{%
   \IfBooleanT{#1}{$}
   {#2},\,{\PCalcBuffer#3}%
   \IfBooleanT{#1}{$\!}
}

\NewDocumentCommand{\PSesEnvTypeQueue}{s E1{{\SesEnv}} E2{{2{\Queue}}}}{%
   \IfBooleanT{#1}{$}
   {#2},\,{\PCfgQueue#3}%
   \IfBooleanT{#1}{$\!}
}

\NewDocumentCommand{\PSesMsg}{s E1{{\PrcMsg}} E2{{\MsgType}}}{%
\IfBooleanT{#1}{$}
{#2}:{#3}\,%
\IfBooleanT{#1}{$\!}
}

\NewDocumentCommand{\PSesTimer}{s E1{{\Tix}} E2{{\Const}}}{%
\IfBooleanT{#1}{$}
{#2}:{#3}\,%
\IfBooleanT{#1}{$\!}
}

\NewDocumentCommand{\PVarEnvTypeTimerVal}{s E1{{\VarEnv}} E2{{1{\Tix}2{\Const}}}}{%
\IfBooleanT{#1}{$}
{#2},\,\PSesTimer#3%
\IfBooleanT{#1}{$\!}
}

\NewDocumentCommand{\PSesSetTypeEnvs}{s O{\SesEnv} O{\SesSet}}{%
   \IfBooleanT{#1}{$}
   {#3}\cup\mkSet[#2]\,%
   \IfBooleanT{#1}{$\!}
}

\NewDocumentCommand{\PSesSetCfgs}{s E1{\ValClocks} E2{\TypeS}}{%
   \IfBooleanT{#1}{$}
   \mkTup[\boldsymbol{#2}][\boldsymbol{#3}]\,%
   \IfBooleanT{#1}{$\!}
}

\NewDocumentCommand{\PVarTypeJudge}{s E1{\VarEnv} E2{\PrcMsg} E3{\DataType}}{%
\IfBooleanT{#1}{$}
{#2}\;\Entails\,{#3}:{#4}%
\IfBooleanT{#1}{$\!}
}

\NewDocumentCommand{\SesDomain}{s O{\SesEnv}}{%
   \IfBooleanT{#1}{$}
   \SesDom\left({#2}\right)%
   \IfBooleanT{#1}{$\!}
}

\NewDocumentCommand{\RedPrc}{s O{\RedTimers} t' e+ m}{%
   \IfBooleanT{#1}{$}
   \Tup({{#2\IfBooleanT{#3}{'}}\IfValueT{#4}{\ListToEnvItems{#4}}},{#5})%
   \IfBooleanT{#1}{$}
}

\NewDocumentCommand{\PrcRecDef}{s O{\RecVar} t' E={\Prc} E>{\Qrc}}{%
\IfBooleanT{#1}{$}
\TTPrcDef~%
{#2}\IfBooleanTF{#3}{\RecPrcVar'}{\RecPrcVar}={#4}%
~\TTPrcDefIn~{#5}%
\IfBooleanT{#1}{$}
}

\NewDocumentCommand{\RecPrcVar}{s E1{\RecSetMsg} E2{\RecSetTimers} E3{\RecSetRoles} t'}{
   \IfBooleanT{#1}{$}
   ({#2\IfBooleanT{#5}{'}};{#3\IfBooleanT{#5}{'}};{#4\IfBooleanT{#5}{'}})%
   \IfBooleanT{#1}{$}
}

\NewDocumentCommand{\RecPrcCall}{s E1{\RecSetMsg} E2{\RecSetTimers} E3{\RecSetRoles} t'}{
\IfBooleanT{#1}{$}
\RecVar\langle{#2\IfBooleanT{#5}{'}};{#3\IfBooleanT{#5}{'}};{#4\IfBooleanT{#5}{'}}\rangle%
\IfBooleanT{#1}{$}
}

\NewDocumentCommand{\RecPrcSubst}{s o o t' t/ t'}{%
   \IfBooleanT{#1}{$}
   \IfValueTF{#3}{
      \Subst[#2][#3]%
   }{
      \IfValueTF{#2}{
         \IfBooleanTF{#4}{
            \Subst[#2'][\RecSetMsg,\RecSetTimers,\RecSetRoles]%
         }{
            \IfBooleanTF{#6}{
               \Subst[#2][\RecSetMsg',\RecSetTimers',\RecSetRoles']%
            }{
               \Subst[#2][\RecSetMsg,\RecSetTimers,\RecSetRoles]%
            }%
         }%
      }{
         \IfBooleanTF{#4}{
            \Subst[\RecSetMsg',\RecSetTimers',\RecSetRoles'][\RecSetMsg,\RecSetTimers,\RecSetRoles]%
         }{
            \IfBooleanTF{#6}{
               \Subst[\RecSetMsg,\RecSetTimers,\RecSetRoles][\RecSetMsg',\RecSetTimers',\RecSetRoles']%
            }{
               \Subst[\RecSetMsg,\RecSetTimers,\RecSetRoles][\RecSetMsg,\RecSetTimers,\RecSetRoles]%
            }%
         }%
      }%
   }%
   \IfBooleanT{#1}{$}
}

\NewDocumentCommand{\EnvPrcVar}{s E1{\RecPrcMsg} E2{\RecPrcTimers} E3{\SesSet} t'}{%
   \IfBooleanT{#1}{$}
   ({#2\IfBooleanT{#5}{'}};{#3\IfBooleanT{#5}{'}};{#4\IfBooleanT{#5}{'}})%
   \IfBooleanT{#1}{$}
}

\let\TypChoiceSep\IListNewLineShort%
\let\PrcRecvSep\IListNewLine%
\let\PrcBranchAfterSep\IListNewLineLong%

\newcommand*\circled[1]{\tikz[baseline=(char.base)]{
    \node[shape=circle,draw,inner sep=1pt] (char) {#1};}}

\newcommand*\squared[1]{\tikz[baseline=(char.base)]{
    \node[shape=rectangle,draw,inner sep=1.5pt] (char) {#1};}}

    \NewDocumentCommand{\Squared}{s m}{%
    \IfBooleanT{#1}{$}
    \tikz[baseline=(char.base)]{%
        \node[shape=rectangle,draw,inner sep=1.5pt] (char) {$\mathtt{#2}$};%
    }%
    \IfBooleanT{#1}{$}
}

\NewDocumentCommand{\Circled}{s m}{%
    \IfBooleanT{#1}{$}
    \tikz[baseline=(char.base)]{%
        \node[shape=circle,draw,inner sep=1pt] (char) {$\mathtt{#2}$};%
    }%
    \IfBooleanT{#1}{$}
}

\NewDocumentCommand{\Rounded}{s m}{%
    \IfBooleanT{#1}{$}
    \tikz[baseline=(char.base)]{%
        \node[shape=rectangle,rounded corners,draw,inner sep=2pt] (char) {$\mathtt{#2}$};%
    }%
    \IfBooleanT{#1}{$}
}

\NewDocumentCommand\ListToBuffContents{>{\SplitList{,}}m}
{%
    {\ProcessList{#1}{\HeadBuffContents}}%
}

\newcommand\HeadBuffContents[1]{%
    \circled{$\mathtt{#1}$}%
    \let\HeadBuffContents\TailBuffContents%
}

\newcommand\TailBuffContents[1]{%
    \!\;\cdot\,\circled{$\mathtt{#1}$}%
}

\NewDocumentCommand{\IBuff}{s O{\ISes} e:}{%
    \IfBooleanT{#1}{$}
    {#2}:%
    \IfNoValueTF{#3}{
        \emptyset%
    }{
        {\ListToBuffContents{#3}}%
    }%
    \IfBooleanT{#1}{$}
}

\NewDocumentCommand\ListToQueContents{>{\SplitList{,}}m}
{%
    {\ProcessList{#1}{\HeadQueContents}}%
}

\newcommand\HeadQueContents[1]{%
    \squared{$\mathtt{#1}$}%
    \let\HeadQueContents\TailQueContents%
}

\newcommand\TailQueContents[1]{%
    \!\;;\,\squared{$\mathtt{#1}$}%
}

\NewDocumentCommand{\IQue}{s O{\ISes} e:}{%
    \IfBooleanT{#1}{$}
    {#2}:%
    \IfNoValueTF{#3}{
        \emptyset%
    }{
        {\ListToQueContents{#3}}%
    }%
    \IfBooleanT{#1}{$}
}

\NewDocumentCommand{\CIso}{s t~ O{\ValClocks} e+ t; t~ O{\TypeS} t' e_}{%
    \IfBooleanT{#1}{$}%
    \IfValueTF{#9}{
        \Tup(\IfBooleanTF{#2}{\Dual[{#3}_{#9}\IfBooleanT{#8}{'}]\IfValueT{#4}{#4}}{{#3}_{#9}\IfBooleanT{#8}{'}\IfValueT{#4}{#4}},\IfBooleanTF{#6}{\Dual[{#7}_{#9}\IfBooleanT{#8}{'}]}{{#7}_{#9}\IfBooleanT{#8}{'}})%
    }{
        \Tup(\IfBooleanTF{#2}{\Dual[{#3}\IfBooleanT{#8}{'}]\IfValueT{#4}{#4}}{{#3}\IfBooleanT{#8}{'}\IfValueT{#4}{#4}},\IfBooleanTF{#6}{\Dual[{#7}\IfBooleanT{#8}{'}]}{{#7}\IfBooleanT{#8}{'}})%
    }%
    \IfBooleanT{#1}{$}%
}

\NewDocumentCommand{\CSoc}{s O{\ValClocks} e+ t; O{\TypeS} E:{\Queue} e+ t' e_}{%
    \IfBooleanT{#1}{$}
    \Tup(
    {{#2\IfValueT{#9}{_{#9}}\IfBooleanT{#8}{'}}\IfValueT{#3}{#3}},%
    {#5\IfValueT{#9}{_{#9}}\IfBooleanT{#8}{'}},%
    {{#6\IfValueT{#9}{_{#9}}\IfBooleanT{#8}{'}}\IfValueT{#7}{;#7}})%
    \IfBooleanT{#1}{$}%
}

\NewDocumentCommand{\CSys}{s O{\VSoc_1} O{\VSoc_2}}{%
    \IfBooleanT{#1}{$}%
    \Parl{{#2},{#3}}%
    \IfBooleanT{#1}{$}%
}

\NewDocumentCommand{\VIso}{s t' e_}{%
    \IfBooleanT{#1}{$}%
    \IfBooleanTF{#2}{
        \IfValueTF{#3}{
            {\IsoCfg}_{#3}'%
        }{
            {\IsoCfg'}%
        }%
    }{
        \IfValueTF{#3}{
            {\IsoCfg}_{#3}%
        }{
            {\IsoCfg}%
        }%
    }%
    \IfBooleanT{#1}{$}%
}

\NewDocumentCommand{\VSoc}{s t' t' e_}{%
    \IfBooleanT{#1}{$}%
    \IfBooleanTF{#2}{
        \IfBooleanTF{#3}{
            \IfValueTF{#4}{
                {\SocCfg}_{#4}''%
            }{
                {\SocCfg''}%
            }%
        }{
            \IfValueTF{#4}{
                {\SocCfg}_{#4}'%
            }{
                {\SocCfg'}%
            }%
        }%
    }{
        \IfValueTF{#4}{
            {\SocCfg}_{#4}%
        }{
            {\SocCfg}%
        }%
    }%
    \IfBooleanT{#1}{$}%
}

\NewDocumentCommand{\VSys}{s t' e_ t' e_}{%
    \IfBooleanT{#1}{$}%
    \IfBooleanTF{#4}{
        \IfBooleanTF{#2}{
            \IfValueTF{#3}{
                \IfValueTF{#5}{
                    \Parl{\VSoc'_{#3},\VSoc'_{#5}}%
                }{
                    \Parl{\VSoc'_{#3},\VSoc'_{2}}%
                }%
            }{
                \IfValueTF{#5}{
                    \Parl{\VSoc'_{1},\VSoc'_{#5}}%
                }{
                    \Parl{\VSoc''_1,\VSoc''_2}%
                }%
            }%
        }{
            \Parl{%
                \IfValueTF{#3}{\VSoc_{#3}}{\VSoc_1},%
                \IfValueTF{#5}{\VSoc'_{#5}}{\VSoc'_2}%
            }%
        }%
    }{
        \IfBooleanTF{#2}{
            \IfValueTF{#3}{
                \Parl{%
                    \VSoc'_{#3},%
                    \IfValueTF{#5}{\VSoc_{#5}}{\VSoc_2}%
                }%
            }{
                \IfValueTF{#5}{
                    \Parl{%
                        \VSoc'_1,%
                        \VSoc'_{#5}%
                    }%
                }{
                    \Parl{%
                        \VSoc'_1,%
                        \VSoc'_2%
                    }%
                }%
            }%
        }{
            \Parl{%
                \IfValueTF{#3}{\VSoc_{#3}}{\VSoc_1},%
                \IfValueTF{#5}{\VSoc_{#5}}{\VSoc_2}%
            }%
        }%
    }%
    \IfBooleanT{#1}{$}%
}

\NewDocumentCommand{\FmtJudgement}{s t| m m o}{%
    \IfBooleanT{#1}{$}%
    \begin{array}[t]{l}{#3}\end{array}%
    \IfBooleanTF{#2}{\ \\\ }{\!\;}%
    {\Entails}%
    \IfBooleanTF{#2}{\ \\\ }{\!\,}%
    \begin{array}[t]{l}{#4}\end{array}%
    \IfValueT{#5}{
        \IfBooleanTF{#2}{\ \\\ }{\!\;}%
        {\PrcTyped}%
        \IfBooleanTF{#2}{\ \\\ }{\!\,}%
        \begin{array}[t]{l}{#5}\end{array}%
    }%
    \IfBooleanT{#1}{$}%
}

\NewDocumentCommand{\ISes}{s t| O{\PrcQ} O{\PrcP}}{%
    \IfBooleanT{#1}{$}
    \IfBooleanTF{#2}{
        {#4}{#3}%
    }{
        {#3}{#4}%
    }%
    \IfBooleanT{#1}{$}
}

\NewDocumentCommand\ListToEnvItems{>{\SplitList{,}}m}{%
    {\ProcessList{#1}{\ListEnvItem}}%
}

\newcommand\ListEnvItem[1]{%
,~{#1}%
}

\NewDocumentCommand{\IConstOr}{s t| O{\Const} E_{i} E:{\SetI}}{%
\IfBooleanT{#1}{$}
\IfBooleanTF{#2}{
    \Past[{\SetOr_{\hspace{-0.5ex}\vspace{0.5ex}\scriptstyle{\AsSet[#4][#5]}}}{#3_{#4}}]%
}{
    {\SetOr_{\hspace{-0.5ex}\vspace{0.5ex}\scriptstyle{\AsSet[#4][#5]}}}{#3_{#4}}%
}%
\IfBooleanT{#1}{$}
}

\NewDocumentCommand{\ActFE}{s t! t? t' m t|}{%
    \IfBooleanT{#1}{$}
    {#5}\!\stackrel{\IfBooleanTF{#2}{\TypSend}{\IfBooleanTF{#3}{\TypRecv}{\TypComm\IfBooleanT{#4}{'}}}}{\IfBooleanT{#6}{\not}\Rightarrow}%
    \IfBooleanT{#1}{$}
}

\NewDocumentCommand{\Parl}{s m}{%
    \IfBooleanT{#1}{$}%
    \ListToParl{#2}%
    \IfBooleanT{#1}{$}%
}

\NewDocumentCommand\ListToParl{>{\SplitList{,}}m}
{%
    {\ProcessList{#1}{\HeadParlContents}}%
}

\newcommand\HeadParlContents[1]{%
    {#1}%
    \let\HeadParlContents\TailParlContents%
}

\newcommand\TailParlContents[1]{%
    \!~\!\mid\!~\!{#1}%
}

\NewDocumentCommand{\Tup}{s D(){\emptyset} t|}{%
    \IfBooleanT{#1}{$}%
    \IfBooleanTF{#3}{
        \left(\ListToTupleContentsTypes{#2}\right)%
    }{
        ({\ListToTupleContents{#2}})%
    }%
    \IfBooleanT{#1}{$}%
}

\NewDocumentCommand\ListToTupleContents{>{\SplitList{,}}m}{%
    {\ProcessList{#1}{\HeadTupleContents}}%
}

\newcommand\HeadTupleContents[1]{%
    \!\;{#1}%
    \let\HeadTupleContents\TailTupleContents%
}

\newcommand\TailTupleContents[1]{%
    \!\,,~{#1}%
    \let\HeadTupleContents\TailTupleContents%
}

\NewDocumentCommand\ListToTupleContentsTypes{>{\SplitList{,}}m}{%
    {\ProcessList{#1}{\HeadTupleContentsTypes}}%
}

\newcommand\HeadTupleContentsTypes[1]{%
    \!\;{#1}%
    \let\HeadTupleContentsTypes\TailTupleContentsTypes%
}

\newcommand\TailTupleContentsTypes[1]{%
    \!\,,~{#1}%
    \let\HeadTupleContentsTypes\TailTupleContentsTypes%
}

\NewDocumentCommand{\IMsgType}{s o d<> t' e_}{%
    \IfBooleanT{#1}{$}
    \IfValueTF{#2}{
        {#2\IfBooleanT{#4}{'}\IfValueT{#5}{_{#5}}}%
    }{
        {\MsgLabel\IfBooleanT{#4}{'}\IfValueT{#5}{_{#5}}}%
    }%
    \left\langle%
    \IfValueTF{#3}{
        {#3\IfBooleanT{#4}{'}\IfValueT{#5}{_{#5}}}%
    }{
        {\DataType\IfBooleanT{#4}{'}\IfValueT{#5}{_{#5}}}%
    }%
    \right\rangle%
    \IfBooleanT{#1}{$}
}

\NewDocumentCommand{\DelType}{s o t' e_}{%
    \IfBooleanT{#1}{$}
    \IfValueTF{#2}{
        \Tup(#2)%
    }{
        \Tup({{\Const\IfBooleanT{#3}{'}}\IfValueT{#4}{_{#4}}},{{\TypeS\IfBooleanT{#3}{'}}\IfValueT{#4}{_{#4}}})%
    }%
    \IfBooleanT{#1}{$}
}

\NewDocumentCommand{\ITypeAct}{s t! t? t! t| O{\MsgType} e_}{%
    \IfBooleanT{#1}{$}
    \begin{array}[t]{>{\hfill$}p{1ex}<{$\hfill} c}%
        \IfBooleanTF{#2}{
            \IfBooleanTF{#3}{
                \mathllap{\TypComm\IfValueT{#7}{_{#7}}}%
            }{
                \mathllap{\TypSend\IfValueT{#7}{_{#7}}}%
            }%
        }{
            \IfBooleanTF{#3}{
                \IfBooleanTF{#4}{
                    \mathllap{\TypComm\IfValueT{#7}{_{#7}}}%
                }{
                    \mathllap{\TypRecv\IfValueT{#7}{_{#7}}}%
                }%
            }{
                \mathllap{\TypComm\IfValueT{#7}{_{#7}}}%
            }%
        } & %
            {#6}\IfValueT{#7}{_{#7}}%
    \end{array}%
    \IfBooleanT{#1}{$}
}

\NewDocumentCommand{\ITypeCond}{s O{\mathtt{true}} t| o e_}{%
    \IfBooleanT{#1}{$}
    \IfBooleanTF{#3}{
        \IfNoValueTF{#4}{
            \Tup({\mathtt{true}\IfValueT{#5}{_{#5}}},{{#2}\IfValueT{#5}{_{#5}}})|%
        }{
            \Tup({{#2}\IfValueT{#5}{_{#5}}},{{#4}\IfValueT{#5}{_{#5}}})|%
        }%
    }{
        \IfNoValueTF{#4}{
            \Tup({{#2}\IfValueT{#5}{_{#5}}})|%
        }{
            \Tup({{#2}\IfValueT{#5}{_{#5}}},{{#4}\IfValueT{#5}{_{#5}}})|%
        }%
    }%
    \IfBooleanT{#1}{$}
}

\NewDocumentCommand{\IType}{s e! e? O{\TypeTrue} t| o e. e_}{%
    \IfBooleanT{#1}{$}
    \!\rule{0pt}{1.5ex}%
    \IfValueTF{#2}{
        \IfValueTF{#8}{\ITypeAct![#2]_{#8}}{\ITypeAct![#2]}%
    }{
        \IfValueTF{#3}{
            \IfValueTF{#8}{\ITypeAct?[#3]_{#8}}{\ITypeAct?[#3]}%
        }{
            \!\hspace{1ex}%
            \IfValueTF{#8}{\ITypeAct|[\MsgType]_{#8}}{\ITypeAct|[\MsgType]}%
        }%
    }%
    \IfValueTF{#6}{
        \IfBooleanTF{#5}{
            \IfValueTF{#8}{\ITypeCond[#4]|[#6]_{#8}}{\ITypeCond[#4]|[#6]}%
        }{
            \IfValueTF{#8}{\ITypeCond[#4][#6]_{#8}}{\ITypeCond[#4][#6]}%
        }%
    }{
        \IfBooleanTF{#5}{
            \IfValueTF{#8}{\ITypeCond[#4]|_{#8}}{\ITypeCond[#4]|}%
        }{
            \IfValueTF{#8}{\ITypeCond[#4]_{#8}}{\ITypeCond[#4]}%
        }%
    }%
    \IfValueTF{#7}{.\,{#7}\IfValueT{#8}{_{#8}}}{.\TypeEnd\IfValueT{#8}{_{#8}}}%
    \IfBooleanT{#1}{$}
}

\NewDocumentCommand{\IChoice}{s t^ t| O{\IType}}{%
   \IfBooleanT{#1}{$}
   \IfBooleanTF{#3}{
      \begin{array}[\IfBooleanTF{#2}{t}{c}]\{{l}\}{#4}\end{array}%
   }{
      \begin{array}[\IfBooleanTF{#2}{t}{c}]\{{l}\}%
         \ListToArray{#4}%
      \end{array}%
   }%
   \IfBooleanT{#1}{$}
}

\NewDocumentCommand{\ListToArray}{>{\SplitList{,}}m}{%
   \ProcessList{#1}{\HeadItemToArrayElement}%
}

\NewDocumentCommand{\HeadItemToArrayElement}{m}{%
   \rule{0pt}{1.5ex}{#1}%
   \let\HeadItemToArrayElement\TailItemToArrayElement%
}
\NewDocumentCommand{\TailItemToArrayElement}{m}{%
   \TypChoiceSep\rule{0pt}{1.5ex}{#1}%
   \let\HeadItemToArrayElement\TailItemToArrayElement%
}

\NewDocumentCommand{\IRecDef}{s o E:{\TypRecLabel} e.}{%
\IfBooleanT{#1}{$}
{\mu{#3}\IfValueT{#2}{^{#2}}}%
\IfValueTF{#4}{.\,{#4}}{.\TypeEnd}%
\IfBooleanT{#1}{$}
}

\NewDocumentCommand{\IRecCall}{s o E:{\TypRecLabel}}{%
\IfBooleanT{#1}{$}
{{#3}\IfValueT{#2}{^{#2}}}%
\IfBooleanT{#1}{$}
}

\NewDocumentCommand{\IScope}{s O{\ISes} e.}{%
    \IfBooleanT{#1}{$}
    \begin{array}[t]{l l}%
        \begin{array}[t]({l})%
            {\PrcScope}{#2}%
        \end{array}%
        & %
        \begin{array}[t]({l})%
            \IfValueTF{#3}{#3}{\PrcEnd}%
        \end{array}%
    \end{array}%
    \IfBooleanT{#1}{$}
}

\NewDocumentCommand{\PDelay}{s O{0} t/ e.}{%
    \IfBooleanT{#1}{$}%
    \begin{array}[t]{l l}%
        \IfBooleanTF{#3}{
            \mathrlap{\FuncDelay\left({#2}\right)\;\!.} & %
            \\ \mbox{ } & {#4}%
        }{
            \FuncDelay\left({#2}\right)\;\!.\,%
            {#4}%
        }%
    \end{array}%
    \IfBooleanT{#1}{$}%
}

\NewDocumentCommand{\PSet}{s m t/ e.}{%
    \IfBooleanT{#1}{$}%
    \begin{array}[t]{l}%
        \IfBooleanTF{#3}{
        \mathrlap{\TTPrcSet\left(\mathtt{#2}\right)} %
        \\ .{#4}%
        }{
        \TTPrcSet\left(\mathtt{#2}\right).{#4}%
        }%
    \end{array}%
    \IfBooleanT{#1}{$}%
}

\NewDocumentCommand{\PFail}{s}{%
    \IfBooleanT{#1}{$}%
    \PrcFail%
    \IfBooleanT{#1}{$}%
}

\NewDocumentCommand{\IMJudgement}{s O{\VarEnv} m}{%
    \IfBooleanTF{#1}{$\FmtJudgement}{\FmtJudgement|}%
    {
        {#2}%
    }{
        {\ListToTupleContents{#3}}%
    }%
    \IfBooleanT{#1}{$}%
}

\NewDocumentCommand{\IMsg}{s m E:{\DataType}}{%
    \IfBooleanT{#1}{$}%
    {#2}:{#3}%
    \IfBooleanT{#1}{$}%
}

\NewDocumentCommand{\PSend}{s E_{\PrcP} t| O{\PrcMsg} t/ e.}{%
    \IfBooleanT{#1}{$}%
    \rule{0pt}{1.5ex}%
        \begin{array}[t]{l l}%
            \mathtt{on}\;{#2}\;%
            \bm{\mathtt{send}}\;%
            \IfBooleanTF{#3}{#4}{\Rounded{#4}}%
            \IfBooleanTF{#5}{
                \;\!\,. \\ \mbox{ }%
                \begin{array}[t]{l}%
                    \IfValueTF{#6}{#6}{\PrcEnd}%
                \end{array}%
            }{
                & .\!\,%
                \begin{array}[t]{l}%
                    \IfValueTF{#6}{#6}{\PrcEnd}%
                \end{array}%
            }
    \end{array}%
    \IfBooleanT{#1}{$}%
}

\NewDocumentCommand{\PRecv}{s E_{\PrcP} t| O{\PrcMsg} t/ E.{\PrcEnd} e< E>{\PrcFail}}{%
    \IfBooleanT{#1}{$}%
    \IfBooleanTF{#3}{
        \begin{array}[t]{>{\hfill$}p{3ex}<{ $} c l}%
            \rule{0pt}{1.5ex}\Rounded{#4}%
            & :%
            & \!\begin{array}[t]{l}{#5}\end{array}%
        \end{array}%
    }{
        \begin{array}[t]{r l}%
            \IfValueT{#7}{\mathllap{{\tikzmark{RecvNode}}}}%
            \mathtt{on}\;{#2}\;%
            \IfValueTF{#7}{
                \bm{\mathtt{recv}}^{#7}%
            }{
                \bm{\mathtt{recv}}^{\infty}%
            }%
            \IfBooleanTF{#5}{
                \begin{array}[t]{r c l}%
                    \rule{0pt}{1.5ex}\Rounded{#4}%
                    & \;\!\,: \\ &%
                    & \hspace{-8ex}\begin{array}[t]{l}{#6}\end{array}%
                    \IfValueT{#7}{
                        \\ \mathllap{\tikzmark{AfterNode}{\mathtt{after}^{#7}}}%
                        & :%
                        & \!\begin{array}[t]{l}{#8}\end{array}%
                        \\[0.75ex]%
                    }%
                \end{array}%
            }{
                \begin{array}[t]{r c l}%
                    \rule{0pt}{1.5ex}\Rounded{#4}%
                    & :%
                    & \!\begin{array}[t]{l}{#6}\end{array}%
                    \IfValueT{#7}{
                        \\ \mathllap{\tikzmark{AfterNode}{\mathtt{after}^{#7}}}%
                        & :%
                        & \!\begin{array}[t]{l}{#8}\end{array}%
                        \\[0.75ex]%
                    }%
                \end{array}%
            }%
        \end{array}%
    }%
    \IfValueT{#7}{%
    }%
    \IfBooleanT{#1}{$}%
}

\NewDocumentCommand{\PBranch}{s E_{\PrcP} O{\PRecv|} e< E>{\PFail}}{%
    \IfBooleanT{#1}{$}%
    \begin{array}[t]{r l}%
        \mathtt{on}\;{#2}\;%
        \IfValueTF{#4}{
            \bm{\mathtt{recv}}^{#4}%
        }{
            \bm{\mathtt{recv}}^{\infty}%
        } & %
        \begin{array}[t]\{{l}\}%
            \PrcListToArray{#3}%
        \end{array}%
        \IfValueT{#4}{
            \PrcBranchAfterSep \rule{0pt}{3ex}{\mathtt{after}}%
            & \begin{array}[t]{l l}%
                :%
                & \!\begin{array}[t]{l}{#5}\end{array}%
            \end{array}%
        }%
    \end{array}%
    \IfBooleanT{#1}{$}%
}

\NewDocumentCommand{\PrcListToArray}{>{\SplitList{,}}m}{%
    \ProcessList{#1}{\HeadPrcItemToArrayElement}%
}

\NewDocumentCommand{\HeadPrcItemToArrayElement}{m}{%
    \rule{0pt}{1.5ex}{#1}%
    \let\HeadPrcItemToArrayElement\TailPrcItemToArrayElement%
}
\NewDocumentCommand{\TailPrcItemToArrayElement}{m}{%
    \PrcRecvSep\rule{0pt}{1.5ex}{#1}%
}

\NewDocumentCommand{\PIf}{s O{true} E:{\PThen} E:{\PElse}}{%
    \IfBooleanT{#1}{$}%
    \begin{array}[t]{l l}%
        \TTPrcIf\;\mathrlap{\mathtt{(#2)}}%
        \\ #3%
        \\ #4%
    \end{array}%
    \IfBooleanT{#1}{$}%
}

\NewDocumentCommand{\PThen}{s O{\PrcEnd}}{%
    \IfBooleanT{#1}{$}%
    {\TTPrcThen} & {: \begin{array}[t]{l}{{#2}}\end{array}}%
    \IfBooleanT{#1}{$}%
}

\NewDocumentCommand{\PElse}{s O{\PFail}}{%
    \IfBooleanT{#1}{$}%
    {\TTPrcElse} & {: \begin{array}[t]{l}{{#2}}\end{array}}%
    \IfBooleanT{#1}{$}%
}

\NewDocumentCommand{\PDefRec}{s E:{\RecPrcVar} o E={\TypRecLabel} e.}{%
    \IfBooleanT{#1}{$}%
    \begin{array}[t]{l l}%
        \TTPrcDef & %
        \IfValueTF{#3}{{#2}^{#3}={#4}^{#3}}{{#2}={#4}}\;%
        \TTPrcDefIn:%
        \\ & %
        \rule{0pt}{1ex}%
        \IfValueTF{#5}{
            {#5}
        }{\PrcEnd}%
        \;%
    \end{array}%
    \IfBooleanT{#1}{$}%
}

\NewDocumentCommand{\PRec}{s E:{\TypRecLabel} o}{%
    \IfBooleanT{#1}{$}%
    \IfValueTF{#3}{{#2}^{#3}}{{#2}}%
    \IfBooleanT{#1}{$}%
}

\NewDocumentCommand{\ICalcBranch}{s E1{{\SesP}} E2{{\PrcAfter}} E3{{\PrcMsg}} E4{{\Prc}} O{i} O{I} t| O{\Qrc}}{%
\IfBooleanT{#1}{$}
{#2}^{#3}\PrcRecv\,\InSet[{#4}_{#6}:{#5}_{#6}][#6][#7]%
\IfBooleanF{#8}{\;\TTPrcAft\,{#9}}
\IfBooleanT{#1}{$\!}
}

\NewDocumentCommand{\ICalcRecv}{s E_{\PrcP} O{\PrcMsg} E.{\Prc} e<}{%
\IfBooleanT{#1}{$}
{#2}^{\IfValueTF{#5}{#5}{\infty}}%
\PrcRecv\,{#3}:{#4}%
\IfBooleanT{#1}{$\!}
}

\NewDocumentCommand{\ITJRec}{s O{\emptyset} e:}{%
    \IfBooleanT{#1}{$}%
    {#2}%
    \IfValueT{#3}{
        \ListToEnvItems{#3}%
    }%
    \IfBooleanT{#1}{$}%
}

\NewDocumentCommand{\ITJCon}{s t| O{\TypeTrue} e_ E:{\SetI}}{%
    \IfBooleanT{#1}{$}%
    \begin{array}[t]{l}%
        \IfValueTF{#4}{
            \IfBooleanTF{#2}{
                \IConstOr|[#3]_{#4}:{#5}%
            }{
                \IConstOr[#3]_{#4}:{#5}%
            }%
        }{
            \IfBooleanTF{#2}{
                \Past[#3]%
            }{
                {#3}%
            }%
        }%
    \end{array}%
    \IfBooleanT{#1}{$}%
}

\NewDocumentCommand{\TypEnvCond}{s E:{\TypEnv} O{\TypeS}}{%
    \IfBooleanT{#1}{$}%
    \FmtJudgement{#2}{#3}%
    \IfBooleanT{#1}{$}%
}

\NewDocumentCommand{\ITJTyp}{s t| O{\TypeEnd} e_ E<{\SetI}}{%
    \IfBooleanT{#1}{$}%
    \IfBooleanTF{#2}{
        \ListToTupleContents{#3}%
    }{
        \IfValueTF{#4}{
            {\begin{array}[t]\{{l}\}%
                \!\ListToTupleContents{#3}%
            \end{array}}_{\AsSet[#4][#5]}%
        }{
            \begin{array}[t]\{{l}\}%
                \!\ListToTupleContents{#3}%
            \end{array}%
        }%
    }%
    \IfBooleanT{#1}{$}%
}

\NewDocumentCommand{\ITJudgement}{s O{\RecEnv} e: t| E;{\TypeTrue} o e- e_ E<{\SetI}}{%
\IfBooleanTF{#1}{$\FmtJudgement}{\FmtJudgement|}%
{
    \IfNoValueTF{#3}{
        \ITJRec%
    }{
        \ITJRec[#2]:{#3}%
    }%
    \!\,;\,%
    \IfBooleanTF{#4}{
        \IfValueTF{#8}{
            \ITJCon|[#5]_{#8}:{#9}%
        }{
            \ITJCon|[#5]%
        }%
    }{
        \IfValueTF{#8}{
            \ITJCon[#5]_{#8}:{#9}%
        }{
            \ITJCon[#5]%
        }%
    }%
}{
    \IfNoValueTF{#7}{
        \IfValueTF{#6}{
            \IfValueTF{#8}{
                \ITJTyp|[#6]_{#8}<{#9}%
            }{
                \ITJTyp|[#6]%
            }%
        }{
            \IfValueTF{#8}{
                \ITJTyp|[\TypeEnd]_{#8}<{#9}%
            }{
                \ITJTyp|[\TypeEnd]%
            }%
        }%
    }{
        \IfValueTF{#8}{
            \ITJTyp[#7]_{#8}<{#9}%
        }{
            \ITJTyp[#7]%
        }%
    }%
}%
\IfBooleanT{#1}{$}%
}

\NewDocumentCommand{\IPJEnv}{s O{\VarEnv} e:}{%
    \IfBooleanT{#1}{$}%
    {#2}%
    \IfValueT{#3}{
        \ListToEnvItems{#3}%
    }%
    \IfBooleanT{#1}{$}%
}

\NewDocumentCommand{\IPJPrc}{s o}{%
    \IfBooleanT{#1}{$}%
    \IfNoValueTF{#2}{\Prc}{
        \Parl{#2}%
    }%
    \IfBooleanT{#1}{$}%
}

\NewDocumentCommand{\IPJSes}{s O{\SesEnv} e:}{%
    \IfBooleanT{#1}{$}%
    {#2}%
    \IfValueT{#3}{
        \ListToEnvItems{#3}%
    }%
    \IfBooleanT{#1}{$}%
}

\NewDocumentCommand{\IPJudgement}{s O{\VarEnv} e: t< O{\Prc} t> O{\SesEnv} e:}{%
\IfBooleanTF{#1}{$\FmtJudgement}{\FmtJudgement|}%
{
\IfNoValueTF{#3}{\IPJEnv[#2]}{
\IPJEnv[#2]:{#3}%
}%
}{
\IPJPrc[#5]%
}[
\IfNoValueTF{#8}{\IPJSes[#7]}{
\IPJSes[#7]:{#8}%
}%
]%
\IfBooleanT{#1}{$}%
}

\NewDocumentCommand{\IEval}{s t< o e: E:{\Premise}}{%
\IfBooleanT{#1}{$}%
\IfValueTF{#3}{
    \IfBooleanTF{#2}{
        {\inferrule*[Left={[#3]}]{%
                    \rule{0pt}{2ex}{#5}%
                }{%
                    {#4}%
                }}%
    }{
        {\inferrule*[Right={[#3]}]{%
                    \rule{0pt}{2ex}{#5}%
                }{%
                    {#4}%
                }}%
    }%
}{
    {\inferrule*{%
                \rule{0pt}{2ex}{#5}%
            }{%
                {#4}%
            }}%
}%
\IfBooleanT{#1}{$}%
}

\NewDocumentCommand{\GenericJudgement}{s O{\emptyset\Entails\Prc\PrcTyped\SesEnv}}{%
    {\ListToInlineArgs{#2}}%
}

\NewDocumentCommand{\Premise}{s t| O{\dots}}{%
    \IfBooleanT{#1}{$}%
    \IfBooleanTF{#2}{
        {#3}%
    }{
        {\ListToInlineArgs{#3}}%
    }%
    \IfBooleanT{#1}{$}%
}

\NewDocumentCommand\ListToInlineArgs{>{\SplitList{,}}m}
{%
    \ProcessList{#1}{\HeadItemToInlineArgs}%
}

\newcommand\HeadItemToInlineArgs[1]{%
    {#1}%
    \let\HeadItemToInlineArgs\TailItemToInlineArgs%
}

\newcommand\TailItemToInlineArgs[1]{%
    \quad{#1}%
    \let\HeadItemToInlineArgs\TailItemToInlineArgs
}

\NewDocumentCommand{\TypRuleEnd}{s}{%
    \IfBooleanT{#1}{$}
    \infer{
    \TypJudge2{\TypeTrue}3{\TypeEnd}
    }{
    \mbox{\rule{0pt}{1ex}}
    }
    \IfBooleanT{#1}{$}
}

\NewDocumentCommand{\TypRuleVar}{s}{%
    \IfBooleanT{#1}{$}
    \infer{
        \TypJudge1{\RecEnv,\TypRecLabel:\Const}3{\TypRecLabel}
    }{
    \mbox{\rule{0pt}{1ex}}
    }
    \IfBooleanT{#1}{$}
}

\NewDocumentCommand{\TypRuleRec}{s}{%
    \IfBooleanT{#1}{$}
    \infer{
        \TypJudge3{\TypRecDef.\TypeS}
    }{
        \TypJudge1{\RecEnv,\TypRecLabel:\Const}3{\TypeS}
    }
    \IfBooleanT{#1}{$}
}

\NewDocumentCommand{\TypRuleChoice}{s}{%
\IfBooleanT{#1}{$}
\infer{
\TypJudge|[i][i]
}{
\begin{array}[t]{c}
    \forall i\in\SetI
    \quad
    \TypJudge2{\TypEnv_i}3{\TypeS_i}
    \quad
    \RSet_i\models\gamma_i
    \qquad
    \DataType_i = \TmpTypBTDelegate''
    \implies
    \TypJudge1{\emptyset}2{\TypEnv'}3{\TypeS'}
    \quad
    \Const'\models\TypEnv'
    \\[1ex]
    \forall\AsSet[i,j][i]: i\neq j
    \quad
    \Const_i\cap\Const_j = \emptyset
    \lor
    \TypComm_i = \TypComm_j
\end{array}
}
\IfBooleanT{#1}{$}
}

\NewDocumentCommand{\CfgIsoRuleInteract}{s}{%
\IfBooleanT{#1}{$}
\infer{
\CfgIso2{\TypInteract}
\Act[\TypComm_j\IMsgType_j]
\CfgIso1{\Resets1{\ValClocks}2{\RSet_j}}2{\TypeS_j}
}{
\ValClocks\models\Const_j
&
j\in\SetI
}
\IfBooleanT{#1}{$}
}

\NewDocumentCommand{\CfgIsoRuleRecursion}{s}{%
    \IfBooleanT{#1}{$}
    \infer{
        \CfgIso2{\CfgRecDef.\TypeS}
        \Act
        \CfgIso''
    }{
        \CfgIso2{\TypeS\Subst[\CfgRecDef.\TypeS][\CfgRecLabel]}
        \Act
        \CfgIso''
    }
    \IfBooleanT{#1}{$}
}

\NewDocumentCommand{\CfgIsoRuleIsoTime}{s}{%
    \IfBooleanT{#1}{$}
    \CfgIso
    \Act[\ValTime]
    \CfgIso1{\ValClocks+\ValTime}
    \IfBooleanT{#1}{$}
}

\NewDocumentCommand{\CfgSocRuleSend}{s}{%
    \IfBooleanT{#1}{$}
    \infer{
        \CfgSoc%
        \Act[\TypSend\Msg]%
        \CfgSoc''%
    }{
        \CfgIso%
        \Act[\TypSend\Msg]%
        \CfgIso''%
    }%
    \IfBooleanT{#1}{$}
}

\NewDocumentCommand{\CfgSocRuleEnqu}{s}{%
    \IfBooleanT{#1}{$}
    \CfgSoc%
    \Act[\TypRecv\Msg]%
    \CfgSoc3{\Queue;\Msg}%
    \IfBooleanT{#1}{$}
}

\NewDocumentCommand{\CfgSocRuleRecv}{s}{%
    \IfBooleanT{#1}{$}
    \infer{
        \CfgSoc3{\Msg;\Queue}%
        \Act[\SiltAction]%
        \CfgSoc''%
    }{
        \CfgIso%
        \Act[\TypRecv\Msg]%
        \CfgIso''%
    }%
    \IfBooleanT{#1}{$}
}

\NewDocumentCommand{\CfgSocRuleTime}{s}{%
\IfBooleanT{#1}{$}
\infer{
\CfgSoc%
\Act[\ValTime]%
\CfgSoc'%
}{
\CfgIso%
\Act[\ValTime]%
\CfgIso'%
&%
\FutureEn1{\CfgIso}[\TypComm]%
\implies%
\FutureEn1{\CfgIso'}[\TypComm']%
&%
\forall \ValTime'<\ValTime:%
\CfgSoc1{\ValClocks+\ValTime'}%
\Act|[\SiltAction]%
}%
\IfBooleanT{#1}{$}
}

\NewDocumentCommand{\CfgSysRuleComm}{s}{%
\IfBooleanT{#1}{$}
\infer[\LblCfgSysLComm]{
\Trans{\Parl{\VSoc_1,\VSoc_2}}:{\SiltAction}[\Parl{\VSoc'_1,\VSoc'_2}]
}{
\Trans{\VSoc_1}:{\TypSend\Msg}[\VSoc'_1]
&
\Trans{\VSoc_2}:{\TypRecv\Msg}[\VSoc'_2]
}
\IfBooleanT{#1}{$}
}

\NewDocumentCommand{\CfgSysRuleDequ}{s}{%
\IfBooleanT{#1}{$}
\infer[\LblCfgSysLPar]{
\Trans{\Parl{\VSoc_1,\VSoc_2}}:{\SiltAction}[\Parl{\VSoc'_1,\VSoc_2}]
}{
\Trans{\VSoc_1}:{\SiltAction}[\VSoc'_1]
}
\IfBooleanT{#1}{$}
}

\NewDocumentCommand{\CfgSysRuleWait}{s}{%
\IfBooleanT{#1}{$}
\infer[\LblCfgSysWait]{
\Trans{\Parl{\VSoc_1,\VSoc_2}}:{\ValTime}[\Parl{\VSoc'_1,\VSoc'_2}]
}{
\Trans{\VSoc_1}:{\ValTime}[\VSoc'_1]
&
\Trans{\VSoc_2}:{\ValTime}[\VSoc'_2]
}
\IfBooleanT{#1}{$}
}

\NewDocumentCommand{\PRedRuleARed}{s}{%
\IfBooleanT{#1}{$}
\infer{
\RedPrc{\Prc}%
~\PrcRed~%
\RedPrc'{\Qrc}%
}{
\RedPrc{\Prc}%
~\ARed~%
\RedPrc'{\Qrc}%
}%
\IfBooleanT{#1}{$}
}

\NewDocumentCommand{\PRedRuleAStr}{s}{%
\IfBooleanT{#1}{$}
\infer{
\RedPrc{\Prc}%
~\rightarrow~%
\RedPrc'{\Qrc}%
}{
\Prc\equiv\Prc'
& %
\RedPrc{\Prc'}%
~\rightarrow~%
\RedPrc'{\Qrc'}%
& %
\Qrc\equiv\Qrc'%
}%
\IfBooleanT{#1}{$}
}

\NewDocumentCommand{\PRedRuleScope}{s}{%
\IfBooleanT{#1}{$}
\infer{
\RedPrc{\PCalcScope}%
~\ARed~%
\RedPrc'{\PCalcScope[\Prc']}%
}{
\RedPrc{\Prc}%
~\ARed~%
\RedPrc'{\Prc'}%
}%
\IfBooleanT{#1}{$}
}

\NewDocumentCommand{\PRedRuleTRed}{s}{%
\IfBooleanT{#1}{$}
\infer{
\RedPrc{\Prc}%
~\PrcRed~%
\RedPrc'{\Qrc}%
}{
\RedPrc{\Prc}%
~\TRed~%
\RedPrc'{\Qrc}%
}%
\IfBooleanT{#1}{$}
}

\NewDocumentCommand{\PRedRuleTStr}{s}{%
\IfBooleanT{#1}{$}
\infer{
\RedPrc{\Prc}%
~\TRed~%
\RedPrc'{\Qrc}%
}{
\Prc\equiv\Prc'
& %
\RedPrc{\Prc'}%
~\TRed~%
\RedPrc'{\Qrc'}%
& %
\Qrc\equiv\Qrc'%
}%
\IfBooleanT{#1}{$}
}

\NewDocumentCommand{\PRedRuleDelay}{s}{%
   \IfBooleanT{#1}{$}
   \RedPrc{\Prc}
   ~\TRed~%
   \RedPrc[\RedTimers+\ValTime]{\PFuncTime[{\Prc}]}%
   \IfBooleanT{#1}{$}
}

\NewDocumentCommand{\PRedRulePar}{s}{%
\IfBooleanT{#1}{$}
\infer{
\RedPrc{%
   \Parl{%
      \Prc,%
      \Qrc%
   }%
}%
~\ARed~%
\RedPrc'{%
\Parl{%
   \Prc',%
   \Qrc%
}%
}%
}{
\RedPrc{\Prc}%
~\ARed~%
\RedPrc'{\Prc'}%
}%
\IfBooleanT{#1}{$}
}

\NewDocumentCommand{\PRedRuleDef}{s}{%
\IfBooleanT{#1}{$}
\infer{
\RedPrc{\PrcRecDef}%
~\ARed~%
\RedPrc'{\PrcRecDef>{\Qrc'}}%
}{
\RedPrc{\Qrc}%
~\ARed~%
\RedPrc'{\Qrc'}%
}%
\IfBooleanT{#1}{$}
}

\NewDocumentCommand{\PRedRuleIf}{s}{%
   \IfBooleanT{#1}{$}
   \infer{%
      \RedPrc{\PCalc{\If*{\Const}~\Then{P}~\Else{Q}}}%
      ~\ARed~%
      \RedPrc{\Prc}%
   }{%
      \RedTimers\models\Const%
   }%
   \IfBooleanT{#1}{$}
}

\NewDocumentCommand{\PRedRuleElse}{s}{%
   \IfBooleanT{#1}{$}
   \infer{%
      \RedPrc{\PCalc{\If*{\Const}~\Then{P}~\Else{Q}}}%
      ~\ARed~%
      \RedPrc{Q}%
   }{%
      \neg \RedTimers\models\Const%
   }%
   \IfBooleanT{#1}{$}
}

\NewDocumentCommand{\PRedRuleRec}{s}{%
   \IfBooleanT{#1}{$}
   \begin{array}[t]{l r}%
      \RedPrc{%
      \Parl{%
      \PrcRecDef'>{\RecPrcCall},%
      \Qrc%
      }%
      }%
      ~\ARed~%
      & %
      \mbox{\hspace{24ex}}%
      \\[0.5ex]%
      & %
      \mathllap{%
      \RedPrc{%
      \Parl{%
      \PrcRecDef'>{\Prc\RecPrcSubst/'},%
      \Qrc%
      }%
      }%
      }%
   \end{array}%
   \IfBooleanT{#1}{$}
}

\NewDocumentCommand{\PRedRuleSetTimer}{s}{%
   \IfBooleanT{#1}{$}
   \RedPrc{\PCalcSetTimer}%
   ~\ARed~%
   \RedPrc[\RedTimers{}{}{}[x\mapsto 0]]{\Prc}
   \IfBooleanT{#1}{$}
}

\NewDocumentCommand{\PRedRuleSend}{s}{%
\IfBooleanT{#1}{$}
\RedPrc{%
\Parl{%
\PCalc{\On{p}\Send{l}[\pl].{\Prc}},%
{{pq}:{h}}%
}%
}%
~\ARed~%
\RedPrc{%
\Parl{%
\Prc,%
{{pq}:{h \cdot {l\pl}}}%
}%
}%
}

\NewDocumentCommand{\PRedRuleRecvAfter}{s}{%
\infer{%
\RedPrc{%
\Parl{%
\PCalc{\On{p}\Recv{l}[\pl]:{P}_{i}[i]~\After<{e}:{Q}},%
{{qp}:{{l\pl} \cdot h}}%
}%
}%
~\ARed~%
\RedPrc{%
\Parl{%
\Prc_j\Subst[\pl][\pl_j],%
{{qp}:{h}}%
}%
}%
}{%
j\in I%
& l={l_j}%
}%
\IfBooleanT{#1}{$}
}

\NewDocumentCommand{\PRedRuleDet}{s}{%
   \IfBooleanT{#1}{$}
   \infer{
      \RedPrc{\PFuncDelay}%
      ~\ARed~%
      \RedPrc{\PFuncDelay[\ValTime]}%
   }{
      ~\models~%
      \Const\Subst[\ValTime][x]%
   }%
   \IfBooleanT{#1}{$}
}

\NewDocumentCommand{\PTypRuleBranch}{s}{%
\IfBooleanT{#1}{$}
\infer{
\VarEnv%
~\Entails~%
\PCalc{\On{p}\Recv{l}[\pl]_{i}[i]~\After<{n}:{\Qrc}}%
~\PrcTyped~%
\SesEnv,%
~\PrcP:\CIso;[\TypInteract[j][j]]%
}{
\begin{array}[t]{c}
    \Delta \textit{ not $n$-reading}
    \\[0.5ex]
    \forall j\in\SetJ:
    \ValClocks\models\Const_j
    ~\Implies~
    \begin{array}[t]{l}
        (1)~\TypComm_j=\TypRecv
        \qquad\qquad (2)~\forall  t \leq n: \ValClocks+\ValTime\models\Const_j
        \\[0.5ex] (3)~\exists i\in\SetI:{\VarEnv%
    ~\Entails~%
    \PCalc{\On{p}\Recv*<{n}{l}[\pl]_{i}}%
    ~\PrcTyped~%
    \SesEnv,%
    ~\PrcP:\CIso;[\TypInteract|_[j][j]]}%
    \end{array}
    \\[0.5ex]
    {\mathtt{AE}(p,n)
    ~\Implies~
    {\VarEnv+n%
    ~\Entails~%
    \Qrc%
    ~\PrcTyped~%
    \SesEnv+n,%
    ~\PrcP:\CIso+{+n};[\TypInteract[j][j]]}}%
\end{array}
}
\IfBooleanT{#1}{$}
}

\NewDocumentCommand{\PTypRuleVRecv}{s}{%
\IfBooleanT{#1}{$}
\infer{
\VarEnv%
~\Entails~%
\PCalc{\On{p}\Recv*<{n}{l}[\pl]}%
~\PrcTyped~%
\SesEnv,%
~\PrcP:\CIso;[\IType?{\IMsgType}[\Const][\RSet].{\TypeS}]%
}{
\begin{array}[t]{c}
    \DataType \textit{ base type }
    \\
    \forall t \leq n:
    {\ValClocks+\ValTime\models\Const}
    ~\land~
    {\VarEnv+\ValTime,~{\pl:\DataType}%
    ~\Entails~%
    \Prc%
    ~\PrcTyped~%
    \SesEnv+\ValTime,%
    ~\PrcP:\CIso+{+\ValTime\ReSet[]};}%
\end{array}
}
\IfBooleanT{#1}{$}
}

\NewDocumentCommand{\PTypRuleDRecv}{s}{%
\IfBooleanT{#1}{$}
\infer{
\VarEnv%
~\Entails~%
\PCalc{\On{p}\Recv*<{n}{l}[q]}%
~\PrcTyped~%
\SesEnv,%
~\PrcP:\CIso;[\IType?{\IMsgType}[\Const][\RSet].{\TypeS}]%
}{
\begin{array}[t]{c}
    \DataType = \CIso[\Const]'
    ~\land~%
    \ValClocks'\models\Const'
    \\
    \forall t \leq n:
    {\ValClocks+\ValTime\models\Const}
    ~\land~
    {\VarEnv+\ValTime%
    ~\Entails~%
    \Prc%
    ~\PrcTyped~%
    \SesEnv+\ValTime,%
    ~\PrcP:\CIso+{+\ValTime\ReSet[]};,
    ~q:\CIso'}%
\end{array}
}
\IfBooleanT{#1}{$}
}

\NewDocumentCommand{\PTypRuleVSend}{s}{%
\IfBooleanT{#1}{$}
\infer{
\VarEnv%
~\Entails~%
\PCalc{\On{p}\Send{l}[\pl]}%
~\PrcTyped~%
\SesEnv,%
~\PrcP:\CIso;[\TypInteract[i][i]]%
}{
\begin{array}[t]{l}
    \exists i\in I: 
    \TypComm_i=\TypSend
    ~\land~
    \ValClocks\models\Const_i
    ~\land~
    \PrcLabel=\MsgLabel_i
    ~\land~
    \DataType_i \textit{ base type }
    ~\land~
    {\VarEnv~\Entails~{\pl:\DataType_i}}
    \\[0.5ex]\qquad\qquad
    ~\land~
    {\VarEnv%
    ~\Entails~%
    \Prc%
    ~\PrcTyped~%
    \SesEnv,%
    ~\PrcP:\CIso+{\ReSet[]_i};[\TypeS_i]}%
\end{array}
}
\IfBooleanT{#1}{$}
}

\NewDocumentCommand{\PTypRuleDSend}{s}{%
\IfBooleanT{#1}{$}
\infer{
\VarEnv%
~\Entails~%
\PCalc{\On{p}\Send{l}[\pl]}%
~\PrcTyped~%
\SesEnv,%
~\PrcP:\CIso;[\TypInteract[i][i]],%
~\pl:\CIso'%
}{
\begin{array}[t]{l}
    \exists i\in I: 
    \TypComm_i=\TypSend
    ~\land~
    \ValClocks\models\Const_i
    ~\land~
    \PrcLabel=\MsgLabel_i
    ~\land~
    \DataType_i= \CIso[\Const]'
    ~\land~
    \ValClocks'\models\Const'
    \\[0.5ex]\qquad\qquad
    ~\land~
    {\VarEnv%
    ~\Entails~%
    \Prc%
    ~\PrcTyped~%
    \SesEnv,%
    ~\PrcP:\CIso+{\ReSet[]_i};[\TypeS_i]}%
\end{array}
}
\IfBooleanT{#1}{$}
}

\NewDocumentCommand{\PTypRuleIfTrue}{s}{%
    \IfBooleanT{#1}{$}
    \infer{
        \VarEnv%
        ~\Entails~%
        \PCalcIf%
        ~\PrcTyped~%
        \SesEnv%
    }{
        \VarEnv\models\Const%
        &
        \VarEnv%
        ~\Entails~%
        \Prc%
        ~\PrcTyped~%
        \SesEnv%
    }
    \IfBooleanT{#1}{$}
}

\NewDocumentCommand{\PTypRuleIfFalse}{s}{%
    \IfBooleanT{#1}{$}
    \infer{
        \VarEnv%
        ~\Entails~%
        \PCalcIf%
        ~\PrcTyped~%
        \SesEnv%
    }{
       \neg \VarEnv\models\Const%
        &
        \VarEnv%
        ~\Entails~%
        \Qrc%
        ~\PrcTyped~%
        \SesEnv%
    }
    \IfBooleanT{#1}{$}
}

\NewDocumentCommand{\PTypRuleTimer}{s}{%
\IfBooleanT{#1}{$}
\infer{
\TmpPrcJudge2{\PCalcSetTimer1{\Tix}}|3{\SesEnv}
}{
\Gamma\, [x\mapsto 0] \vdash P \triangleright \Delta
}
\IfBooleanT{#1}{$}
}

\NewDocumentCommand{\PTypRuleDelDelta}{s}{%
\IfBooleanT{#1}{$}
\infer{
\TmpPrcJudge2{\PFuncDelay}|3{\SesEnv}
}{
\forall\ValTime\in\Const:
\TmpPrcJudge2{\PFuncDelay[\ValTime]}|3{\SesEnv}
}
\IfBooleanT{#1}{$}
}

\NewDocumentCommand{\PTypRuleDelTime}{s}{%
\IfBooleanT{#1}{$}
\infer{
\TmpPrcJudge2{\PFuncDelay[\ValTime]}|3{\SesEnv}
}{
{\VarEnv+\ValTime%
        ~\Entails~%
        \Prc%
        ~\PrcTyped~%
        \SesEnv+\ValTime}%
& 
\Delta \textit{ not $t$-reading}
}
\IfBooleanT{#1}{$}
}

\NewDocumentCommand{\PTypRuleRes}{s}{%
\IfBooleanT{#1}{$}
\IEval:{
\IPJudgement<[\PCalcScope]>
}:{\Premise[
\Compat[\CSoc_1][\CSoc_2],
\IPJudgement<[\Prc]>:{%
\PrcP:\CIso_1,
\PrcQ:\CIso_2,
\PrcQ\PrcP:\Queue_1,%
\PrcP\PrcQ:\Queue_2%
}
]}
\IfBooleanT{#1}{$}
}

\NewDocumentCommand{\PTypRuleVar}{s}{%
    \IfBooleanT{#1}{$}
    \infer{
        \VarEnv,~\RecVar:\EnvPrcVar%
        ~\Entails~%
        \RecPrcCall%
        ~\PrcTyped~%
        \SesEnv%
    }{
        \SesEnv\in\SesSet
        &
        \forall i: {\VarEnv~\Entails~{%
        \RecSetMsg_i:\RecPrcMsg_i%
        }}
        &
        \forall j: {\VarEnv~\Entails~{%
        \RecSetTimers_j\models\RecPrcTimers_j%
        }}
    }
    \IfBooleanT{#1}{$}
}

\NewDocumentCommand{\PTypRulePar}{s}{%
\IfBooleanT{#1}{$}
\infer{
\TmpPrcJudge2{\Par[\Prc][\Qrc]}|3{\SesEnv_1,\SesEnv_2}
}{
\TmpPrcJudge2{\Prc}|3{\SesEnv_1}
&
\TmpPrcJudge2{\Qrc}|3{\SesEnv_2}
}
\IfBooleanT{#1}{$}
}

\NewDocumentCommand{\PTypRuleRec}{s}{%
    \IfBooleanT{#1}{$}
    \infer{
        \VarEnv%
        ~\Entails~%
        \PrcRecDef%
        ~\PrcTyped~%
        \SesEnv%
    }{
\begin{array}[t]{r}
        \forall \PSesSetCfgs\in\SesSet:%
        \VarEnv,~\RecSetMsg:\RecPrcMsg,~\RecSetTimers:\RecPrcTimers,~\RecVar:\EnvPrcVar%
        ~\Entails~%
        \Prc%
        ~\PrcTyped~%
        \RecSetRoles:\PSesSetCfgs%
        \quad
        \VarEnv,~{\RecVar:\EnvPrcVar}%
        ~\Entails~%
        \Qrc%
        ~\PrcTyped~%
        \SesEnv%
        \end{array}
    }
    \IfBooleanT{#1}{$}
}

\NewDocumentCommand{\PTypRuleWeak}{s}{%
\IfBooleanT{#1}{$}
\infer{
\TmpPrcJudge3{2{2{2{\TypeEnd}}}}
}{
\TmpPrcJudge|3{\SesEnv}
}
\IfBooleanT{#1}{$}
}

\NewDocumentCommand{\PTypRuleEmptyQ}{s}{%
\IfBooleanT{#1}{$}
\TmpPrcJudge2{\PCalcBuffer2{\emptyset}}|3{\PCalcBuffer2{\emptyset}}
\IfBooleanT{#1}{$}
}

\NewDocumentCommand{\PTypRuleVQue}{s}{%
    \IfBooleanT{#1}{$}
    \infer{
        \VarEnv%
        ~\Entails~%
        {{pq}:{l\pl}\cdot\QHead}%
        ~\PrcTyped~%
        \SesEnv,~{{pq}:{{l'\left\langle T \right\rangle};\Queue}}%
    }{
        \DataType \neq \CIso[\Const]
        & l=l'%
        &
        \VarEnv%
        ~\Entails~%
        {\pl:\DataType}%
        &
        \VarEnv%
        ~\Entails~%
        {\PrcP\PrcQ:\QHead}%
        ~\PrcTyped~%
        \SesEnv,~{\PrcP\PrcQ:\Queue}%
    }%
    \IfBooleanT{#1}{$}
}

\NewDocumentCommand{\PTypRuleDQue}{s}{%
    \IfBooleanT{#1}{$}
    \infer{
        \VarEnv%
        ~\Entails~%
        {{pq}:{lp}\cdot\QHead}%
        ~\PrcTyped~%
        \SesEnv,~{{pq}:{{l'\left\langle T \right\rangle};\Queue}},~{p:\CIso}%
    }{
        \DataType = \CIso[\Const]
        & l=l'%
        &
        \VarEnv%
        ~\Entails~%
        {pq:\QHead}%
        ~\PrcTyped~%
        \SesEnv,~{pq:\Queue}%
        &
        \ValClocks\models\Const
    }
    \IfBooleanT{#1}{$}
}

\NewDocumentCommand{\PTypRuleEnd}{s}{%
\IfBooleanT{#1}{$}
\TmpPrcJudge2{\PrcEnd}|3{\emptyset}
\IfBooleanT{#1}{$}
}

\begin{document}

\title{Safe asynchronous mixed-choice for timed interactions
   \thanks{This work
    has been partially supported by EPSRC project EP/T014512/1 (STARDUST) and the BehAPI project funded by the EU H2020 RISE under the Marie Sklodowska-Curie action (No: 778233). 
    We thank Simon Thompson and Maurizio Murgia for their insightful comments on an early version of this work.
    }}

\providecommand{\orcidID}[1]{$^{\text{[#1]}}$}
\author{%
   Jonah Pears\inst{1}\orcidID{0000-0003-4492-4072}%
   \and Laura Bocchi\inst{1}\orcidID{0000-0002-7177-9395}%
   \and \\ Andy King\inst{1}\orcidID{0000-0001-5806-4822}%
}
\authorrunning{J. Pears, L. Bocchi, and A. King}
\institute{University of Kent, Canterbury, UK
   \\ \email{\{%
   jjp38,
   ~l.bocchi,
   ~a.m.king
   \}@kent.ac.uk}%
}

\maketitle
%
\def\middot{\textperiodcentered~}
\begin{abstract}
    Mixed-choice has long been barred from models of asynchronous communication since it compromises key properties of communicating finite-state machines.
    Session types inherit this restriction, which precludes them from fully modelling timeouts -- a key programming feature to handle failures.
    To address this deficiency, we present (binary) TimeOut Asynchronous Session Types ({TOAST}) as an extension to (binary) asynchronous timed session types to permit mixed-choice.
    {TOAST} deploy timing constraints to regulate the use of mixed-choice so as to preserve communication safety.
    %
    %
    We provide a new behavioural semantics for {TOAST} which guarantees progress in the presence of mixed-choice.
    Building upon {TOAST}, we provide a calculus featuring process timers which is capable of modelling timeouts using a $\mathtt{receive\text{-}after}$ pattern, much like Erlang,
    and informally illustrate the correspondence with TOAST specifications.

\end{abstract}

\section{Introduction}\label{sec:intro}

Mixed-choice is an inherent feature of models of communications such as communicating finite-state machines (CFSM)~\cite{Brand1983} where actions are classified as either send or receive. In this setting, a state of a machine is said to be mixed if there exist both a sending action and a receiving action from that state.
When considering an asynchronous model of communication, absence of deadlocks is undecidable in general~\cite{Gouda1984} but can be guaranteed in presence of three sufficient and decidable conditions: determinism, compatibility, and \emph{absence} of mixed-states~\cite{Gouda1984,Denielou2013}. Intuitively, determinism means that it is not possible, from a state, to reach two different states with the same kind of action, and compatibility requires that for each send action of one machine, the rest of the system can eventually perform a complementary receive action.

In the desire to ensure deadlock freedom, mixed-choice has been given up, even though this curtails the descriptive capabilities of CFSM and its derivatives. Despite the rapid evolution of session types, even to
the point of deployment in Java~\cite{Hu2008}, Python~\cite{Neykova2013,Neykova2013a},
Rust~\cite{Lagaillardie2020},
{F\#}~\cite{Neykova2018} and {Go}~\cite{Castro2019}, thus far mixed-choice has only been introduced into the synchronous binary setting~\cite{Vasconcelos2020}. 
In fact, the exclusion of mixed-choice pervades work on asynchronous communication which guarantee deadlock-freedom, both for
communicating timed automata~\cite{Bocchi2015,Krcal2006} and session types~\cite{Bettini2008,Carbone2008,Honda2008,Yoshida2007}.
Determinism and the absence of mixed-states is baked into the very syntax of session types (the correspondence between session types and so-called safe CFSM is explained in~\cite{Denielou2013}).

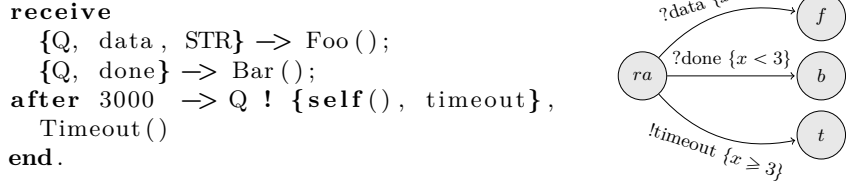
\begin{figure}[t!]
\begin{center}
\begin{tabular}{@{}c@{\qquad\qquad}c@{}}
\begin{tabular}{@{}c@{}}
\begin{lstlisting}[basicstyle=\small,language=Erlang]
receive 
  {Q, data, STR} -> Foo();
  {Q, done} -> Bar();
after 3000  -> Q ! {self(), timeout}, 
  Timeout()
end.
\end{lstlisting}
\end{tabular}
&
\hspace{-0.5cm}\begin{tabular}{@{}c@{}}
\resizebox{3cm}{!}{\begin{tikzpicture}[shorten >=1pt,node distance=1.5cm,auto]
\tikzstyle{every state}=[fill={rgb:black,1;white,10}]

\node[state] (0) at (0,0) {$ra$};
\node[state] (1) at (3,1)  {$f$};
\node[state] (2) at (3,0)  {$b$};
\node[state] (3) at (3,-1) {$t$};

\path[->] (0) edge [bend left]  node[above, sloped, midway] {?data $\{x<3\}$} (1);

\path[->] (0) edge  node {?done $\{x<3\}$} (2);

\path[->] (0) edge [bend right] node[below, sloped, midway] {!timeout $\{x \geq 3\}$} (3);
\end{tikzpicture}}
\end{tabular}
\vspace{-0.5cm}
\end{tabular}
\end{center}
\caption{An Erlang snippet and its mixed-state machine representation}\label{fig:timeout_snippets}
\end{figure}

Timed session types~\cite{Bartoletti2014,Bocchi2019,Bocchi2014}, which extend session types with time constraints, inherit the same syntactic restrictions of session types, and hence rule out mixed-states.
This is unfortunate since in the timed setting, mixed-states are a useful abstraction for timeouts. 
Illustrated in~\Cref{fig:timeout_snippets}, the mixed-state CFSM (right) can be realised using a \texttt{receive-after} statement in Erlang (left).
In the Erlang snippet, the process waits to receive either a `data' or `done' message. If neither are received within 3 seconds, then a timeout message is issued.

Timeouts are important for handling failure and unexpected delays, for instance, the SMTP protocol stipulates:
\emph{``An SMTP client \emph{must} provide a timeout mechanism''}~\cite[Section~4.5.3.2]{RFC5321}.
Mixed-states would allow, for example, states where a server is waiting to \emph{receive} a message from the client and, if nothing is received after a certain amount of time, \emph{send} a notification that ends the session.
Current variants of timed session types allow deadlines to be expressed but cannot, because of the absence of mixed-states, characterise (and verify) the behaviour that should follow a missed deadline, e.g., a restart or retry strategy.
In this paper, we argue that time makes mixed-states more powerful (allowing timeouts to be expressed), while just adding sufficient synchonisation to ensure that mixed-states are safe in an asynchronous semantics (cannot produce deadlocks).

\paragraph{Contributions}
This work makes three orthogonal contributions to the theory of binary session types, with a focus on improving their descriptive capabilities:
\begin{itemize}
      \item We introduce TimeOut Asynchronous (binary) Session Types (TOAST) to support timeouts.
            Inspired by asynchronous timed binary session types~\cite{Bocchi2019}, TOAST shows how timing constraints provide an elegant solution for guaranteeing the safety of mixed-choice. Technically, we provide a semantics for TOAST and a well-formedness condition. We show that well-formedness is sufficient to guarantee progress for TOAST (which may, instead, get stuck in general). 
      \item We provide a new process calculus whose functionality extends to support programming motifs such as the widely used \texttt{receive\text{-}after} pattern of Erlang for expressing timeouts. 
    \item We introduce timers in our process calculus to structure the counterpart of a timeout (i.e., a process that interacts with one other process displaying a timeout), as well as time-sensitive conditional statements, where the selection of a branch may be determined by timers. Time-sensitive conditional statements provide processes with knowledge that can be used to decide which branch should be followed e.g., helping understanding whether the counterpart may have timed out or not. 
    \item We provide an informal discussion on the correspondence between TOAST and the aforementioned primitives of our new process calculus. 
\end{itemize}

%
\section{Timeout Asynchronous Session Types (TOAST)}\label{sec:types}

This section presents the syntax, semantics and formation rules for TimeOut Asynchronous Session Types (TOAST), which extend asynchronous binary timed session types~\cite{Bocchi2019} with a well-disciplined (hence safe) form of mixed-choice.

\paragraph*{Clocks \& Constraints}
We start with a few preliminary definitions borrowed from timed automata~\cite{Alur1994}. Let \ClockSet*\ be a finite set of clocks denoted \Clx*, \Cly*\ and \Clz*.
A (clock) valuation \ValClocks*\ is a map $\ValClocks: \ClockSet\to\RatSet$.
The initial valuation is $\ValClocks_0$ where $\ValClocks_0=\{x\mapsto0\mid x\in\ClockSet\}$.
Given a time offset $\ValTime\in\RatSet$ and a valuation \ValClocks*, $\ValClocks+\ValTime=\{x\mapsto\ValClocks(x)+\ValTime\mid x\in\ClockSet\}$.
Given $\RSet\subseteq\ClockSet$ and \ValClocks*,
$\ReSet[\ValClocks]=\{\text{if\ }(x\in\RSet)\,0\,\text{else\ }\ValClocks(x)\mid x\in\ClockSet\}$.
Observe $\ReSet[\ValClocks]:{\emptyset}=\ValClocks$.
\GuardSet*\ denotes the set of clock constraints, where a clock constraint \Const*\ takes the form:
\begin{equation}\label{eq:types_constraints}
    \Const ::= \TypeTrue%
    \mid \Clx>\ValTime%
    \mid \Clx=\ValTime%
    \mid \Clx-\Cly>\ValTime%
    \mid \Clx-\Cly=\ValTime%
    \mid \neg\Const%
    \mid \Const_1\land\Const_2%
\end{equation}

\noindent We write $\nu\models\delta$ for a constraint $\delta$ that is satisfied by the valuations of clocks within $\nu$.
We write $\Past$ (the past of $\delta$) for a constraint $\delta'$ such that $\nu \models \delta'$ if and only if $\exists t  : \nu + t \models \delta$.
For example $\Past[\Tup(3<x<5)]={x<5}$ and $\Past[\Tup(x>2)]=\TypeTrue$.

\subsection{Syntax of TOAST}\label{ssec:types_syntax}

The syntax of TOAST (or just \emph{types}) is given in~\cref{eq:types_syntax}.
A type \TypeS*\ is a choice $\simplechoice$, recursive definition $\TypRecDef.\TypeS$, call $\TypRecCall$, or termination type $\TypeEnd$.
%
%
\begin{minieq}\label{eq:types_syntax}
  \begin{array}[c]{lcl p{3ex} lcl}
    \TypeS %
     & ::=
     & \simplechoice
    ~\mid~ \TypRecDef.\TypeS
    ~\mid~ \TypRecCall
    ~\mid~ \TypeEnd
     &                                      & %
    \TypeChoice %
     & ::=
     & \TypComm\,\IMsgType (\delta,\lambda)
    \\[0.3cm]
    \DataType %
     & ::=
     & \TmpTypBTDelegate
    ~\mid~ \BTypeNat
    ~\mid~ \BTypeBool
    ~\mid~ \BTypeString
    ~\mid~ \BTypeNone
    ~\mid~ \ldots
     &                                      &
    \TypComm
     & ::=
     & \TypSend~\mid~\TypRecv
  \end{array}
\end{minieq}
%

\noindent Type $\simplechoice$ models a choice among options $i$ ranging over a non-empty set $I$. Each option $i$ is a selection/send action if $\TypComm=\TypSend$, or alternatively a branching/receive action if $\TypComm=\TypRecv$. 
An option sends (resp. receives) a label $l$ and a message of a specified data type \DataType*\ is delineated by $\langle\cdot\rangle$.
The send or receive action of an option is guarded by a time constraint \Const*. After the action, the clocks within \RSet*\ are reset to 0.
Data types, ranged over by $T$, $T_i$, $\ldots$ can be sorts (e.g., natural, boolean), or higher order types $(\delta,S)$ to model session delegation.
Only the message label is exchanged when the data type is \BTypeNone*.
Labels of the options in a choice are pairwise distinct.
Recursion and terminated types are standard.

\paragraph{Remarks on the notation}
One convention is to model the exchange of payloads as a separated action with respect to the communication of branching labels. 
In this paper we follow~\cite{Bocchi2015,Yoshida2021}, and model them as unique actions. 
When irrelevant, we omit the payload, yielding a notation closer to that of timed automata.
%
%
%
%

\subsection{Semantics of TOAST}\label{ssec:types_semantics}

We present the semantics of TOAST, building on those given in~\cite{Bocchi2019}; any changes are highlighted.
Following~\cite{Bocchi2019}, we define the semantics using three layers: (1) \emph{configurations}, (2) \emph{configurations with queues} that model asynchronous interactions, and (3) \emph{systems} that model the parallel composition of configurations with queues.
%
The semantics are defined over the labels $\ell$ given below:
%
%
\begin{minieq}\label{eq:type_labels}
  \ProgAction::=\TypComm\Msg
  ~\mid~\ValTime
  ~\mid~\SiltAction
  \qquad
  \Msg::=\DefMsgType
  \qquad
  \TypComm ::= \TypSend~\mid~\TypRecv
  %
  %
\end{minieq}
\noindent where $\ell$ is either a communication, time, or silent action, and $\Msg$ is a message.

\subsubsection{Configurations} 
A configuration \VIso*\ is a pair $(\nu, \TypeS)$. 
%
The semantics for configurations are defined by a Labelled Transition System (LTS) over configurations, the labels in~\cref{eq:type_labels}, and the rules given in~\Cref{fig:typesemantics_tuple}.
%
%
%
{
\begin{figure}[h]\centering
    $
    \begin{array}[c]{c}
        \begin{array}[c]{c}
          \begin{array}[c]{c}
            \begin{array}[c]{c c}
              \ColBoxNew{\CfgIsoRuleInteract} & \LblCfgIsoInteract
            \end{array}
          \end{array}
          \\[-1.5ex]\\
          \begin{array}[c]{c}
            \begin{array}[c]{c @{\qquad} c}
              \begin{array}[c]{c c}
                \ColBoxOld{\CfgIsoRuleRecursion} & \LblCfgIsoUnfold
              \end{array}
            &
              \begin{array}[c]{c c}
                \ColBoxOld{\CfgIsoRuleIsoTime} & \LblCfgIsoTick
              \end{array}
            \end{array}
          \end{array}
        \end{array}
    \end{array}
    $
  \caption[LTS of Configurations]{Semantics of Configurations.}
  \label{fig:typesemantics_tuple}
\end{figure}}

Rule \LblCfgIsoInteract*\ deviates from~\cite{Bocchi2019} and handles choice types. By this rule, a configuration performs one action with index $j\in I$ provided that the constraint $\delta_j$ is satisfied in the current valuation $\nu$ ($\nu\models\delta_j$).
All clocks in $\lambda_j$ are reset to 0 in the resulting configuration's valuation of clocks.
Rule \LblCfgIsoUnfold*\ unfolds recursive types.
Rule \LblCfgIsoTick*\ describes time passing.
A transition \Trans*{\VIso}:{\ValTime,\CommMsg}[\VIso']\ indicates \Trans*{\Trans{\VIso}:{\ValTime}[{\VIso}'']}:{\CommMsg}[\VIso'], where ${\VIso}''$ is some intermediate configuration. We write \Trans*{\VIso}:{\ValTime,\CommMsg}[] if there exists $\VIso'$ such that \Trans*{\VIso}:{\ValTime,\CommMsg}[\VIso']. 

\subsubsection{Configurations with queues}
A configuration with queues \VSoc*\ is a triple $(\nu, \TypeS, \Queue)$ where $\Queue$ is a FIFO queue of messages which have been received but not yet processed.
A queue takes the form $\Queue ::= \emptyset \mid \Msg;\Queue$ thus is either empty, or has a message at its head.
%
%
The transition \Trans*{\VSoc}:{\ValTime,\CommMsg}[{\VSoc}']\ is defined analogously to \Trans*{\VIso}:{\ValTime,\CommMsg}[\VIso'].
%
The semantics of configurations with queues is defined by an LTS over the labels in~\cref{eq:type_labels} and the rules in~\Cref{fig:typesemantics_triple}. 

\begin{figure}[t]
\[
    \begin{array}[]{c}
            \ColBoxOld{\CfgSocRuleSend} ~ \LblCfgSocSend
          \qquad
            \ColBoxOld{\CfgSocRuleRecv} ~ \LblCfgSocRecv
          \\[0.6cm]
            \ColBoxOld{\CfgSocRuleEnqu} ~ \LblCfgSocEnqu 
        \\[0.4cm]
         \begin{array}[c]{c c}
            \ColBoxNew{
              \infer{
                (\nu,\TypeS,\Queue) \Act[\ValTime] (\nu',\TypeS,\Queue)
              }
              {\begin{array}{c p{2ex} l}
                  (\nu,\TypeS) \Act[\ValTime] (\nu',\TypeS)                                  &  & \text{(configuration)} \\
                  %
                  %
                  (\nu,\TypeS)\isFE \implies (\nu',\TypeS)\isFE \qquad                       &  & \text{(persistency)}   \\
                  %
                  %
                  \forall t' < t : (\nu + t',\TypeS,\Queue)\stackrel{\tau}{\not \rightarrow} &  & \text{(urgency)}
                \end{array}
              }} ~~ \LblCfgSocTime
          \end{array}
           \end{array}
\]
  \caption{Semantics of Configurations with queues.}
  \label{fig:typesemantics_triple}
\end{figure}

Rule \LblCfgSocSend*\ is for sending a message. Message reception is handled by two rules: \LblCfgSocEnqu*\ inserts a message at the back of \Queue*, and \LblCfgSocRecv*\ removes a message from the front of the queue.
Rule \LblCfgSocTime*\ is for time passing which is formulated in terms of a future-enabled configuration, given in~\Cref{def:configs_fe}.
The second condition in the premise for rule \LblCfgSocTime*\ ensures the latest-enabled action is never missed by advancing the clocks.
The third condition models an urgent semantics, ensuring messages are processed as they arrive.
Urgency is critical for reasoning about progress.

\begin{definition}[Future-enabled Configurations (\isFE*)]\label{def:configs_fe}
For some $\Msg$, a configuration \VIso*\ (resp. a configuration with queues \VSoc*) is \emph{future-enabled (\isFE*)} if
  $\exists\ValTime\in\RatSet : %
  \Trans{\VIso}:{\ValTime,\CommMsg}$ 
  (resp. \Trans*{\VSoc}:{\ValTime,\CommMsg}).
\end{definition}

\subsubsection{Systems} Systems are the parallel composition of two configurations with queues,
written as \Parl*{\CSoc_1,\CSoc_2} or \VSys*.
The semantics of systems is defined by an LTS over the labels in~\cref{eq:type_labels} and the transition rules in~\Cref{fig:typesemantics_sys}.
{
\begin{figure}[]\centering
  \[  \begin{array}[c]{c}
          \ColBoxOld{\CfgSysRuleComm}  
        %
        %
          \ColBoxOld{\CfgSysRuleDequ} \\
          \ColBoxOld{\CfgSysRuleWait}
          %
          %
    \end{array}
\]
  \caption{Semantics of Systems.}
  \label{fig:typesemantics_sys}
\end{figure}}

Rule \LblCfgSysLComm*\ handles asynchronous communication where \VSoc*_1\ sends a message \Msg*\ via rule \LblCfgSocSend*, which arrives at the queue of \VSoc*_2\ via rule \LblCfgSocEnqu*.
Rule \LblCfgSysRComm*\ is symmetric, allowing for \VSoc*_2\ to be the sending party, and is omitted.
%
Rule \LblCfgSysLPar*\ allows \VSoc*_1\ to process the message at the head of \Queue*_1\ via \LblCfgSocRecv*.
Rule \LblCfgSysRPar*\ is symmetric, allowing \VSoc*_2\ to receive messages, and is omitted.
%
%
By rule \LblCfgSysWait*\ time passes consistently across systems.
%

\begin{example}[Weak Persistency]\label{ex:weak_persistency}
In language-based approaches to timed semantics~\cite{Krcal2006}, time actions are always possible, even if they bring the model into a stuck state by preventing available actions. Execution traces are then filtered a posteriori, removing all ‘bad’ traces (defined on the basis of final states). 
In contrast, and to facilitate the reasoning on process behaviour, we adopt a process-based approach, e.g.,~\cite{Bartoletti2018,Bocchi2019}, that only allows for actions that characterise \emph{intended} executions of the model.
Precisely, we build on the semantics in~\cite{Bartoletti2018} for asynchronous timed automata with mixed-choice, where time  actions are possible only if they do not disable: (1) the latest-enabled sending action, and (2) the latest-enabled receiving action if the queue is not empty. This ensures that time actions preserve the viability of at least one action (\emph{weak-persistency)}. In our scenario, constraint (1) is too strict. Consider type $S$ and its dual below:
  %
  \begin{minieq}*
    \begin{array}[c]{c p{0.5ex} c}
      \begin{array}[c]{lcl}
        \TypeS & = & %
        \IChoice[%
        \IType!{data\text{<}String\text{>}}[x<3].{\TypeS'},%
        \IType?{timeout}[x>4]%
        ]%
      \end{array}
      &  & %
      \begin{array}[c]{lcl}
        \Dual & = & %
        \IChoice[%
        \IType?{data\text{<}String\text{>}}[y<3].{\Dual'},%
        \IType!{timeout}[y>4]%
        ]%
      \end{array}
    \end{array}
  \end{minieq}
  \noindent According to (1), it would never be possible for $S$ to take the timeout branch since a time action of $t\geq 3$ would disable the latest-enabled send. 
  This is reasonable in~\cite{Bartoletti2018} because, in their general setting, there is no guarantee that a timeout will indeed be received. 
  Unlike~\cite{Bartoletti2018}, we can rely on duality of $\overline{S}$ (introduced later in this section),  which guarantees that $\overline{S}$ will send a timeout message when $y>4$. 
  Our new rule \LblCfgSocTime*\ -- condition (persistency) --  implements a more general constraint than (1), requiring that one latest-enabled (send or receive) action is preserved. 
  Constraint (2) remains to implement urgency and, e.g., prevents $\overline{S}$ from sending a timeout if a message is waiting in the queue when $y<3$.
  %
  %
  %
  %
  %
\end{example}

\subsection{Duality, Well-formedness, and Progress}\label{ssec:types_formation}


In the untimed scenario, the composition of a binary type with its dual characterises a \emph{protocol}, which specifies the ``correct'' set of interactions between a party and its co-party. The dual of a type, formally defined below, is obtained by swapping the directions ($!$ or $?$) of each interaction:
%
%
\begin{definition}[Dual Types]\label{def:types_dual}
   The dual type \Dual*\ of type \TypeS*\ is defined as follows:
   \begin{minieq}*
      \begin{array}[t]{r c l p{7ex} r c l p{7ex} r c l p{7ex} r c l}
         \Dual[\TypRecDef]
         & =
         & \TypRecDef.\Dual[\TypeS]
         &
         &                                                              %
         \Dual[\TypRecLabel]
         & =
         & \TypRecLabel
         &
         &                                                              %
         \Dual[\TypeEnd]
         & =
         & \TypeEnd
         &
         &                                                              %
         \Dual[\TypSend]
         & =
         & \TypRecv
         \\[1ex]
         &
         &
         &
         &                                                              %
         \mathllap{\TypInteract~}
         & =
         & \mathrlap{\TypInteract1{1{\Dual[\TypComm]}}3{\Dual[\TypeS]}}
         &
         &
         &
         &
         &
         &                                                              %
         \Dual[\TypRecv]
         & =
         & \TypSend
      \end{array}
   \end{minieq}
   \vspace{-2ex}
   %
\end{definition}
\noindent Unfortunately, when annotating session types with time constraints, one may obtain protocols that are infeasible, as shown in~\Cref{ex:junk_types}.
This is a known problem, which has been addressed by providing additional conditions or constraints on timed session types, for example compliance~\cite{Bartoletti2017}, feasibility~\cite{Bocchi2014}, interaction enabling~\cite{Bocchi2015}, and well-formedness~\cite{Bocchi2019}.

Building upon~\cite{Bocchi2019}, well-formedness is given as a set of formation rules for types. The rules check that in every reachable state (which includes every possible clock valuation) it is possible to perform the next action immediately or at some point in the future, unless the state is final. (This is formalized as the progress property in~\Cref{def:types_progress}.) By these rules, the type in~\Cref{ex:junk_types} would not be well-formed.
%
The use of mixed-choice in asynchronous communications may result in infeasible protocols or, more concretely, systems (or types) that get stuck, even if they are well-formed in the sense of~\cite{Bocchi2019} (discussed in~\Cref{ex:mixed_bad}).

\begin{example}[Junk Types]\label{ex:junk_types}
   %
   Consider the \emph{junk type} defined below: 
   %
   \begin{minieq}*
      \begin{array}[c]{lcl}
         \TypeS & = & %
        !a~(x>3,~\emptyset).~\{%
        !b~(y=2,~\emptyset).\TypeEnd,~%
        ?c~(2<x<5,~\emptyset).\TypeEnd%
        \}
      \end{array}
   \end{minieq}
   Assume all clocks are 0 before $a$ is sent.
   After $a$ is sent, all clocks hold values greater than 3.
   The constraint on sending $b$ is never met, and the one on receiving $c$ may not be met.
   Types with unsatisfiable constraints are called \emph{junk types}~\cite{Bocchi2019}.
   \TypeS*\ can be amended to obtain, for example, $S'$ or $S''$ below.
\begin{minieq}*
      \begin{array}[c]{lcl}
         S' & = & %
        !a~(x>3,~\{y\}).~\{%
        !b~(y=2,~\emptyset).\TypeEnd,~%
        ?c~(2<x<5,~\emptyset).\TypeEnd%
        \}\\
        S'' & = & %
        !a~(3<x<5,~\emptyset).~\{%
        !b~(y=2,~\emptyset).\TypeEnd,~%
        ?c~(2<x<5,~\emptyset).\TypeEnd%
        \}
      \end{array}
   \end{minieq}
   $S'$ makes both options of the choice satisfiable by resetting clock $y$, while $S''$ makes at least one option ($?c$) always satisfiable by changing the first constraint.   
\end{example}

\begin{example}[Unsafe Mixed-choice]\label{ex:mixed_bad}
   A mixed-choice is considered \emph{unsafe} if actions of different directions compete to be performed (i.e., they are both viable at the same point in time). 
   Consider system \VSys*, where $\VSoc_1 = (\nu_0,~{S_1},~\emptyset)$, $\VSoc_2=(\nu_0,~{S_2},~\emptyset)$, and types $S_1$ and $S_2$ are dual as defined below:
   %
   %
   \begin{minieq}*
      \begin{array}[c]{c p{4ex} c}
         \begin{array}[c]{lcl}
            \TypeS_1 & = & %
            \IChoice[%
            \IType?{a}[x<5],%
            \IType!{b}[x=0].{\TypeS'_1}%
            ]%
         \end{array}
         &  & %
         \begin{array}[c]{lcl}
            S_2 & = & %
            \IChoice[%
            \IType!{a}[y<5],%
            \IType?{b}[y=0].{\TypeS'_2}%
            ]%
         \end{array}
      \end{array}
   \end{minieq}
   
   \noindent In the system \VSys*\ it is possible for both \Trans*{\VSoc_1}:{!,b}[\VSoc'_1]\ and \Trans*{\VSoc_2}:{!,a}[\VSoc'_2]\ to occur at the same time.
   In the resulting system \Parl*{{(\nu_0,~{S'_1},~{a})},{(\nu_0,~{\TypeEnd},~{b})}}
    neither message can be received, and $S'_1$ may be stuck waiting for interactions from $S'_2$ indefinitely. 
   %
   %
   %
\end{example}

\subsubsection*{Well-formedness}
In this work we extend well-formedness of~\cite{Bocchi2019} so that progress is guaranteed in the presence of mixed-choice. 
The formation rules for types are given in~\Cref{fig:types_rule}; rules differing from \cite{Bocchi2019} are highlighted.
Types are evaluated against judgements of the form: \TypEnvCond*:{\RecEnv;\,\Const}\ where
\RecEnv*\ is an environment containing recursive variables, and \Const*\ is a constraint over all clocks characterising the times in which state \TypeS*\ can be  reached.
%
%
\begin{figure}[h]%
    \centering$%
        \begin{array}[t]{c}%
            \begin{array}[c]{c c}%
                \ColBoxNew{
                \infer{A; \downarrow \bigvee_{i \in I}{\delta_i} \vdash \left\{\TypComm_i \,\MsgLabel_i\langle T_i\rangle \left(\delta_i,\lambda_i\right).\TypeS_i\right\}_{i\in I}}{%
                \begin{array}{c lc}
                    \forall i\in I :  A;\gamma_i\vdash \TypeS_i ~\land~ \delta_i[\lambda_i\mapsto 0]\models \gamma_i
                    & \text{(feasibility)}  \\[0.5ex]
                    \forall i, j \in I : i\neq j \implies \delta_i \land \delta_j \models \TypeFalse ~\lor~ \TypComm_i = \TypComm_j
                    & \text{(mixed-choice)} \\[0.5ex]
                    \forall i\in I : T_i = (\delta',\TypeS') \implies \emptyset; \gamma' \vdash \TypeS' ~\land~ \delta'\models \gamma'~~
                    & \text{(delegation)}
                \end{array}%
                }}
                & %
                \LblTypChoice
            \end{array}
            \\[-1ex]\\
            \begin{array}[t]{c p{2ex} c p{2ex} c}
                \begin{array}[c]{c c}
                    \ColBoxOld{\TypRuleEnd}
                    & %
                    \LblTypEnd
                \end{array}
                &  & %
                \begin{array}[c]{c c}
                    \ColBoxOld{\TypRuleRec}
                    & %
                    \LblTypRec
                \end{array}
                &  & %
                \begin{array}[c]{c c}
                    \ColBoxOld{\TypRuleVar}
                    & %
                    \LblTypVar
                \end{array}
            \end{array}
        \end{array}
    $
    \caption[Well-formedness rules for types.]{Well-formedness rules for types.}%
    \label{fig:types_rule}%
    \vspace{-10pt}
\end{figure}%

Rule \LblTypChoice*\ checks well-formedness of choices with three conditions:
the first and third conditions are from the branching and delegation rules in~\cite{Bocchi2019}, respectively; the second condition is new and critical to ensure progress of mixed-choice.
By using the weakest past of all constraints ($\downarrow \bigvee_{i \in I}{\delta_i}$) only one of the options within the choice is required to be \emph{always} viable, for the choice to be well-formed.
The first condition (feasibility) ensures that, for each option in a choice, there exists an environment $\gamma$ such that the continuation $\TypeS_i$ is well-formed, given the current constraints on clocks $\delta_i$ (updated with resets in $\lambda_i$).
%
This ensures that in every choice, there is always at least one viable action; it would, for example, rule out the type in~\Cref{ex:junk_types}.
%
The second condition (mixed-choice) requires all actions that can happen at the same time to have the same (send/receive) direction.
This condition allows for types modelling timeouts, as in~\Cref{ex:weak_persistency}, and rules out scenarios as the one in~\Cref{ex:mixed_bad}.
The third condition (delegation) checks for well-formedness of each delegated session with respect to their corresponding initialization constraint $\delta'$.
Rule \LblTypEnd*\ ensures termination types are \emph{always} well-formed.
Rule \LblTypRec*\ associates, in the environment, a variable $\alpha$ with an invariant $\delta$.
%
%
Rule \LblTypVar*\ ensures recursive calls are defined.
\begin{definition}[Well-formedness]\label{def:types_wf}
   %
   A type $\TypeS$ is well-formed with respect to $\nu$ if there exists $\delta$ such that $\nu\models \delta$ and $\emptyset;\delta\vdash \TypeS$. A type $\TypeS$ is well-formed if it is well-formed with respect to $\nu_0$.
\end{definition}
\noindent Well-formedness, together with the urgent receive features of the semantics (rule $\LblCfgSocTime$ in~\Cref{fig:typesemantics_triple}) ensures that the composition of a well-formed type $\TypeS$ with its dual $\overline \TypeS$ enjoys progress.
A system enjoys progress if its configurations with queues can continue communicating until reaching the end of the protocol, formally:
\begin{definition}[Type Progress]\label{def:types_progress}
   A configuration with queues $\SocCfg$ is \emph{final} if $\VSoc=\CSoc;[\TypeEnd]:{\emptyset}$.
   A system \VSys*\ \emph{satisfies progress} for all \VSys*'\ reachable from \VSys*, either:
   \begin{itemize}
       \item \VSoc*'_1\ and \VSoc*'_2\ are \emph{final}, or
       \item there exists a $t\in\RatSet$ such that \Trans*{\VSys'}:{\ValTime,\SiltAction}.
   \end{itemize}
\end{definition}
%

%
%
\begin{restatable}[Progress of Systems]{theorem}{ThmProgress}\label{thm:progress}
   If $\TypeS$ is \emph{well-formed} against \ValClocks*_0\ then
   \linebreak
   \CSys*[\CSoc[\ValClocks_0]:{\emptyset}][\CSoc[\ValClocks_0];[\Dual]:{\emptyset}]\ \emph{satisfies progress}.
\end{restatable}

\noindent The main result of this section is that, for a system composed of \emph{well-formed} and dual types, any state reached is either \emph{final}, or allows for further progress.
By ensuring a system will make progress, it follows that such a system is free from communication mismatches and will not reach deadlock.

%
The main differences with~\cite{Bocchi2019} is not in the formulation of the theory (e.g., \Cref{def:types_progress,def:configs_compat}, and the statement of~\Cref{thm:progress} are basically unchanged) but in the proofs that, now, have to check that the conditions of rule \LblTypChoice*\ are sufficient to ensure progress of asynchronous mixed-choice.
Additionally, the proof of progress in~\cite{Bocchi2019} relies on receive urgency. Because of mixed-choice, it is necessary to reformulate (and relax) urgency in the semantic rule \LblCfgSocTime*\ in~\Cref{fig:typesemantics_triple}. 
Despite generalising the notion of urgency the desired progress property can still be attained (see~\Cref{ex:weak_persistency} for a discussion).

The proof of~\Cref{thm:progress} proceeds by showing that system \emph{compatibility}~\cite{Bocchi2019} is preserved by transitions. The formal definition of compatibility is given in~\Cref{def:configs_compat}.
\begin{definition}[Compatible Systems]\label{def:configs_compat}
    Let $\VSoc_1 = \CSoc_1$ and $\VSoc_2 = \CSoc_2$. 
    System \VSys*\ is \emph{compatible} (written $\VSoc_1\bot\, \VSoc_2$) if:
    \begin{enumerate}
    \item\label{itm:configs_compat_non_empty_queues} $\Queue_1=\emptyset%
            ~\lor~%
            \Queue_2=\emptyset$%
        \\
       \item\label{itm:configs_compat_dual_types} $\Queue_1=\Queue_2=\emptyset%
            ~\implies~%
            \ValClocks_1=\ValClocks_2%
            ~\land~%
            \TypeS_1=\Dual_2$
        \\
  \item
    \label{itm:configs_compat_expected_receive} 
    $\Queue_1=\Msg;\Queue'_1
            ~\Implies~%
            \exists\ValClocks'_1,\TypeS'_1:
            \Trans{\CIso_1}:{\RecvMsg}[\CIso'_1]%
            ~\land~%
            \CSoc'_1 \bot\, \VSoc_2$
            \\
\item
    $\Queue_2=\Msg;\Queue'_2
            ~\Implies~%
            \exists\ValClocks'_2,\TypeS'_2:
            \Trans{\CIso_2}:{\RecvMsg}[\CIso'_2]%
            ~\land~%
            \VSoc_1 \bot\, \CSoc'_2$
            \\
        %
    \end{enumerate}
    %
\end{definition}
\noindent Informally, $\VSoc_1\bot~ \VSoc_2$ if:
(1) at most one of their queues is non-empty (equivalent to a half-duplex automaton),
(2) if both queues are empty, then $\VSoc_1$ and $\VSoc_2$ have dual types and same clock valuations, 
and (3) and (4) a configuration is always able to receive any message that arrives in its queue.
%
%
%

%
\section{A Calculus for Processes with Timeouts}\label{sec:prc_calc}\label{ssec:prc_reduct}

We present a new calculus for timed processes which extends existing timed session calculi~\cite{Bocchi2015,Bocchi2019} with: (1) timeouts, and (2) time-sensitive conditional statements. 
Timeouts are defined on receive actions and may be immediately followed by sending actions, hence providing an instance of mixed-choice -- which is normally not supported. 
Time-sensitive conditional statements (i.e., \texttt{if\text{-}then\text{-}else} with conditions on program clocks/timers) provide a natural counterpart to the timeout construct and enhance the expressiveness of the typing system in~\cite{Bocchi2019}. By counterpart, we intend a construct to be used by the process communicating with the one that sets the timeout. 
\newcommand{\pl}[0]{\mathtt v}
Processes are defined by the grammar below.
To better align processes with TOAST, send and select actions have been streamlined by each message consisting of both a label $l$ and some message value $\pl$, which is either data or a delegated session; the same holds for receive/branch actions where $q$ is a variable for data or delegated sessions. We assume a set of timer names \TimerSet*, ranged over by $x,y$ and $z$.

%
%
\begin{processcalculus}\label{eq:prc_calc}
   \begin{array}[c]{l}\begin{array}[t]{lcl p{2ex} lcl}
      {\Prc,\Qrc} %
      & ::= & {\PSet{\mathit{x}}.{P}}%
      & & & \mid & {\RecPrcCall}%
      \\
      & \mid & {\On{p}\Send{l}[\pl].{P}}
      & & & \mid & {\PrcEnd}%
      \\
      & \mid & {\On{p}\Recv{l}[\mathit{q}]:{P}_{i}~\After<{e}:{Q}}
      & & & \mid & {\PrcErr}%
      \\
      & \mid & {\If*{\Const}~\Then{P}~\Else{Q}}%
      & & & \mid & {(\nu pq)\Prc}%
      \\
      & \mid & {\PFuncDelay[\Const]}%
      & & & \mid & {\Parl{P,Q}}%
      \\
      & \mid & {\PFuncDelay[\ValTime]}%
      & & & \mid & {{qp}:{h}}
      \\
      & \mid & {\PrcRecDef}%
      & & 
      %
      {h} & {::=} & {\emptyset~\mid~{h\cdot {l\pl}}}
      %
   \end{array}\end{array}
\end{processcalculus}
%
%
\noindent Process $\PSet{\mathit{x}}.{P}$ creates a timer $x$, initialises it to 0 and continues as $P$. If $x$ already exists it is reset to $0$.
For simplicity, we assume that the timers set by each process $P$ and $Q$ in a parallel composition $P\mid Q$ are pair-wise disjoint.

Process \PCalc*{\On{p}\Send{l}[\pl]}\ is the select/send process: it selects label $l$ and sends payload $\pl$ to endpoint $p$, and continues as $P$.
Its dual is the branch/receive process \PCalc*{\On{p}\Recv{l}[\pl]:{P}_{i}[I]~\After<{e}:{Q}}. It receives one of the labels $l_i$, instantiates $\mathit{q}_i$ with the received payload, and continues as $P_i$. 
(Note that a similar construct has been used to model timeouts in the Temporal Process Language~\cite{Bocchi2022a}, outside of session types.) 
Parameter $e$ is a linear expression over the timers and numeric constants drawn from $\mathbb{N}_{\geq 0} \cup \{\infty\}$.
The expression $e$ determines the duration of a timeout, after which $Q$ is executed.
Once a process with an \texttt{after} branch is reached, its expression $e$ is evaluated against the values of the timers, to derive a timeout value $n$ where $n\in\mathbb{N}_{\geq 0} \cup \{\infty\}$.
Setting $e=\infty$ models a blocking receive primitive that waits potentially forever for a message. Setting $e=0$ models a non-blocking receive action. 
To retain expressiveness from~\cite{Bocchi2019} where non-blocking receive actions would trigger an exception (i.e., modelling deadlines that must not be missed) we allow $Q$ to be $\PrcErr$.
For simplicity: (i) we write \PCalc*{\On{p}\Recv{l}[\mathit{q}]:{P}_{i}[I]} when $e=\infty$; (ii) we omit the brackets in the case of a single  option; (iii) for options with no payloads we omit $q_i$.
The advantage of using an expression $e$ to express the value of a timeout, rather than a fixed constant, is illustrated in~\Cref{ex:delayed_timeout}.

Process \PCalc*{\If*{\Const}~\Then{P}~\Else{Q}}\ is a conditional statement, except that the condition $\delta$ is on timers. 
Syntactically, the condition is expressed as a time constraint $\delta$ in~\cref{eq:types_constraints}, but instead of clocks, defined on the timers previously set by that process.
%
%
Process $\PFuncDelay$ models time passing for an unknown duration described by $\delta$, and is at runtime reduced to process $\PFuncDelay[t]$ if $\models\Const\Subst[\ValTime][x]$. 
In $\PFuncDelay$ we assume $\delta$ is a constraint on a single clock $x$. 
The name of the clock here is immaterial, where $x$ is a syntactic tool used to determine the duration of a time-consuming (delay) action at run-time. 
In this sense, assume $x$ is bound within $\PFuncDelay$.
%
Recursive processes are defined by a process variable $\RecVar$ and parameters \RecSetMsg*, \RecSetTimers*\ and \RecSetRoles*\ containing \emph{base type} values, timers and session channels, respectively.  
%
%
%
%
The end process is \PrcEnd*.
The error process is \PrcErr*.

As standard~\cite{Bocchi2014,Honda2008,Yoshida2007}, the process calculus allows parallel processes \Parl*{\Prc,\Qrc} and scoped processes $(\nu pq)\Prc$ between endpoints $p$ and $q$. 
Endpoints communicate over pairs of channels $pq$ and $qp$, each with their own unbounded FIFO buffers $h$.\footnote{Similar to the queues used by configurations in~\Cref{ssec:types_semantics}.}
Within a session, $p$ sends messages over $pq$ and receives messages from $qp$ (and vica versa for $q$). 
We have adopted the simplifying assumption in~\cite{Bocchi2019} that sessions are already instantiated. 
Therefore, rather than relying on reduction rules to produce correct session instantiation, we rely on a syntactic well-formedness assumption. 
A \emph{well-formed process} consists of sessions of the form  $(\nu pq)(\Parl{{P},{Q},{{pq}:{h}},{{qp}:{h'}}})$, which can be checked syntactically as in~\cite{Bocchi2019}.

\begin{example}[Parametric Timeouts]\label{ex:delayed_timeout}
Consider the process below:
\[\PCalc{%
    {\mathtt{delay}(z<2)}.%
    \On{p}~\Recv{msg}:{P}~\After<{3}:{Q}%
}\]
It expresses a timeout of $3$ after a delay with a duration between $0$ and $2$ time units (whatever this delay turns out to be). To express a timeout of $3$ \emph{despite} prior execution of a time-consuming action, we need a way to tune the timeout with the actual delay of the time-consuming action. 
%
A \emph{parametric timeout} can model this behaviour:
\[\PCalc{%
    {\mathtt{set}(x)}.%
    {\mathtt{delay}(z<2)}.%
    \On{p}~\Recv{msg}:{P}~\After<{3-z}:{Q}%
}\]
By setting the timeout as an expression, $3-x$, with parameter $x$ reflecting the passage of time, we allow the process to compensate for the exact delay occurred. 
%
%
%
\end{example}

%
%
%
%
%
%

\subsection{Process Reduction}


A \emph{timer environment} $\RedTimers$ is a map from a set of timer names \TimerSet* to \RatSet*.
We define $\RedTimers+\ValTime = \{ x\mapsto\RedTimers(x)+\ValTime \mid x\in\TimerSet \}$ and ${\RedTimers} [x\mapsto 0]$ to be the map ${\RedTimers} [x\mapsto 0](y) = ~\text{if\ } (x = y)~0 ~\text{else\ } \RedTimers(y)$.

The semantics of processes are given in~\Cref{fig:prc_reduc}, as a reduction relation on pairs of the form \RedPrc*{\Prc}.
The reduction relation is defined on two kinds of reduction: instantaneous communication actions $\ARed$, and time-consuming actions $\TRed$. We write $\PrcRed$ to denote a reduction that is either by $\ARed$ or $\TRed$.
%

%
\begin{figure}[p]
   \centering
   \resizebox{\textwidth}{!}{$
         \begin{array}[c]{c}
            \begin{array}[c]{c c}
               \begin{array}[c]{c c}
                 \ColBoxOld{\PRedRuleSetTimer} & \LblPrcRedSetTimer\end{array}
& %
               \begin{array}[c]{c c}
                  \ColBoxOld{\PRedRuleIf} & \LblPrcRedIf
                \end{array}
             \end{array}
            \\[-1.45ex]\\
            \begin{array}[c]{c}
               \begin{array}[c]{c c}
                  \ColBoxOld{\PRedRuleRecvAfter} & \LblPrcRedRecv\end{array}
            \end{array}
            \\[-1.45ex]\\%
            \begin{array}[c]{c}
               \begin{array}[c]{c c}
                  \ColBoxOld{\PRedRuleSend} & \LblPrcRedSend\end{array}
            \end{array}
            \\[-1.45ex]\\
            \begin{array}[c]{c p{0ex} c}
               \begin{array}[c]{c c}
                  \ColBoxOld{\PRedRuleDet} & \LblPrcRedDet\end{array}
               & & %
               \begin{array}[c]{c c}
                  \ColBoxOld{\PRedRuleDelay} & \LblPrcRedDelay\end{array}
            \end{array}
            \\[-1.45ex]\\%
            \begin{array}[c]{c p{0ex} c}
               \begin{array}[c]{c c}
                  \ColBoxOld{\PRedRuleScope} & \LblPrcRedScope\end{array}
               & & %
               \begin{array}[c]{c c}
                  \ColBoxOld{\PRedRulePar} & \LblPrcRedParL\end{array}
                \end{array}
            \\[-1.45ex]\\%
            \begin{array}[c]{c}
               \begin{array}[c]{c c}
                  \ColBoxOld{\PRedRuleDef} & \LblPrcRedDef\end{array}
            \end{array}
            \\[-1.45ex]\\%
            \begin{array}[c]{c}
               \begin{array}[c]{c c}
                  \ColBoxOld{\PRedRuleRec} & \LblPrcRedRec\end{array}
            \end{array}
            \\[-1.45ex]\\%
            \begin{array}[c]{c}
\begin{array}[c]{c c}
                  \ColBoxOld{\PRedRuleAStr} & \LblPrcRedAStr\end{array}
            \end{array}
         \end{array}
      $}
   \caption[Reduction Rules]{Reduction Rules for Processes}
   \label{fig:prc_reduc}
\end{figure}
%
\begin{figure}[p]
\begin{equation*}
\begin{array}{r c l}
  \PFuncWait & = &
  \begin{cases}
     \mkSet[\SesP]
      & \text{if}\,
      \Prc=\PCalc{\On{p}\Recv{l}[q]:{P}_{i}[i]~\After<{e}:{Q}}
     %
     \\[0ex]
     \PFuncWait[\Qrc]\setminus\mkSet[\PrcP,\PrcQ]
      & \text{if}\,
     \Prc=\PCalcScope[\Qrc]
     \\[0ex]
     \PFuncWait[\Qrc]
      & \text{if}\,
     \Prc=\PrcRecDef={P'}
     \\[0ex]
     \PFuncWait[\Prc']\cup\PFuncWait[\Qrc]
      & \text{if}\,
     \Prc=\Par[\Prc'][\Qrc]
     \\[0ex]
     \emptyset
      & \text{otherwise}
  \end{cases}
  \\[-0.5ex]\\%
		\PFuncNEQ & = &
		\begin{cases}
			\mkSet[\SesQ]
			 & \text{if}\,
			\Prc=\PCalcBuffer\land\QHead\neq\emptyset
			\\[0ex]
			\PFuncNEQ[\Qrc]\setminus\mkSet[\PrcP,\PrcQ]
			 & \text{if}\,
			\Prc=\PCalcScope[\Qrc]
			\\[0ex]
			\PFuncNEQ[\Qrc]
			 & \text{if}\,
			\Prc=\PrcRecDef={P'}
			\\[0ex]
			\PFuncNEQ[\Prc']\cup\PFuncNEQ[\Qrc]
			 & \text{if}\,
			\Prc=\Parl{P',Q}
			\\[0ex]
			\emptyset
			 & \text{otherwise}
		\end{cases}
  \end{array}
	\end{equation*}
	\caption[Definition of \PFuncWait*\ and \PFuncNEQ*]{Definition of \PFuncWait*\ and \PFuncNEQ*}
	\label{fig:prc_func_neq_wait}
\end{figure}
%

Rule $\LblPrcRedSetTimer$ creates a new timer $x$ if $x$ is undefined, and otherwise resets $x$ to 0.
Rule $\LblPrcRedIf$ selects a branch $P$ depending on time-sensitive condition $\delta$. The symmetric rule selects branch $Q$ if the condition is not met, and is omitted. 
Rules $\LblPrcRedRecv$ and $\LblPrcRedSend$ are standard~\cite{Milner1999}.
Rule \LblPrcRedDet*\ determines the actual duration $t$ of the delay $\delta$ (which is a constraint on a single clock).
Rule \LblPrcRedDelay*\ outsources time-passing to function \PFuncTime*\ (see~\Cref{fig:prc_func_time}), which returns process $P$ after the elapsing of $t$ units of time, and updates \RedTimers*\ accordingly.
Rules $\LblPrcRedScope$ and $\LblPrcRedParL$ are as standard and the only instant reductions, which may update $\RedTimers$, if any timers are introduced by \LblPrcRedSetTimer*.
We omit the symmetric rule for $\LblPrcRedParL$.
The rule for structural congruence \LblPrcRedAStr*\ applies to both instantaneous and time-consuming actions. 
Structural equivalence of $P$ and $Q$, denoted $P\equiv Q$ is as standard with the addition of rule 
${\mathtt{delay}(0).{P}}\equiv {P}$ following~\cite{Bocchi2019,Bocchi2014}.

\begin{definition}\label{fig:prc_func_time}
The time-passing function \PFuncTime*\ is a partial function on process terms, defined only for the cases below, where we use $\mathtt C_I$ as a short notation for $\{l_i(\pl_i):P_i\}_{i\in I}$:  
\small\[
\begin{array}[t]{lll}
 \PFuncTime[\PCalc{\On{p}~\PrcRecv{~\mathtt C_I}~\After<{e}:{Q}}]= 
    \begin{cases}
               \PCalc{\On{p}~\PrcRecv{~\mathtt C_I}~\After<{e}:{Q}}
                &
               { e = \infty}
               \\[0.5ex]
               \PCalc{\On{p}~\PrcRecv{~\mathtt C_I}~\After<{e-t}:{Q}}
                &
               {e\in\RatSet \textit{ and  } e \geq t}
               \\[0.5ex]
               \PFuncTime_{t-e}[Q]
                &
               \text{otherwise}
\end{cases}
\\[0.8cm]
               \PFuncTime[\PFuncDelay[\ValTime']]=
               \begin{cases}
               \PFuncDelay[\ValTime'-\ValTime]
               & 
               \text{if\ }\ValTime'\geq\ValTime
               \\[0.5ex]
               \PFuncTime_{t-t'}[P]
               & 
               \text{otherwise}
               \end{cases}
            %
\\[0.55cm]
               \PFuncTime[\Par[\Prc_1][\Prc_2]]=
               \Par[\PFuncTime[\Prc_1]][\PFuncTime[\Prc_2]]
               \qquad
               \text{if}\;
               \PFuncWait[\Prc_i]\cap\PFuncNEQ[\Prc_j]=
               \emptyset,
               i\neq j\in\left\{1,2\right\}%
\\[0.3cm]
\PFuncTime[\PrcEnd]=\PrcEnd 
\qquad
\PFuncTime[\PrcErr]=\PrcErr
\\[0.3cm]
\PFuncTime[\PCalcBuffer]= \PCalcBuffer
\qquad
\PFuncTime[\PCalcScope]= \PCalcScope|\;\PFuncTime[\Prc]
\\[0.3cm]
            {
            \PFuncTime[\PrcRecDef]=
            \PrcRecDef>{\PFuncTime[\Qrc]}
            }
         \end{array}
\]
\end{definition}

\noindent The first case in~\Cref{fig:prc_func_time} models the effect of time passing on timeouts. The second case is for time-consuming processes. The third case distributes time passing in parallel compositions and ensures that time passes for all parts of the system equally. 
The auxiliary functions \PFuncWait*\ and \PFuncNEQ*\
ensure time does not pass while a process is waiting to receive a message already in their queue, similar to rule \LblCfgSocTime*\ in~\Cref{fig:typesemantics_triple} for configuration transitions. Informally, \PFuncWait*\ returns the set
of channels on which P is waiting to receive
a message, and \PFuncNEQ*\ returns the set of endpoints with a non-empty inbound queue.
Full definitions are given in~\Cref{fig:prc_func_neq_wait}.
The remaining cases allow time to pass.

%
%

%

\begin{example}
%
Consider the process below:
\[
   P = (\nu pq) \left(\Parl{\PCalc{\On{p}\Recv{l}[\pl]:{P}_{i}[i]~\After<{e}:{Q}},{{qp}:\emptyset},{Q'}}\right)%
\]
For a time-consuming action of $t$ to occur on $P$ it is required that $\Phi_t$ is defined for all parallel components in $P$. Note that, if  ${qp}$ was not empty, then time could not pass since $\PFuncNEQ[P] = \PFuncWait[P] = \{p\}$. 
Set $t = n+1$ so that we can observe the expiring of the timeout. The evaluation of \PFuncTime*_{t}[{\PCalc{\On{p}\Recv{l}[\pl]:{P}_{i}[i]~\After<{e}:{Q}}}]\ results in the evaluation of \PFuncTime*_{1}[Q]. 
If $Q= \PrcEnd$ (or similarly $\PrcErr$, a delay, or a timeout with $n>0$) and \PFuncTime*_{t}[Q'] is defined then time passing of $t$ is possible and: 
 \[\PFuncTime_{t}[P]{}{} = (\nu pq) (\PrcEnd \mid {{qp}:\emptyset}\mid \PFuncTime_{t}[Q'] )\]
If $Q$ is a sending process then \PFuncTime*_{t}[Q]\ would be undefined, and hence \PFuncTime*_{t}[P]. 
\end{example}

\section{Expressiveness}\label{sec:expressiveness}
In this section we reflect on the expressiveness of our mixed-choice extension, particularly in regard to \cite{Bocchi2019}, using examples to illustrate differences.
Furthermore, given the increase in expressiveness, we discuss how type-checking becomes more interesting with the inclusion of \texttt{receive\text{-}after}.

\subsection{Missing deadlines}
The process corresponding to $?a~(\mathtt{true},\emptyset).S$ is merely \PCalc*{\On{p}~\Recv{a}:{P}}\!, which waits to receive $a$ forever.
By way of contrast, $?a~(x<3,~\emptyset).S$, cannot receive when $x\geq 3$, requiring the process to take the form: $\PCalc{\On{p}~\Recv{a}:{P}~\After<{3}:{Q}}$, where $Q=\PrcErr$.
More generally, if an action is enabled when $x\geq 3$: \[\{?a~(x<3,~\emptyset).S,~?b~(3< x<5,~\emptyset).{S}'\}\] then, amending the previous process, \PCalc*{Q=\On{p}~\Recv{b}:{{P}'}~\After<{2}:{\PrcErr}}.



\subsection{Ping-pong protocol}
The example in this section illustrates the usefulness of time-sensitive conditional statements. The ping-pong protocol consists of two participants exchanging messages between themselves on receipt of a message from the other~\cite{Lagaillardie2022}.
One interpretation of the protocol is the following:
\[
\mu\alpha.%
\begin{array}[c]\{{l}\}
    {!ping(x\leq3,~\{x\})}.
    \begin{array}[c]\{{l}\}
        ?ping(x\leq3,~\{x\}).\alpha\\
        ?pong(x>3,~\{x\}).\alpha
    \end{array} \\[-1ex]\\
    {!pong(x>3,~\{x\})}.
    \begin{array}[c]\{{l}\}
        ?ping(x\leq3,~\{x\}).\alpha\\
        ?pong(x>3,~\{x\}).\alpha
    \end{array} 
\end{array}
\]
where each participant exchanges the role of sender, either sending ping early, or pong late.
Without time-sensitive conditional statements, the setting in~\cite{Bocchi2019} only allows implementations where the choice between the `ping' and the `pong' branch are hard-coded. 
In presence of non-deterministic delays (e.g., $\mathtt{delay}(z<6)$), the hard-coded choice can only be for the latest branch to `expire', and the highlighted fragment of the ping-pong protocol above could be naively implemented as follows (omitting $Q$ for simplicity):
\[
\mathtt{def\ }{\RecVar\left(\RecSetMsg;\RecSetTimers;\RecSetRoles\right)}={P}~\mathtt{in\ } P \qquad
P = \PCalc{\mathtt{delay}(z<6).~\On{p}\Send{pong}.{Q}}
\]
The choice of sending ping is \emph{always} discarded as it may be unsatisfied in \emph{some} executions. 
The calculus in this paper, thanks to the time-awareness deriving from a program timer $y$, allows us to \emph{potentially} honour each branch, as follows:
\[
\mathtt{def\ }{\RecVar\left({\RecSetMsg;\RecSetTimers;\RecSetRoles}\right)}={P}~\mathtt{in\ } P%
\qquad 
P = {\mathtt{set}(y)}.\PCalc{\mathtt{delay}(z<6).\begin{array}[t]{l}\If{y\leq 3}\\\Then{\On{p}\Send{ping}.{Q}}\\\Else{\On{p}\Send{pong}.{Q'}}\end{array}}
\]

\subsection{Mixed-choice Ping-pong protocol}\label{ssec:mixed-choice_ping_pong}
An alternative interpretation of the ping-pong protocol can result in an implementation with mixed-choice, as shown below:
\[
\mu\alpha.%
\begin{array}[c]\{{l}\}
    ?ping(x\leq3,~\{x\}).
    \begin{array}[c]\{{l}\}
        !pong(x\leq3,~\{x\}).\alpha
        \\ ?timeout(x>3,~\emptyset).end
    \end{array} \\[-1ex]\\
    !pong(x>3,~\{x\}).
    \begin{array}[c]\{{l}\}
        ?ping(x\leq3,~\{x\}).\alpha
        \\ !timeout(x>3,~\emptyset).end
    \end{array} \\
\end{array}
\]
where pings are responded by pongs and vica versa.
Notice that if a timely ping is not received, a pong is sent instead, which if not responded to by a ping, triggers a timeout.
Similarly, once a ping has been received, a pong must be sent on time to avoid a timeout.
Such a convoluted protocol can be fully implemented:
\[\resizebox{\textwidth}{!}{$\PCalc{%
    \begin{array}[t]{l}%
       \mathtt{def\ }{\RecVar\left({\RecSetMsg;\RecSetTimers;\RecSetRoles}\right)}={P}~\mathtt{in\ } P%
\\\qquad%
P =  {\mathtt{set}(x)}.%
        \begin{array}[t]{l}
            \On{p}~\Recv*{ping}:{%
                {\mathtt{set}(x)}.%
                \begin{array}[t]{l}%
                    \If{x\leq 3}%
                    \\\Then{%
                        \On{p}~\Send{pong}.{\RecVar\left\langle{\RecSetMsg';\RecSetTimers';\RecSetRoles'}\right\rangle}%
                    }\\\Else{%
                        \On{p}~\Recv*{timeout}:{\PrcEnd}%
                    }%
                \end{array}}%
            \\\After<{3-x}:{%
                \On{p}~\Send{pong}.{%
                    \begin{array}[t]{l}%
                        {\mathtt{set}(x)}.%
                        \begin{array}[t]{l}%
                            \On{p}~\Recv*{pong}:{\RecVar\left\langle{{\RecSetMsg}'';{\RecSetTimers}'';{\RecSetRoles}''}\right\rangle}%
                            \\\After<{3-x}:{%
                                \On{p}~\Send{timeout}.{\PrcEnd}%
                            }%
                        \end{array}%
                    \end{array}%
                }%
            }%
        \end{array}%
    \end{array}%
}$}\]
%

\subsection{Message throttling}
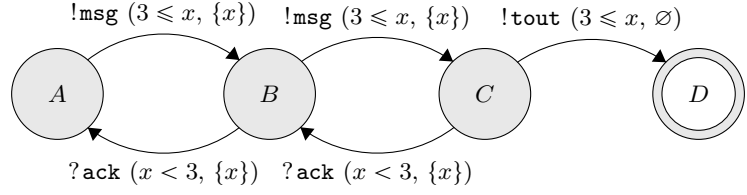
\begin{figure}[t]
    \centering
\resizebox{0.8\textwidth}{!}{%
	\begin{tikzpicture}[scale=0.2]
		\tikzstyle{every node}+=[inner sep=0pt]
		\draw [black,fill={rgb:black,1;white,10}] (7.2,-7.8) circle (3);
		\draw (7.2,-7.8) node {$A$};
		\draw [black,fill={rgb:black,1;white,10}] (21.1,-7.8) circle (3);
		\draw (21.1,-7.8) node {$B$};
		\draw [black,fill={rgb:black,1;white,10}] (35.2,-7.8) circle (3);
		\draw (35.2,-7.8) node {$C$};
		\draw [black,fill={rgb:black,1;white,10}] (49.2,-7.8) circle (3);
		\draw [black,fill=white] (49.2,-7.8) circle (2.4);
		\draw (49.2,-7.8) node {$D$};
		\draw [black] (9.169,-5.559) arc (128.05694:51.94306:8.08);
		\fill [black] (19.13,-5.56) -- (18.81,-4.67) -- (18.19,-5.46);
		\draw (14.15,-3.34) node [above] {$\TypSend\!\;{\mathtt{msg}}\mbox{ }(3\leq x,\,\{x\})$};
		\draw [black] (19.111,-10.023) arc (-52.39041:-127.60959:8.129);
		\fill [black] (9.19,-10.02) -- (9.52,-10.91) -- (10.13,-10.12);
		\draw (14.15,-12.21) node [below] {$\TypRecv\!\;{\mathtt{ack}}\mbox{ }(x<3,\,\{x\})$};
		\draw [black] (23.183,-5.663) arc (125.6468:54.3532:8.522);
		\fill [black] (33.12,-5.66) -- (32.76,-4.79) -- (32.18,-5.6);
		\draw (28.15,-3.57) node [above] {$\TypSend\!\;{\mathtt{msg}}\mbox{ }(3\leq x,\,\{x\})$};
		\draw [black] (33.195,-10.00) arc (-52.57665:-127.42335:8.302);
		\fill [black] (23.11,-10.00) -- (23.44,-10.89) -- (24.04,-10.1);
		\draw (28.15,-12.22) node [below] {$\TypRecv\!\;{\mathtt{ack}}\mbox{ }(x<3,\,\{x\})$};
		\draw [black] (37.33,-5.709) arc (124.48082:55.51918:8.602);
		\fill [black] (47.07,-5.71) -- (46.69,-4.84) -- (46.13,-5.67);
		\draw (42.2,-3.7) node [above] {$\TypSend\!\;{\mathtt{tout}}\mbox{ }(3\leq x,\,\emptyset)$};
	\end{tikzpicture}
}%
    \caption{Message throttling protocol for $m=2$.}
    \label{fig:message_throttling}
    \vspace{-10pt}
\end{figure}
A real-world application of the previous example is \emph{message throttling}.
The rationale behind message throttling is to cull unresponsive processes, which do not keep up with the message processing tempo set by the system.
This avoids a server from becoming overwhelmed by a flood of incoming messages.
In such a protocol, upon receiving a message, a participant is permitted a grace period to respond before receiving another.
The grace period is specified as a number of unacknowledged messages, rather than a period of time.
Below we present a fully parametric implementation of this behaviour, where $m$ is the maximum number of messages that can go unacknowledged before a timeout is issued.
\[
\begin{array}[c]{lcl}
S_0 & = & \mu\alpha^{0}.!msg(x \geq 3, \{x\}). S_1
\\[-1ex]\\ %
S_i & = & \mu\alpha^{i}.
\begin{array}[c]\{{l}\}
?ack(x < 3, \{ x \}). \alpha^{i-1},
~!msg(x \geq 3, \{ x \}) . S_{i+1}
\end{array}
\\[-1ex]\\ %
S_m & = &
\begin{array}[c]\{{l}\}
?ack(x < 3, \{ x \}). \alpha^{m-1},
~!tout(x \geq 3,~\emptyset) . \TypeEnd
\end{array}
\end{array}
\]

\noindent which has the corresponding processes:
\begin{processcalculus}*
    \begin{array}{lclcl}
    \mathtt{def\ }{\RecVar_0\left({\RecSetMsg_0;\RecSetTimers_0;\RecSetRoles_0}\right)}
    & = & {P_0} & \mathtt{in\ } & \On{p}~\Send{msg}.{P_1}
    \\[-1.5ex]\\ %
   \mathtt{def\ }{\RecVar_i\left({\RecSetMsg_i;\RecSetTimers_i;\RecSetRoles_i}\right)}
   & = & {P_i} & \mathtt{in\ } & 
        \begin{array}[t]{l}
            \On{p}~\Recv*{ack}:{\RecVar_{i-1}\left\langle{\RecSetMsg_{i-1};\RecSetTimers_{i-1};\RecSetRoles_{i-1}}\right\rangle}%
         \\\After<{3}:{%
            \On{p}~\Send{msg}.{P_{i+1}}%
            }
        \end{array}
    \\[-2ex]\\ %
    & & P_m & = &
    \begin{array}[t]{l}
        \On{p}~\Recv*{ack}:{P_{m-1}}
         \\\After<{3}:{\On{p}~\Send{tout}.{\PrcEnd}}
    \end{array}
    \end{array}
\end{processcalculus}\vspace{-0.5ex}

\noindent The system shown in~\cref{fig:message_throttling} illustrates the system for the $m=2$ instance. 
Arguably, instead of sending the message $tout$, it would also be equally valid for the system to simply reach an error state:
\PCalc*{P_m = \On{p} ~\Recv*{ack}:{P_{m-1}} ~\After<{3}:{\PrcErr}}.

\section{Concluding Discussion}\label{sec:conclusion}
We have shown how timing constraints provide an intuitive way of integrating mixed-choice into asynchronous session types. 
The desire for mixed-choice has already prompted work in (untimed) synchronous session types~\cite{Vasconcelos2020}.
Further afield, coordination structures have been proposed that overlap with mixed-choice, for example, fork and join~\cite{Denielou2012a}, which permit messages within a fork (and its corresponding join) to be sent or received in any order; reminiscent of mixed-choice.
Affine sessions~\cite{Lagaillardie2022,Mostrous2018} support exception handling by enabling an end-point to perform a subset of the interactions specified by their type, but there is no consideration of time, hence timeouts.
Before session types gained traction, timed processes~\cite{Berger2007} were proposed for realising timeouts, but lacked any notion of a counterpart for timeouts.

We have integrated the notion of mixed-choice with that of time-constraints.
There are many conceivable ways to realise mixed-choice using programming primitives. 
However, our integration with time, embodied in TOAST, offers new capabilities for modelling timeouts which sit at the heart of protocols and are a widely-used idiom in programming practice. To provide a bridge to programming languages, we provide a timed session calculus enriched with a \texttt{receive\text{-}after} pattern and process timers, the latter providing a natural counterpart to the former. Taken altogether, we have lifted a long-standing restriction on asynchronous session types by allowing for safe mixed-choice, through the judicious application of timing constraints.

Future work will provide type checking against TOAST for our new processes, and establish time-safety (a variant of type-safety which ensures punctuality of interactions via subject reduction) for well-typed processes.
Time-safety for timed session types~\cite{Bocchi2019,Bocchi2014} (without mixed-choice) relies on a progress property called receive-liveness, which is defined on the untimed counterpart of a timed process. Receive-liveness that can be checked with existing techniques for global progress \cite{Bettini2008,DezaniCiancaglini2007}. 
A progress property may seem too strong a precondition for ensuring time-safety. In untimed formulations of session types, type-safety and subject reduction do not depend on progress. Arguably, when considering time and punctuality, the distinction between progress and safety is no longer clear-cut, since deadlocks may cause violation of time constraints.



%

\bibliographystyle{splncs04.bst}
\bibliography{main.bib}




\end{document}